\documentclass[structabstract]{aa}  
\usepackage{amssymb}
\usepackage{amsmath}
\usepackage{amsfonts}
\usepackage[latin1]{inputenc}
\usepackage{graphicx}
\usepackage{epsfig}
\usepackage{subfigure}
\usepackage[authoryear]{natbib}
\usepackage{rotating}
\usepackage{txfonts}
\newcommand{\teff}{\ifmmode T_{\rm eff} \else $T_{\mathrm{eff}}$\fi}
\newcommand{\logg}{\ifmmode \log g \else $\log g$\fi}
\newcommand{\lL}{\ifmmode \log \frac{L}{L_{\odot}} \else $\log \frac{L}{L_{\odot}}$\fi}

\newcommand{\vsini}{\ifmmode v \sin i \else $v \sin i$\fi}
\newcommand{\vinf}{$v_{\infty}$}

\newcommand{\vmac}{\ifmmode v_{\rm mac} \else $v_{\rm mac}$\fi}

\newcommand{\kms}{km~s$^{-1}$}
\newcommand{\msun}{\ifmmode M_{\odot} \else $M_{\odot}$\fi}
\newcommand{\zsun}{\ifmmode Z_{\odot} \else $Z_{\odot}$\fi}
\newcommand{\lsun}{\ifmmode L_{\odot} \else $L_{\odot}$\fi}
\newcommand{\rsun}{\ifmmode R_{\odot} \else $R_{\odot}$\fi}
\newcommand{\qh}{\ifmmode Q_{\rm H} \else $Q_{\rm H}$\fi}
\newcommand{\qhei}{\ifmmode Q_{\ion{He}{i}} \else $Q_{\ion{He}{i}}$\fi}

\begin{document}

   \title{A spectroscopic investigation of the O-type star population in four Cygnus OB associations\thanks{Based on observations collected at the Haute-Provence and San Pedro M{\`a}rtir Observatories and with FUSE and IUE missions.}}

   \subtitle{ II. Determination of the fundamental parameters}

   \author{L. Mahy\inst{1}\fnmsep\thanks{Postdoctoral Researcher F.R.S-FNRS}
          \and
          G. Rauw\inst{1}
          \and
          M. De Becker\inst{1}
          \and
          P. Eenens\inst{2}
          \and
          C. A. Flores\inst{2}
          }

   \offprints{L. Mahy}

   \institute{
     Institut d'Astrophysique et de G\'eophysique, Universit\'e de Li\`ege, B\^at. B5C, All\'ee du 6 Ao\^ut 17, B-4000, Li\`ege, Belgium\\
     \email{mahy@astro.ulg.ac.be}
     \and 
     Departamento de Astronom\'ia, Universidad de Guanajuato, Apartado 144, 36000 Guanajuato, GTO, Mexico
   }
   
   \date{Received ...; accepted ...}
   
 
  \abstract
   {}
   {Having established the binary status of nineteen O-type stars located in four Cygnus OB associations, we now determine their fundamental parameters to constrain their properties and their evolutionary status. We also investigate their surface nitrogen abundances, which we compare with other results from the literature obtained for galactic O-type stars. }
   {Using optical spectra collected for each object in our sample and some UV data from the archives, we apply the CMFGEN atmosphere code to determine their main properties. For the binary systems, we have disentangled the components to obtain their individual spectra and  investigate them as if they were single stars.}
   {We find that the distances of several presumably single O-type stars seem poorly constrained because their luminosities are not in agreement with the "standard" luminosities of stars with similar spectral types. The ages of these O-type stars are all less than 7~Myrs. Therefore, the ages of these stars agree with those, quoted in the literature, of the four associations, except for Cyg\,OB8 for which the stars seem older than the association itself. However, we point out that the distance of certain stars is debatable relative to values found in the literature. The N content of these stars put in perspective with N contents of several other galactic O-type stars seems to draw the same five groups as found in the "Hunter" diagram for the O and B-type stars in the LMC even though their locations are obviously different. We determine mass-loss rates for several objects from the H$\alpha$ line and UV spectra. Finally, we confirm the "mass discrepancy" especially for O stars with masses smaller than $30~M_{\odot}$. }
   {}

   \keywords{Stars: early-type - Stars: binaries: spectroscopic - Open clusters and associations: individual: Cygnus OB1 - Open clusters and associations: individual: Cygnus OB3 - Open clusters and associations: individual: Cygnus OB8 - Open clusters and associations: individual: Cygnus OB9}
   \titlerunning{The fundamental parameters of O-type star population in four Cygnus OB associations}
   \authorrunning{L. Mahy et al.}
   \maketitle


\section{Introduction}
\label{sect:intro}

Massive stars are the engines of the galaxies. This small part of the stellar population has a considerable influence on its environment both mechanically and by radiation. Indeed, their strong ionizing fluxes power the \ion{H}{ii} regions. Furthermore, their powerful winds and their deaths as supernovae enrich their surroundings with heavy chemical elements and reshape the interstellar medium, triggering new generations of stars. Because their lifetime is relatively short, massive stars are excellent tracers of star formation. Therefore, constraining their properties gives us the opportunity to better understand how they form and how they evolve. 

The  fates of these stars are mainly governed by their initial mass, but also to some extent by their rotation and their mass loss. In addition to giving more oblate shapes to the stars, the rotation modifies their temperature gradient at the surface and also induces an internal mixing that affects the angular momentum and chemical element transport. The rotation thus influences the surface abundances and the lifetimes of the massive stars on the main sequence \citep{mm00}. Several observational studies analysed these two aspects of rotation. On the one hand, \citet{hun08,hun09} reported that a large part of B-type stars in the Large Magellanic Cloud (LMC) presented a clear trend between their projected rotational velocity (\vsini) and their surface nitrogen content. This trend was detected in about 60\% of the B star population. The remaining 40\% consists of B stars with slow rotation and high nitrogen enrichment and unenriched B stars with fast rotation. These groups are currently not explained by the evolutionary models of single massive stars which include rotational mixing. On the other hand, numerous investigations have shown that the rotation brings modifications to the evolutionary tracks and to the isochrones of massive stars \citep[see e.g.][]{mm00,mm03,brott11}. Besides rotation, the stellar winds also play a crucial role in the evolution of massive stars. Indeed, these objects lose a significant fraction of their mass during their life. However, the exact quantity of ejected mass by a massive star remains difficult to determine because the true structure of these winds, and notably their homogeneity, is still poorly known. Although direct \citep{eversberg98} and indirect \citep{bouret05} evidence highlighted the presence of outward-moving inhomogeneities in the winds of O-type stars, the shapes, the sizes, and the optical depths of these inhomogeneities (or clumps) still remain unanswered questions. Investigations of the structure of these winds seem, however, to indicate that the clumps are spherical \citep{sun11,her12} rather than flattened as proposed by \citet{feld03}. The presence of clumps in the stellar winds thus modifies the determination of mass-loss rates of massive stars since many mass-loss diagnostics are dependent on the density squared (such as notably H$\alpha$).

To bring more constraints on these properties, we have determined and analysed the fundamental parameters of nineteen O-type stars belonging to four OB associations in the Cygnus complex. This area is an active star-forming region that includes a huge number of massive stars. These targets have already been presented in \citet[][hereafter Paper I]{mah13} who  established their multiplicity and analysed the distribution of the orbital parameters of the binary systems. Because we know their multiplicity, in the present paper we can be more accurate on the determination of valuable physical parameters of these objects. Although all these targets are formed in the Cygnus region, they constitute a non-homogeneous sample of main-sequence band, giant, and supergiant massive stars. Among this sample of stars, we also detected in Paper I four binary systems: three SB2s and one SB1, all with orbital periods shorter than 10 days. Therefore, for the SB2s, we disentangled the observed spectra of these systems to obtain the individual spectrum of each component to investigate it with an atmosphere code as we do for single stars. 
 
In the following section, we present the observations of these targets. The atmosphere code and the main UV/optical diagnostics are described in Sect.~\ref{sec:mod}. The results and a discussion are given in Sect.~\ref{sec:res} and in Sect.~\ref{sec:dis}, respectively. Finally, our conclusions are provided in Sect.~\ref{sec:conc}.


\section{Observations and data reduction}
\label{sect:obs}

We  focus on the same data as presented in Paper I. In addition, we retrieve the IUE spectra (SWP with a dispersion of 0.2~\AA) to model the stellar winds of five objects. These data were obtained between 1979 and 1992 with exposure times from 650 s to 8400~s. These spectra cover the [1200--1800]~\AA\ wavelength domain, containing the \ion{C}{iv}~1548--1550 and \ion{N}{v}~1240 resonance P-Cygni profiles as well as the \ion{O}{v}~1371 line, used to estimate the wind clumping in early O stars. We also add to our sample four FUSE spectra retrieved from the archives. Because of the interstellar absorption, we only focus on the [1100--1200]~\AA\ wavelength region, and mainly on the \ion{P}{v}~1118--1128, \ion{C}{iv}~1169, and \ion{C}{iii}~1176 lines. Unfortunately, the FUV and UV spectra are not available for all the stars in our sample. Therefore, the mass-loss rates of the stars for which none of these spectra exists will only be derived on the basis of the H$\alpha$ line.   

\begin{sidewaystable*}[ph!]
\caption{Journal of optical and UV observations}  
\label{tab:obs}
\begin{tabular}{l|cccc|ccc|ccc}
\hline\hline
 & \multicolumn{4}{c|}{Optical}  & \multicolumn{3}{c}{IUE}& \multicolumn{3}{c}{FUSE}\\
\hline
 Star & Instrument & Spectral range & HJD & Obs. date & data ID & HJD & Obs. date & data ID & HJD & Obs. date \\
  &  & [\AA] & [d] &  &  & [d] &  \\
\hline
\multicolumn{11}{c}{Cyg\,OB1}\\
\hline
HD\,193443 & -- & [4450--4900] & \multicolumn{2}{c|}{disentangled spectra} &  -- & -- & -- &  -- & -- & -- \\
HD\,228989 & -- & [4450--4900] & \multicolumn{2}{c|}{disentangled spectra} &  -- & -- & -- &  -- & -- & -- \\
HD\,229234 & Espresso & [4000--6750] & 2\,454\,992.7849 & 10 Jun 2009 &    --    &              --  &         --  &  -- & -- & -- \\
HD\,193514 & Elodie      & [4000--6800] & 2\,453\,600.4087 & 17 Aug 2005 & SWP18145 & 2\,445\,241.2984 & 28 Sep 1982 & E0820701 & 2\,453\,165.3275 & 08 Jun 2004\\
HD\,193595 & Espresso    & [4000--6750] & 2\,454\,989.8582 & 07 Jun 2009 &    --    &              --  &         --  &  -- & -- & -- \\
HD\,193682 & Espresso    & [4000--6750] & 2\,454\,989.9081 & 07 Jun 2009 & SWP09022 & 2\,444\,376.4267 & 16 May 1980 & E0820301 & 2\,453\,365.5448 & 09 Jun 2004\\
HD\,194094 & Espresso    & [4000--6750] & 2\,455\,820.7063 & 16 Sep 2011 &    --    &              --  &         --  &  -- & -- & -- \\
HD\,194280 & Aur{\'e}lie & [4450--4900] & 2\,454\,407.2709 & 02 Nov 2007 &    --    &              --  &         --  &  -- & -- & -- \\
HD\,228841 & Espresso    & [4000--6750] & 2\,454\,989.8092 & 07 Jun 2009 &    --    &              --  &         --  &  -- & -- & -- \\
\hline
\multicolumn{11}{c}{Cyg\,OB3}\\
\hline
HD\,190864 & Elodie      & [4000--6800] & 2\,452\,134.4733 & 12 Aug 2001 & SWP06946 & 2\,44\,4168.7353 & 22 Oct 1979 &  E0820501 & 2\,453\,148.7486 & 23 May 2004\\
HD\,227018 & Aur{\'e}lie & [4450--4900] & 2\,454\,717.3188 & 07 Sep 2008 &    --    &              --  &         --  &  -- & -- & -- \\
HD\,227245 & Espresso    & [4000--6750] & 2\,455\,727.7497 & 15 Jun 2011 &    --    &              --  &         --  &  -- & -- & -- \\
HD\,227757 & Espresso    & [4000--6750] & 2\,455\,726.7373 & 14 Jun 2011 &    --    &              --  &         --  &  -- & -- & -- \\
\hline
\multicolumn{11}{c}{Cyg\,OB8}\\
\hline
HD\,191423 & Elodie      & [4000--6800] & 2\,453\,247.3339 & 29 Aug 2004 & SWP16212 & 2\,445\,000.2270 & 30 Jan 1982 & E0821301 & 2\,453\,165.7979 & 09 Jun 2004\\
HD\,191978 & Elodie      & [4000--6800] & 2\,453\,247.4196 & 29 Aug 2004 & SWP46532 & 2\,448\,976.0517 & 19 Dec 1992 &  -- & -- & -- \\
HD\,193117 & Espresso    & [4000--6750] & 2\,454\,991.9784 & 09 Jun 2009 &    --    &              --  &         --  &  -- & -- & -- \\
\hline
\multicolumn{11}{c}{Cyg\,OB9}\\
\hline
HD\,194334 & Espresso    & [4000--6750] & 2\,455\,053.7053 & 10 Aug 2009 &    --    &              --  &         --  &  -- & -- & -- \\
HD\,194649 & -- & [4450--4900] & \multicolumn{2}{c|}{disentangled spectra} &  -- & -- & -- &  -- & -- & -- \\
HD\,195213 & Aur{\'e}lie & [4450--4900] & 2\,454\,717.5787 & 07 Sep 2008 &    --    &              --  &         --  &  -- & -- & -- \\
\hline
\end{tabular}
\end{sidewaystable*}

For  the targets detected as binary systems in Paper I, we used the disentangled spectra corrected for the brightness ratio which were already presented in Paper I. We note that we used a programme based on the \citet{gl06} technique to compute the individual spectra of both components of a binary system, but also to refine the radial velocities by applying a cross-correlation technique. These spectra can thus be considered as mean spectra for the components of the binary systems. To have as many diagnostic lines as possible, we favour the optical spectra with the widest wavelength coverage, as well as the highest spectral resolution and the highest signal-to-noise ratio. The journal of optical and FUV/UV observations is presented in Table~\ref{tab:obs}.


\section{Modelling}
\label{sec:mod}

We use the code CMFGEN \citep{hm98} for the quantitative analysis of the optical and FUV/UV spectra. CMFGEN provides non-LTE atmosphere models including winds and line-blanketing. CMFGEN needs as input an estimate of the hydrodynamical structure that we construct from TLUSTY models (taken from the OSTAR2002 grid of \citealt{lh03}) connected to a $\beta$ velocity law of the form $v = v_{\infty}(1 - R/r)^{\beta}$, where $v_{\infty}$ is the wind terminal velocity.  For the stars having only optical spectra, we adopted $\beta = 0.8$, which represents a typical value for O dwarfs \citep[see e.g.][]{rep04}. Our final models include the following chemical elements: \ion{H}{i}, \ion{He}{i-ii}, \ion{C}{ii-iv}, \ion{N}{ii-v}, \ion{O}{ii-vi}, \ion{Ne}{ii-iii}, \ion{Mg}{ii}, \ion{Si}{ii-iv}, \ion{S}{iii-vi}, \ion{P}{iv-v}, \ion{Ar}{iii-iv}, \ion{Al}{iii}, \ion{Fe}{ii-vii}, and \ion{Ni}{iii-v} with the solar composition of \citet{gas07} unless otherwise stated. CMFGEN also uses the super-level approach to reduce the memory requirements. On average, we include about 1600 super levels for a total of 8000 levels. For the formal solution of the radiative transfer equation leading to the emergent spectrum, a microturbulent velocity varying linearly with velocity from 10~\kms\ to $0.1 \times v_{\infty}$ was used. We include X-ray emission in the wind since this can affect the ionization balance and the strength of key UV diagnostic lines. In practice, we adopt a temperature of three million degrees and we adjust the flux level so that the X-ray flux coming out of the atmosphere matches the observed $L_{\mathrm{X}}/L_{\mathrm{bol}}$ ratio. We simply adopt the canonical value $L_{\mathrm{X}}/L_{\mathrm{bol}} = 10^{-7}$ \citep{san06b,naze09}.
In practice, we proceed as follows to derive the stellar and wind parameters.
\begin{itemize}
\item Effective temperature: we use the classical ratio between the strengths of \ion{He}{i} and \ion{He}{ii} lines to determine \teff. The main indicators are the \ion{He}{i}~4471 and \ion{He}{ii}~4542 lines, but additional diagnostics can be built with the \ion{He}{i}~4026, \ion{He}{i}~4389, \ion{He}{i}~4713, \ion{He}{i}~4921, \ion{He}{i}~5876, \ion{He}{ii}~4200, and \ion{He}{ii}~5412 lines. When possible we also use  the \ion{C}{iv}~1169 to \ion{C}{iii}~1176 line ratio, which has  been shown to provide a useful temperature diagnostic \citep{heap06}. The typical uncertainty on the \teff\ determination is 1000~K, except for the binary components for which the uncertainty on \teff\ is  2500~K.
\item Gravity: the wings of the Balmer lines H$\beta$, H$\gamma$, and H$\delta$ are the main indicators of \logg. Generally, an accuracy of about 0.1~dex on \logg\ is achieved. However, because of side effects of the disentangling programme, \logg\ of the binary components has an uncertainty of 0.25~dex.
\item Wind terminal velocity: the blueward extension of UV P-Cygni profiles provides $v_{\infty}+v_{\mathrm{max}}$ where $v_{\mathrm{max}}$ is the maximum turbulent velocity. Using the above relation for microturbulent velocity gives us a direct determination of \vinf\ with an accuracy of 100~\kms. For stars without UV spectra, we use as input the standard values given by \citet{pri90} according to their spectral types.
\item  Mass-loss rate: we use two diagnostics to constrain the mass-loss rate, the UV P-Cygni profiles and H$\alpha$. For the UV domain, we focus mainly on \ion{N}{v}~1240, \ion{Si}{iv}~1394--1403, and \ion{C}{iv}~1548--1551, and when FUSE spectra are available, we consider \ion{C}{iv}~1169 and \ion{C}{iii}~1176 as additional diagnostics to those lines for mass-loss rate determination. A single value of $\dot{M}$ should allow a good fit of both types of lines even though a study by \citet{mar12} has shown that this was not always the case. In our analysis, we reach a general agreement between UV and H$\alpha$ lines by playing  on other parameters such as  the clumping or the $\beta$. However, when no FUV/UV data are available, we fix both values to 1.0 (homogeneous model) and 0.8 \citep[see][]{rep04}, respectively. We also stress that some objects were only observed on [4450--4900] \AA\ wavelength domain. Since no sufficient diagnostic exists to determine the mass-loss rate on this region, we also use as input the values provided by \citet{muijres2012} as mass-loss rate values for our models.
\end{itemize}

Given the complexity of the parameter space for CMFGEN models, it is not possible for us to describe all the effects of every parameter change on the synthetic spectra, but we refer to dedicated papers such as \citet{hil03}, \citet{bouret05}, or \citet{mar11}. In our analysis, we decide to vary the different parameters until we obtain the solution which provided the best $\chi^2$ fit. To this end we  generated a non-uniform grid composed of between 10 and 40 models depending on the wavelength coverage of the observed spectra (and thus of the number of parameters to determine). From these best-fit models, we reproduce the UV-Optical-Infrared spectral energy distribution (SED) for each star (Fig.~\ref{SED_1}) from the UBVJHK fluxes (see Table~\ref{tab:phot}) and we include the UV fluxes for the stars that have FUSE and/or IUE spectra to constrain the distance and the extinction. The galactic reddening law of \citet{car89} is used. We also derive $R_V$ when the UV spectrum is available. When this is not the case, we use $R_V$ values from \citet{pat01,pat03} and if no value exists, we fix it to 3.1. We then recompute the $M_V$ with the new extinction and the new distance values and we derive the luminosities on the basis of the $M_V$ and of the bolometric corrections given in \citet{mar06}. Once the new luminosity is obtained, we compute a new model with this value to improve the best-fit models. The uncertainties calculated on the luminosities are mainly due to the poorly-known distances. To quantify these uncertainties, we computed the largest difference between the obtained luminosity and the luminosity computed with the minimum or maximum mean distances of the associations listed in \citet{hum78}.

The  projected rotational velocities of the stars are obtained from the Fourier transform method \citep{sim07}. The synthetic spectrum is then convolved by \vsini. 
The rotationally broadened synthetic line profiles usually provided a good match to the observed profile for the fast rotators. In the case of moderately to slowly rotating objects, some amount of macroturbulence has to be introduced to correctly reproduce the line profiles of several features (e.g. \ion{He}{i}~4713, \ion{C}{iv}~5812, \ion{He}{i}~5876). The need for extra broadening is well documented \citep{how97,ryans2002,howarth07, nieva07,martins10,fraser10}, but its origin is unclear. \citet{aerts09} notably suggested that non-radial pulsations could trigger large-scale motions (hence macroturbulence), but this needs to be confirmed by a study covering a wider parameter space. A study by \citet{sim10} also showed that the amplitude of macroturbulence was correlated to the amplitude of line profile variability in a sample of OB supergiants. In practice, we introduce macroturbulence by convolving our rotationally broadened synthetic spectra with a Gaussian profile, thus mimicking isotropic turbulence. This is obviously a very simple approach, but it significantly improves the quality of the fits. In practice, we used \ion{He}{i}~4713 as the main indicator
of macroturbulence since it is present with sufficient signal-to-noise ratio in all the stars in our sample. Moreover, secondary indicators are the \ion{O}{iii}~5592 and \ion{He}{i}~5876 lines.

To determine the nitrogen content of our stars, we rely mainly on the \ion{N}{iii} lines between 4500 and 4520~\AA. They are present in absorption in the spectra of all stars, they are not affected by winds and they are strong enough for the abundance determination. The uncertainties are of the order of 50\%. They are estimated from the comparison of the selected \ion{N}{iii} lines to models with various N content. The uncertainties do not take any systematics related to atomic data into account.

When UV spectra are available (both FUSE and IUE), the degree of inhomogeneities of the stellar winds is determined. Clumping is implemented in CMFGEN by means of a volume filling factor $f$ following the law $f = f_{\infty} + (1 - f_{\infty})e^{-v/v_{\mathrm{cl}}}$, where $f_{\infty}$ is the maximum clumping factor at the top of the atmosphere and $v_{\mathrm{cl}}$ a parameter indicating the position where the wind starts to be significantly clumped. As shown by \citet{bouret05}, \ion{P}{v}~1118--1128, \ion{O}{v}~1371, and \ion{N}{iv}~1720 are UV features especially sensitive to wind inhomogeneities. We use these lines to constrain $f_{\infty}$ in the stars in our sample. Unfortunately, the last two lines are mainly good indicators for early O-type stars. Therefore, we are not able to use these two lines to constrain the clumping factor for HD\,191978 given the absence of a FUSE spectrum.


\section{Results}
\label{sec:res}
The derived stellar and wind parameters are listed in Table~\ref{cygnus_param}. The CMFGEN best-fit models are given in the Appendix. Below, we briefly comment on each star.

\begin{table*}
\caption{UBVJHK photometric parameters of the stars in our sample}
\label{tab:phot}
\centering          
\begin{tabular}{lcccccccccccc}
\hline\hline
Star    &   Cluster &    U      &      B&      V&      J&      H&      K&      R$_V$&   E(B$-$V)& Dist.  &  M$_V$   &  Dist.  \\
        &           &           &       &       &       &       &       &           &           &  (SED) &         &  \citet{hum78}\\
\hline
\multicolumn{13}{c}{Cyg\,OB1}\\
\hline
HD\,193443&          &   7.08$^{b}$&  7.61&  7.26&  6.40&  6.34&  6.34&     3.10&     0.61&    $-$       & -5.93 &    1.82 \\
HD\,193514&          &   7.32$^{a}$&  7.84&  7.43&  6.35&  6.24&  6.18&     3.00&     0.74&   1.71       & -5.96 &    1.90 \\
HD\,193595& Ber\,86  &   8.50$^{b}$&  9.08&  8.78&  7.92&  7.87&  7.85&     3.02&     0.69&   1.85       & -4.64 &    1.82 \\
HD\,193682&          &   8.39$^{b}$&  8.87&  8.42&  7.30&  7.19&  7.14&     2.82&     0.83&   1.95       & -5.37 &    1.82 \\
HD\,194094&          &   9.18$^{b}$&  9.57&  9.05&  7.74&  7.59&  7.58&     2.99&     0.89&   2.65       & -5.73 &    2.42 \\
HD\,194280&          &   8.94$^{a}$&  9.08&  8.42&  6.74&  6.60&  6.49&     2.95&     1.06&   2.67       & -6.84 &    2.20 \\
HD\,228841& Ber\,86  &   9.08$^{a}$&  9.50&  9.01&  7.91&  7.78&  7.75&     2.78&     0.88&   2.70       & -5.59 &    2.09 \\
HD\,228989& Ber\,86  &  10.29$^{b}$& 10.49&  9.83&  8.15&  7.97&  7.90&     3.10&     0.92&    $-$       & -3.81 &    1.43 \\
HD\,229234& NGC\,6913&   9.35$^{d}$&  9.58&  8.94&  7.34&  7.18&  7.10&     2.87&     1.07&   1.50       & -5.02 &    1.59 \\
\hline
\multicolumn{13}{c}{Cyg\,OB3}\\
\hline
HD\,190864& NGC\,6871&   7.20$^{a}$&  7.93&  7.78&  7.26&  7.28&  7.26&     2.45&     0.58&   2.15       & -5.30 &    2.29 \\
HD\,227018& NGC\,6871&   8.73$^{b}$&  9.35&  9.00&  8.04&  7.95&  7.90&     3.28&     0.71&   2.75       & -5.53 &    3.15 \\
HD\,227245&          &   9.81$^{b}$& 10.25&  9.69&  8.21&  8.07&  7.99&     3.26&     0.91&   2.15       & -4.94 &    2.29 \\
HD\,227757&          &   8.72$^{e}$&  9.41&  9.27&  8.77&  8.78&  8.76&     3.06&     0.53&   2.15       & -4.02 &    2.36 \\
\hline
\multicolumn{13}{c}{Cyg\,OB8}\\
\hline
HD\,191423&          &   7.43$^{a}$&  8.19&  8.03&  7.73&  7.72&  7.79&     3.00&     0.46&   2.90       & -5.66 &    2.82 \\
HD\,191978&          &   7.38$^{a}$&  8.15&  8.03&  7.76&  7.80&  7.80&     2.70&     0.48&   2.75       & -5.46 &    2.84 \\
HD\,193117&          &   8.87$^{c}$&  9.29&  8.77&  7.24&  7.09&  6.97&     3.18&     0.87&   2.25       & -5.76 &    2.37 \\
\hline
\multicolumn{13}{c}{Cyg\,OB9}\\
\hline
HD\,194334&          &   9.38$^{c}$&  9.54&  8.82&  6.99&  6.82&  6.74&     2.94&     1.13&   1.19       & -4.88 &    1.04 \\
HD\,194649&          &   9.82$^{b}$&  9.93&  9.07&  6.86&  6.63&  6.51&     3.10&     1.12&    $-$       & -4.80 &    1.20 \\
HD\,195213&          &   9.29$^{b}$&  9.55&  8.82&  6.67&  6.44&  6.26&     3.27&     1.13&   1.00       & -4.88 &    1.20 \\
\hline
\end{tabular}

\tablebib{
$a$:~\citet{mai04}; $b$:~\citet{wes82}; $c$:~\citet{gua92}; $d$:~\citet{fer83};
$e$:~\citet{kre12}.}
\end{table*}

\subsection{The Cyg\,OB1 association}
\hspace{0.5cm}{\it HD\,193443: }In Paper I, we estimated the brightness ratio to $3.9\pm0.4$. The disentangled spectra corrected for this ratio are relatively well fitted. However, the wavelength coverage does not allow us to determine the wind parameters. The individual stellar parameters of the primary are more reminiscent of a giant star, whilst the secondary has the characteristics of a main-sequence object. This confirms the luminosity classes of both components mentioned in Paper I.

{\it HD\,228989: }We found in Paper I a brightness ratio of $1.2\pm0.1$ between the primary and the secondary. The spectral lines of the disentangled spectra are relatively well fitted even though the \ion{He}{ii}~4686 line of both stars stronger than with the synthetic profiles. This is probably due to a poor estimation of the wind parameters, but, as stressed for HD\,193443, the absence of other diagnostic lines prevents us from  providing a good determination of the wind parameters.

{\it HD\,229234: }This object was reported as SB1 in Paper I. However, since the signature of the putative companion is not detectable in the observed spectra, we consider that the these spectra are only produced by the primary star, suggesting that the derived parameters correspond to an upper limit. The fit is relatively good and indicates stellar parameters  similar to  those of a giant star. This agrees with the spectral classification established in Paper I.

{\it HD\,193514: }The parameters derived for HD\,193514 are similar to those of \citet{rep04}. However, the wind parameters are slightly different. Indeed, we need a clumping factor of $0.01$ to correctly reproduce the FUV lines. This thus affects the mass-loss rate of the star even though $\dot{M}_{\mathrm{H}\alpha}/\sqrt{f}$ remains quite close in both analyses. The emissions of the \ion{N}{iii}~4634--41 lines in the spectrum of HD\,193514 are poorly reproduced, but it is probably not due to the N content, given that the triplet at 4500--4520~\AA\ and  the \ion{N}{iv}~1720 line are well fitted. Moreover, we stress that an increase of the mass-loss rate does not improve the quality of the fit. 

{\it HD\,193595: }The fit of this object is of good quality. The core of the H$\alpha$ line is filled in with some emission that is likely to be partially of nebular origin. The stellar parameters agree with a main-sequence luminosity class as established in Paper I. 

{\it HD\,193682: }We use FUSE and IUE spectra to constrain the wind parameters (see Table~\ref{cygnus_param}). The FUSE spectrum is correctly fitted by our CMFGEN model. The IUE spectrum was observed with a low dispersion and the small aperture. Therefore, the quality of the observation is very low and we must strongly convolve the synthetic spectrum to achieve such a quality. The \ion{C}{iv}~1548-1550 doublet appears too saturated in comparison to the observation, whilst the fit of the \ion{N}{iv}~1718 line seems acceptable. Finally, the fit in the optical range is of excellent quality, even though the \ion{He}{i}~5876 line is too strong compared to the observations. Unlike \citet{rep04},  the red wings of the \ion{He}{ii}~4686 and H$\alpha$ lines are rather well reproduced. The line profiles show an important macroturbulence. We find that a combination of \vsini\ of about 150~\kms\ with a macroturbulent velocity of about 70~\kms\ leads to much better fits than do purely rotationally broadened profiles. 

{\it HD\,194094: }No UV spectrum is available for this star, we therefore restrict ourselves to the optical range. The core of the Balmer lines is too deep in our synthetic spectra except for H$\alpha$. Moreover, the \ion{He}{ii}~4686 is poorly fitted and appears stronger in the synthetic model. 

{\it HD\,228841: }A good fit of the optical wavelength domain is achieved, except for the \ion{He}{ii}~4686 and the \ion{He}{i}~5876 lines. The best fit is obtained with profiles only broadened by rotation, i.e. without any macroturbulence. 

{\it HD\,194280: }The Aur{\'e}lie spectrum only allows us to determine the stellar parameters. The spectral classification is similar to an OC\,9.7I star according to the tables of \citet{mar05}. From the CMFGEN best-fit model, we find a C overabundance and a N depletion, estimated to about 3.3 and 0.2 times the solar abundances\footnote{$(\mathrm{C/H})_{\odot} = 2.45\,10^{-4}$ and $(\mathrm{N/H})_{\odot} = 6.02\,10^{-5}$ \citep[expressed by number,][]{gas07}}, respectively, as reported by \citet{wal00}. To obtain a fit of good quality, we also need to increase the helium abundance to $(\mathrm{He/H}) = 0.18$ (by number) i.e. almost twice the solar value ($(\mathrm{He/H})_{\odot} = 0.10$). However, additional diagnostic lines are required to confirm such  abundances. 

\subsection{The Cyg\,OB3 association}
\hspace{0.5cm}{\it HD\,190864: }A good quality of fit is achieved. We need a volume filling factor of $0.06$ and a $\beta$ of 1.5 to reach a good accuracy for the wind parameters in both UV and optical domains. As for HD\,193514, $\dot{M}_{\mathrm{H}\alpha}/\sqrt{f}$ is consistent with the value determined by \citet{rep04}. Moreover, as for HD\,193514, the synthetic model fails to correctly fit the \ion{N}{iii}~4634--41 lines.

{\it HD\,227018: }The [$4450-4900]~\AA$ wavelength domain is properly fitted, which gives us a very good determination of the stellar parameters. 

{\it HD\,227245: }The synthetic model correctly reproduces the observations which leads to a good accuracy on the determination of the stellar parameters. Furthermore, the \ion{He}{ii}~4686 and H$\alpha$ lines are well fitted, thereby corresponding to a reliable determination of the mass-loss rate of the star on the basis of these diagnostic lines.. 

{\it HD\,227757: }We need a slow rotation rate to correctly reproduce the photospheric lines. We also see that the \ion{O}{ii} lines are present in the spectrum of the star. The stellar parameters, listed in Table~\ref{cygnus_param}, are in agreement with those determined by \citet{mar14}. The luminosity derived for that object and its mass-loss rate determined from the H$\alpha$ line could indicate that HD\,227757 is a weak-wind star (see \citealt{mar05b} and \citealt{mar09} for other examples of weak-wind O-type stars). However, without any UV spectrum for this star, we cannot confirm this assumption. 

\subsection{The Cyg\,OB8 association}
\hspace{0.5cm}{\it HD\,191423: }We need to broaden the synthetic model by a rotational velocity of 410~\kms to reach a sufficient quality of the fit. Moreover, the model also yields, through the adjustment of the carbon lines at 4070 and 4650~\AA, a large depletion in C (about 0.1 times the solar abundance, i.e. $3.5 \times 10^{-5}$ in number) and an enrichment in N (about 9 times the solar abundance, i.e. $5.4 \times 10^{-4}$ in number). The wind parameters derived on the UV P-Cygni profiles indicate a low terminal velocity (about 600~\kms). We also see two weak emissions in the external wings of H$\alpha$. These emissions could be created by a disk of matter ejected by the high rotational velocity of the star. We stress that the \ion{P}{v}~1118--1128 and \ion{N}{iv}~1718 lines appear slightly weak in our model in comparison to the observed spectra.

{\it HD\,191978: }The fit is of an excellent quality, except for the red wing of the \ion{C}{iv}~1548--50 lines and for the \ion{He}{ii}~4686 line which is too weak compared to the observations. As we already stressed, the \ion{O}{v}~1371 and \ion{N}{iv}~1718 lines are not good diagnostic lines to estimate the clumping factor in late O-type stars. Furthermore, given the lack of FUSE spectrum for this star, we adopt $f=1.0$ (homogeneous model) for HD\,191978. We need a rotational velocity  similar to the macroturbulent velocity ($\vsini = 40$~\kms and $\vmac = 32$~\kms) to correctly reproduce the line profiles. 

{\it HD\,193117: }The cores of the Balmer lines are too deep in the best-fit model. This could come from a nebular emission at least for H$\alpha$. The other lines are rather well fitted. We also stress that although this star appears to be evolved, its nitrogen content is solar.

\subsection{The Cyg\,OB9 association}
\hspace{0.5cm}{\it HD\,194649: }In Paper I we determined  a brightness ratio of $4.7 \pm 0.4$. The fit of the primary disentangled spectrum is good, except for the core of the H$\beta$ line. The secondary component being fainter in the composite spectra, its spectrum appears much noisier when corrected for the relative brightness and the disentangling process fails to correctly reconstruct the wings of H$\beta$. Nonetheless, the lines are relatively well reproduced by the synthetic model. Although the spectral classifications indicate an evolved primary star and a main-sequence secondary star, \logg\ determined on the basis of disentangled spectra fail, once again, to characterize this luminosity class. 

{\it HD\,194334: }A good fit is achieved for this object. Only the cores of the Balmer lines and \ion{N}{iii}~4634--41 are less well reproduced. The fits of the \ion{He}{ii}~4686 and H$\alpha$ lines indicate a good determination of the mass-loss rate of the star. 

{\it HD\,195213: }As for several stars in our sample, the cores of the Balmer lines are poorly reproduced. However, H$\alpha$ shows an emission component in its centre which probably originates from nebular emission. Furthermore, to reach a good fit of the helium lines, we have to increase the helium abundance to $0.15$ in number. 

\begin{sidewaystable*}[htbp]
\caption{Stellar and wind parameters derived for the O-type stars in four Cygnus OB associations.}\label{cygnus_param}
\small
\begin{tabular}{llrrrrrrrrrrrrrrrr}
\hline
Source   & SpT  & \teff\ & \lL\ & \logg & \logg$_{\mathrm{true}}$\tablefootmark{a} & R & log($\dot{M}_{\mathrm{theo}}$)\tablefootmark{b} & log($\frac{\dot{M}_{\mathrm{H}\alpha}}{\sqrt(f)}$)\tablefootmark{c} & $f_{\infty}$ &\vinf & $\beta$ & \vsini\ & \vmac\ & M$_{\mathrm{evol}}$\tablefootmark{d} & M$_{\mathrm{spec}}$\tablefootmark{e} & M$_{\mathrm{dyn}}$\tablefootmark{f} & N/H\\
         &      & [kK]   &      &       &                   & [\rsun] &                        &                              &            &[\kms] &  & [\kms]  & [\kms] & [\msun]         & [\msun]      & [\msun]         & [$\times 10^{-4}$]\\
\hline
\multicolumn{18}{c}{Cyg\,OB1}\\
\hline
HD\,193443P&        O\,9III& 31.2 &     $5.35\pm0.13 $ &    3.65   &    3.66 & $16.2_{-4.3}^{+6.0}$ &  $\sim~-6.8$ &              --   &    --  &     -- &     --   &   105 &   38  & $30.0_{-4.8}^{+4.7}$ &  $43.8_{-30.4}^{+103.0}$ & $1.0 \pm 0.1$ & $0.6\pm0.3$\\[3pt]
HD\,193443S&        O\,9.5V& 29.7 &     $4.76\pm0.23 $ &    3.75   &    3.76& $9.1_{-3.2}^{+5.0}$  &       --     &              --   &    --  &     -- &     --   &   100 &   27  & $18.7_{-5.3}^{+4.6}$ &  $17.4_{-13.2}^{+57.2}$ & $0.4 \pm 0.1$ & $0.6\pm0.3$\\[3pt]
HD\,193514& O\,7--7.5III(f)& 34.5 &     $5.65\pm0.09$ &    3.50   &    3.51&  $18.7_{-2.8}^{+3.3}$ & $-6.5$ &            $-5.6$  &  0.03  &   2190 & 1.5   &    90 &   27  & $45.7_{-4.9}^{+6.8}$ &  $41.3_{-17.5}^{+30.7}$ & -- & $2.0\pm1.0$\\[3pt]
HD\,193595&           O\,7V& 35.5 &     $5.11\pm0.05$ &    3.75   &    3.75& $9.5_{-1.0}^{+1.2}$&        --      &         $-8.7$  &   1.0  &     -- &     --   &    50 &   28  & $26.5_{-2.1}^{+2.1}$ & $18.7_{-6.8}^{+10.9}$  & -- & $3.0\pm1.5$\\[3pt]
HD\,193682&      O\,5III(f)& 39.4 &     $5.50\pm0.09$ &    3.75   &    3.77&  $12.1_{-1.7}^{+2.0}$ &   $-5.8$      &   $-5.7$  &  0.01  &   2650 &   1.0   &   150 &   73  & $41.4_{-4.7}^{+5.4}$ &  $31.4_{-13.1}^{+22.5}$ & -- & $6.0\pm3.0$\\[3pt] 
HD\,194094&         O\,8III& 29.9 &     $5.45\pm0.36$ &    3.25   &    3.26& $19.8_{-7.6}^{+12.3}$ &   $-6.7$     &           $-7.0$  &   1.0  &     --  &     --  &    65 &   32  & $33.0_{-10.1}^{+23.8}$ &  $25.9_{-18.0}^{+59.7}$ & -- & $1.0\pm0.5$\\[3pt]
HD\,194280&         O\,9.7I& 26.9 &     $5.61\pm0.36$ &    3.10   &    3.11& $29.4_{-11.4}^{+18.6}$ &   $<~-6.5$     &              --   &    --  &     -- &     --   &    92 &   41  & $39.2_{-12.9}^{+25.4}$ &  $41.1_{-28.8}^{+96.8}$ & -- & $0.1\pm0.1$\\[3pt]
HD\,228841&        O\,7    & 34.5 &     $5.42\pm0.37$ &    3.50   &    3.62& $14.4_{-5.5}^{+9.0}$ &   $\sim~-6.4$  &           $-7.1$  &   1.0  &     -- &     --   &   317 &    0  & $33.7_{-10.1}^{+22.7}$ & $31.4_{-21.9}^{+72.9}$  & -- & $5.4\pm2.7$\\[3pt]
HD\,228989P&        O\,8.5V& 29.4 &     $4.43\pm0.14$ &    3.50   &    3.56& $6.3_{-1.8}^{+2.6}$ &       --      &              --   &    --  &     -- &     --   &   145 &    3  & $15.9_{-3.7}^{+3.7}$ & $ 5.3_{-3.8}^{+13.3}$ & $7.0 \pm 0.2$ & $1.0\pm0.5$\\[3pt]
HD\,228989S&        O\,9.7V& 27.8 &     $4.28\pm0.15$ &    3.75   &    3.78& $6.0_{-1.7}^{+2.6}$ &      --      &              --   &    --  &     -- &     --   &   120 &    3  & $13.0_{-3.4}^{+3.8}$ &  $ 7.7_{-5.6}^{+20.6}$ & $6.1 \pm 0.2$ & $3.0\pm1.5$\\[3pt]
HD\,229234&   O\,9III + ...& 32.0 &     $5.09\pm0.20$ &    3.60   &    3.61& $11.4_{-2.9}^{+3.9}$ &   $\sim~-6.8$     &           $-6.5$  &   1.0  &     --  &     --  &    85 &   38  & $24.8_{-4.1}^{+4.7}$ &  $19.4_{-10.8}^{+24.6}$ & -- & $0.6\pm0.3$\\[3pt]
\hline
\multicolumn{18}{c}{Cyg\,OB3}\\
\hline
HD\,190864&    O\,6.5III(f)& 38.0 &     $5.35\pm0.14$ &    3.75   &    3.76& $10.9_{-2.1}^{+2.6}$  &   $-6.3$      &           $-6.4$  &   0.06  &   2250 & 1.0  &    80 &   43  & $34.1_{-3.8}^{+5.1}$ &  $24.9_{-12.0}^{+23.2}$  & -- & $5.0\pm3.0$\\[3pt]
HD\,227018&    O\,6.5      & 37.3 &     $5.50\pm0.18$ &    4.00   &    4.00& $13.5_{-3.1}^{+4.0}$  &   $-6.9$      &           --  &   1.0  &     --  &     --  &    70 &   35  & $37.2_{-4.8}^{+13.4}$ & $66.7_{-35.2}^{+75.0}$ & -- & $1.0\pm0.5$\\[3pt]
HD\,227245&         O\,7III& 36.0 &     $5.18\pm0.14$ &    3.75   &    3.75& $10.0_{-1.9}^{+2.4}$ &  $\sim~-6.4$ &           $-6.7$  &   1.0  &     --  &     --  &    65 &   19 & $27.7_{-2.3}^{+6.3}$ & $20.8_{-10.1}^{+19.7}$  & -- & $3.0\pm0.5$\\[3pt]
HD\,227757&           O\,9V& 32.0 &     $4.73\pm0.13$ &    3.75   &    3.75& $7.6_{-1.4}^{+1.8}$ &       --      &         $-8.2$  &   1.0  &     -- &     --   &    45 &    3  & $18.6_{-4.8}^{+3.0}$ &  $11.8_{-5.6}^{+10.9}$  & -- & $0.6\pm0.3$\\[3pt]
\hline
\multicolumn{18}{c}{Cyg\,OB8}\\
\hline
HD\,191423&       ON\,9IIIn& 30.6 &     $5.42\pm0.28$ &    3.50   &    3.67& $18.3_{-5.9}^{+8.7}$ &    $\sim~-6.8$      &    $-6.4$  &  0.01  &    600 &    0.9   &   410 &    0  & $34.7_{-10.3}^{+12.5}$  & $54.6_{-34.6}^{+95.1}$ & -- & $5.4\pm2.5$\\[3pt]
HD\,191978&         O\,8III& 33.2 &     $5.35\pm0.23$ &    3.75   &    3.75& $14.3_{-4.0}^{+5.5}$ &   $-6.7$      &         $-8.7$  &   1.0  &   1600 &     --   &    40 &   32  & $32.1_{-7.4}^{+9.4}$ & $42.2_{-24.7}^{+59.8}$  & -- & $3.0\pm1.5$\\[3pt] 
HD\,193117&         O\,9III& 30.1 &     $5.36\pm0.11$ &    3.30   &    3.31& $17.6_{-3.1}^{+3.8}$  &  $\sim~-6.8$  &           $-6.3$  &   1.0  &     -- &     --   &    80 &   28  & $30.4_{-4.1}^{+5.1}$ & $23.2_{-10.7}^{+19.9}$  & -- & $0.6\pm0.3$\\[3pt]
\hline
\multicolumn{18}{c}{Cyg\,OB9}\\
\hline
HD\,194334& O\,7--7.5III(f)& 33.9 &     $5.19\pm0.07$ &    3.50   &    3.51& $11.4_{-1.5}^{+1.7}$&    $\sim~-6.5$     &           $-6.2$  &   1.0  &     -- &     --   &    62 &   32  & $26.6_{-2.4}^{+2.4}$ & $15.3_{-6.1}^{+10.2}$  & -- & $3.0\pm1.5$\\[3pt]
HD\,194649P& O\,6III(f)    & 38.0 &     $5.14\pm0.15$ &    3.75   &    3.75& $8.6_{-2.2}^{+3.1}$ &  $-6.1$     &              --   &    --  &     -- &     --   &    60 &   67  & $28.1_{-4.2}^{+7.8}$ & $15.3_{-10.6}^{+35.1}$ & $4.9 \pm 0.4$ & $3.0\pm1.3$\\[3pt]
HD\,194649S& O\,8V         & 33.0 &     $4.36\pm0.27$ &    3.75   &    3.76& $4.6_{-1.7}^{+2.8}$ &       --      &              --   &    --  &     -- &     --   &    60 &   27  & $9.6_{-9.0}^{+9.4}$  & $4.5_{-3.5}^{+15.9}$ & $1.9 \pm 0.1$ & $1.0\pm0.5$\\[3pt]
HD\,195213&      O\,7III(f)& 35.3 &     $5.19\pm0.22$ &    3.50   &    3.51& $10.5_{-2.8}^{+3.8}$ &    $\sim -6.4$     &           $-6.0$  &   1.0  &     --  &     --  &    85 &   28  & $27.6_{-6.1}^{+5.7}$ &  $13.2_{-7.6}^{+17.8}$  & -- & $6.0\pm3.0$\\[3pt]
\hline
\end{tabular}
\tablefoot{\tablefoottext{a}{$g_{\mathrm{true}}$: gravity corrected from the contribution of the centrifugal force i.e. $g_{\mathrm{true}} = g + (\vsini)^2/R$}
\tablefoottext{b}{log($\dot{M}_{\mathrm{theo}}$: theoretical mass-loss rate estimated by \citet{muijres2012} for a given spectral classification}
\tablefoottext{c}{log($\dot{M}_{\mathrm{H}\alpha}$: mass-loss rate (where the correction of the clumping $\sqrt{f}$ is not applied) computed from the H$_\alpha$ line. }
\tablefoottext{d}{M$_{\mathrm{evol}}$: mass estimated from the HR diagram}
\tablefoottext{e}{M$_{\mathrm{spec}}$: mass computed from $M = g\,R^2/G$}
\tablefoottext{f}{M$_{\mathrm{dyn}}$: minimum mass of a binary component ($M^3\,\sin i$)}}
\end{sidewaystable*}


\section{Discussion}
\label{sec:dis}

\subsection{Distances and ages of the OB associations}

The stellar parameters derived in Sect.~\ref{sec:res} provide the positions of the stars in the Hertzsprung-Russell (HR) diagram. However, these locations are dependent on the real distances of the stars. As we have already mentioned, the luminosities were computed in the present paper from the distance modulus given by \citet{hum78} and refined through the fit of the SED of each object. When no initial value of the distance of the stars is available in the Humphreys catalogue, we have taken the mean value of the corresponding OB association. The observed $V$ and $(B-V)$ values are from \citet{kha09}, whilst the bolometric corrections were taken from \citet{mar06} as a function of the spectral classification of the stars. The error-bars on the luminosities of the presumably single stars are mainly determined by the uncertainties on the distances. For the binaries, an additional uncertainty is also provided by the brightness ratio between both components of the system. 

\begin{figure*}[htbp]
\begin{center}
\subfigure[Cyg\,OB1]{
\includegraphics[scale=0.45,bb=29 8 536 400,clip]{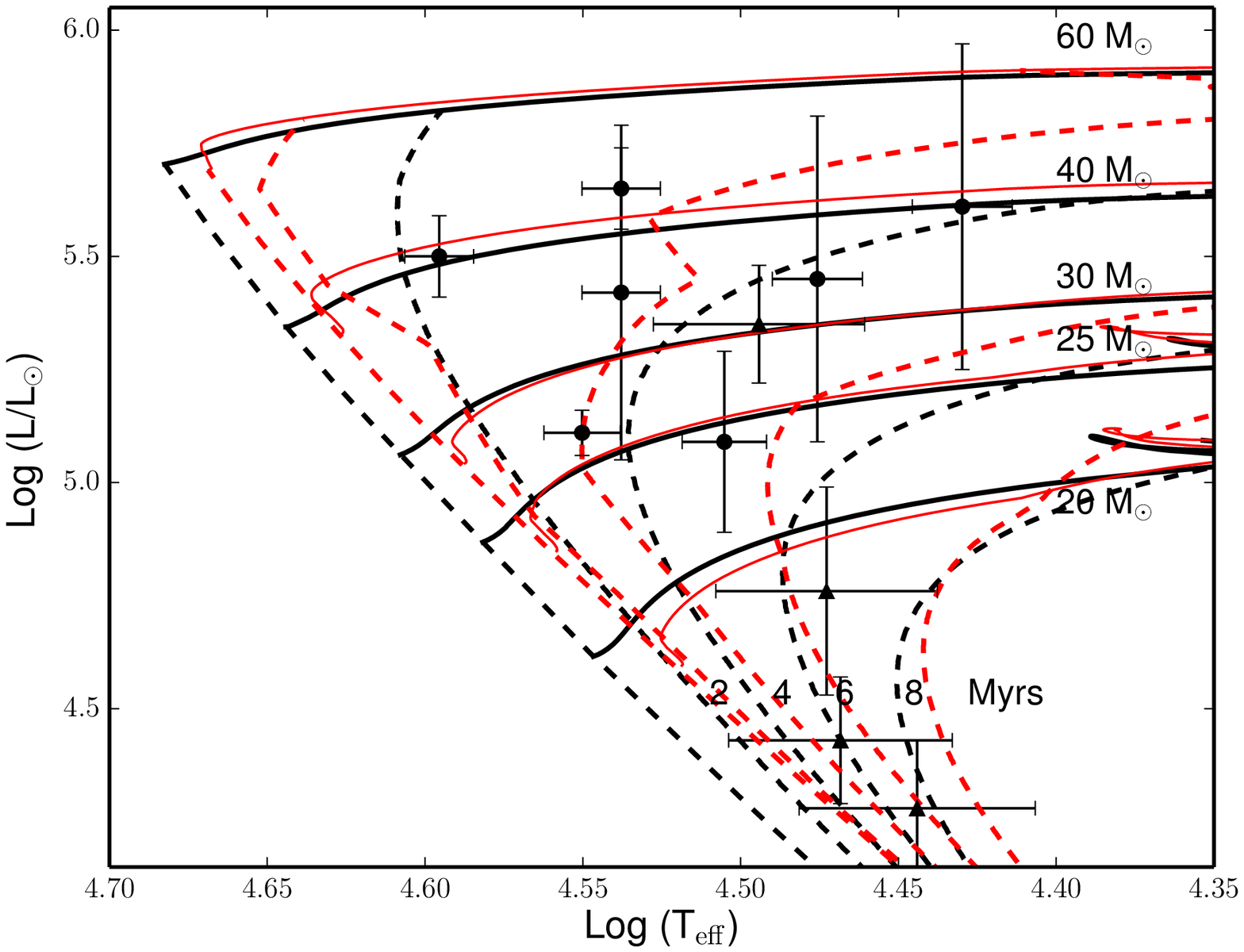}
\label{cygob1_track}}
\subfigure[Cyg\,OB3]{
\includegraphics[scale=0.45,bb=29 8 536 400,clip]{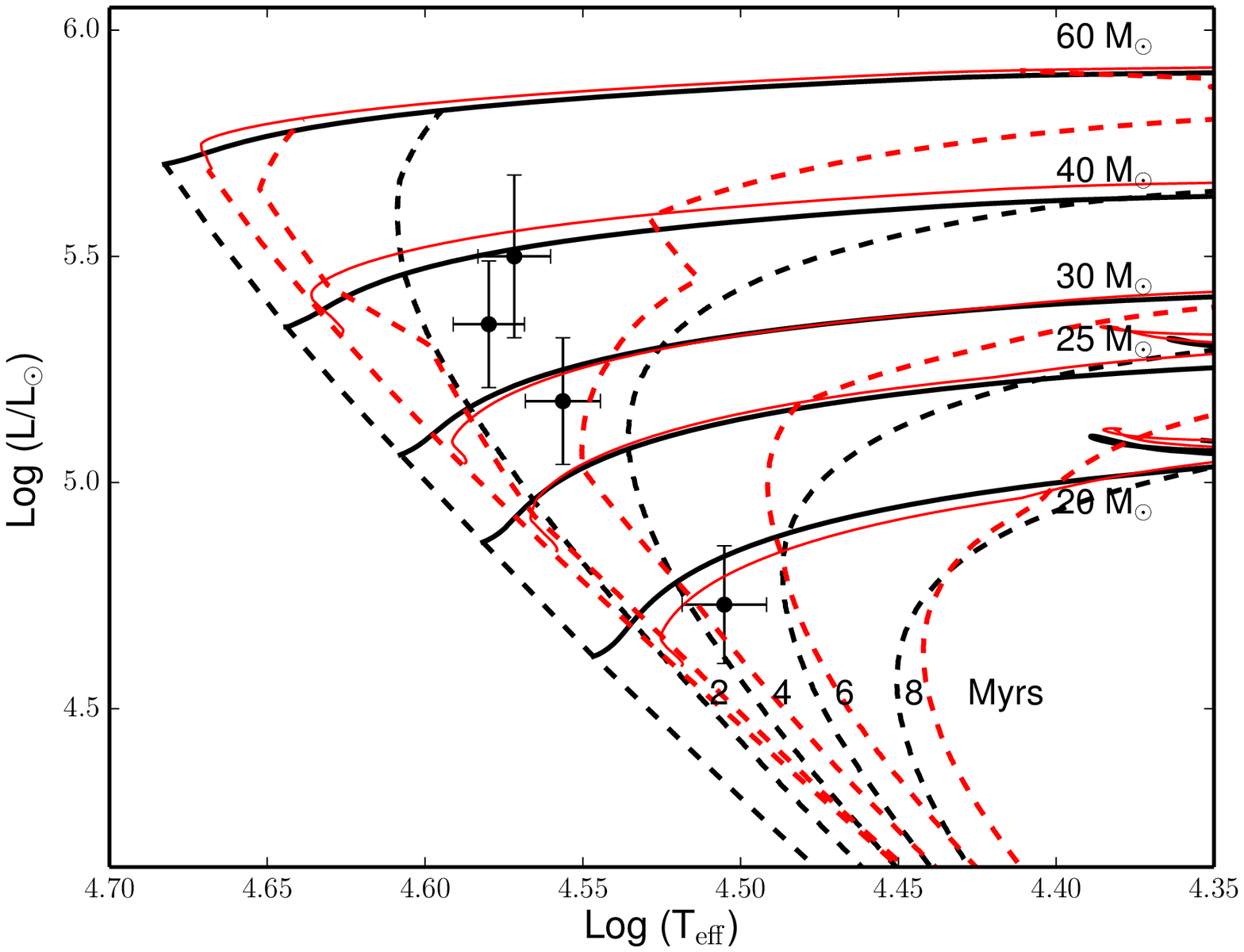}
\label{cygob3_track}}
\subfigure[Cyg\,OB8]{
\includegraphics[scale=0.45,bb=29 8 536 400,clip]{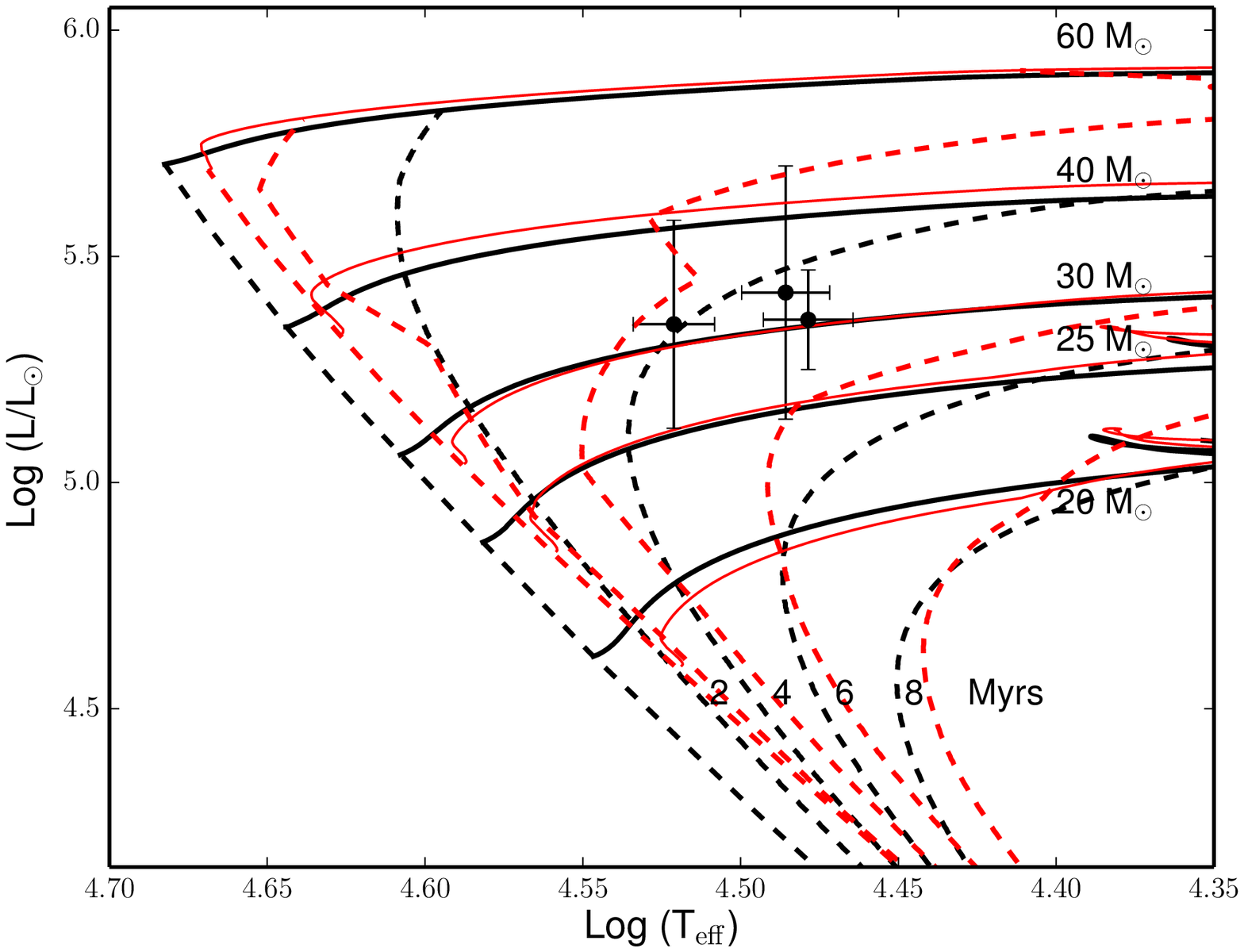}
\label{cygob8_track}}
\subfigure[Cyg\,OB9]{
\includegraphics[scale=0.45,bb=29 8 536 400,clip]{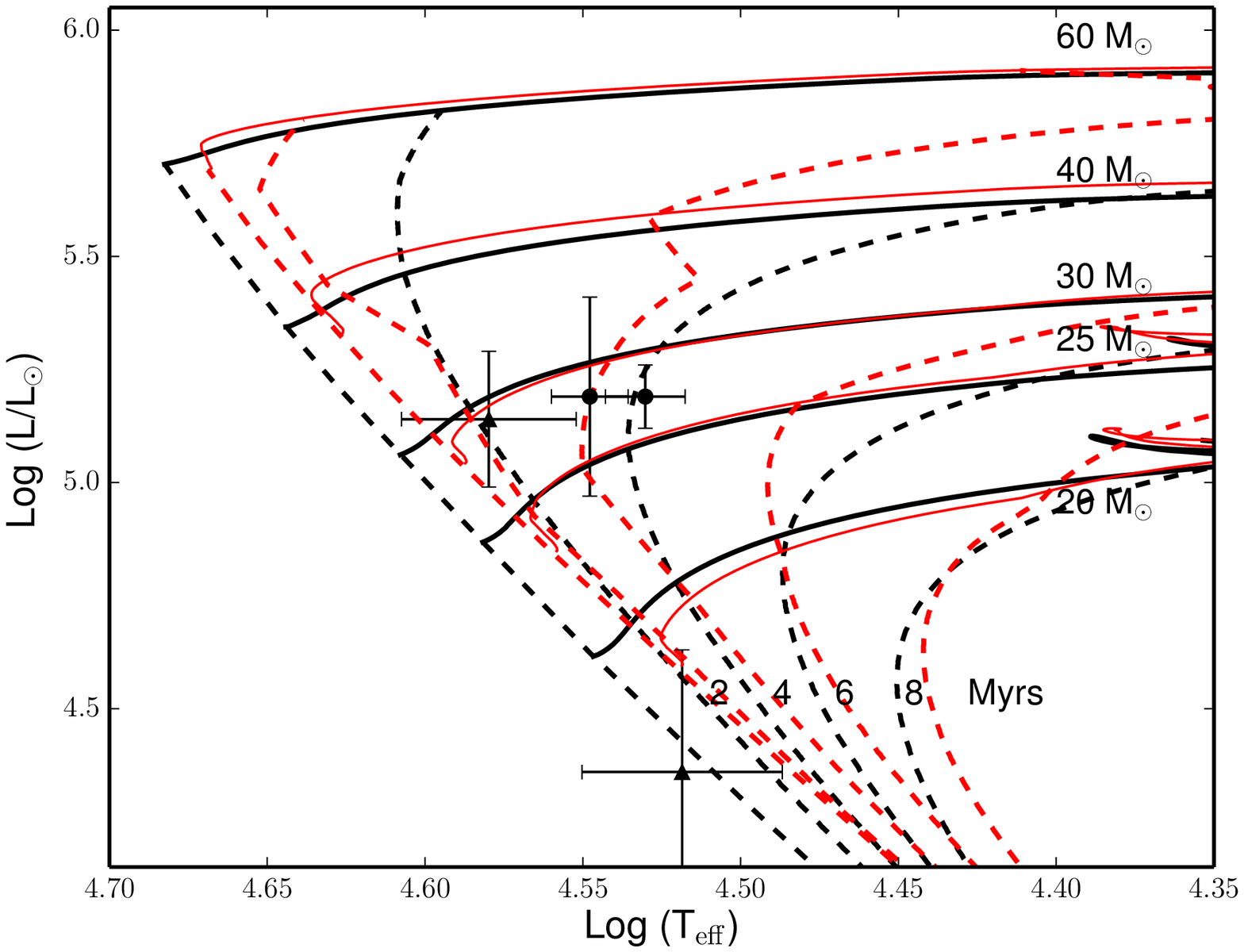}
\label{cygob9_track}}
\caption{Positions of the Cygnus O-type stars in the HR diagram. Evolutionary tracks and isochrones are from \citet{brott11} and were computed for $v_{\mathrm{rot}_{\mathrm{init}}} = 100$~\kms (black lines) and for $v_{\mathrm{rot}_{\mathrm{init}}} = 400$~\kms (red lines). Isochrones, represented by dashed lines, correspond to ages ranging from 2 to 8~Myrs with a step of 2~Myrs.}\label{cygnus_track}
\end{center}
\end{figure*}

For this analysis, we use the evolutionary tracks of \citet{brott11} computed for single stars. Given that rotation plays a major role in the evolution of the stars, we select the tracks computed with initial rotational velocities of about 100~\kms\  (black lines in Fig.~\ref{cygnus_track}) and of about 400~\kms\  (red lines in Fig.~\ref{cygnus_track}). We stress that an initial rotational velocity of 300~\kms \ is generally considered as a "standard" value for O-type stars. In this analysis, we do not investigate all the sample stars together because they are not from the same forming regions and so do not constitute  a homogeneous O-type star population. 
 
The distances taken from \citet{hum78} are close to the largest estimates of \citet{uyaniker2001}. This yields the largest values to the luminosities of the stars. By comparing the luminosities computed in the present paper with the observational values (considered as standard) reported by \citet{mar05}, we see that the stars belonging to Cyg\,OB9 as well as HD\,227245 or both components of HD\,228989 are fainter than the standard expectations. On the contrary, HD\,193117 and HD\,191423 belonging to the Cyg\,OB8 association are brighter than expected. This makes the estimation of their distance questionable. It is clear that these standard values are dependent on other stellar parameters that are not the same for the stars in our sample such as, e.g., $\logg$. To match the standard values for the luminosity, we should increase the mean distance of Cyg\,OB9 to a value of 2.0~kpc and decrease that of Cyg\,OB8 to a mean value of 1.9~kpc. 

These  assumptions could have a great impact on the location of the stars in the HR diagram and could thus affect the real determination of their age. Indeed, from these possible distances for Cyg\,OB8 and Cyg\,OB9 (about 1.9 and 2.0~kpc, respectively), the stars of Cyg\,OB8 could be slightly older (between 4 and 6 Myrs) than what we find in Fig.\,\ref{cygob8_track}, whilst those of Cyg\,OB9 could have an estimated age between 2 and 4 Myrs, i.e. slightly younger. Furthermore, the location of the secondary star of HD,194649 would be more realistic than what we see in Fig.\,\ref{cygob9_track}. This means that the ages of the stars in Cyg\,OB8 do not seem to confirm the estimates of $\sim 3$~Myrs computed from the evolutionary tracks by \citet{uyaniker2001} for this association, but this assumption must still be verified on the basis of a larger sample of stars belonging to this association.

From the distances quoted in Table~\ref{tab:phot}, we estimate the evolutionary ages of O-type stars in Cyg\,OB1 and in Cyg\,OB3 between 2 and 7 Myrs and between 3 and 5 Myrs, respectively, according to the parameters reported in Table~\ref{cygnus_param}.
We must be careful about these HR diagrams, because the real ages of the stars can be slightly different depending on their initial rotational velocity. Therefore, some stars such as HD\,193682 or HD\,228841 (with $\vsini = 150$ and 317~\kms, respectively) could be older than inferred from their position in the HR diagram \citep[see][for further details]{brott11}. 

Determining the real positions of the O stars from the HR diagram is difficult and requires a good knowledge of their real distances. Therefore, our study stresses the importance of astrometrical missions such as Gaia to better understand the evolutionary properties of massive stars.

\subsection{The N content}

Nitrogen is an important element to analyse and to understand the rotational mixing in the evolutionary processes of a massive star. Theoretical studies \citep[e.g.][]{mm00} have shown that rotationally induced mixing can affect the chemical composition of the surface layer of massive stars even though they are still on the main-sequence band. The theory thus predicts that the larger the rotational velocity of a star, the higher its nitrogen abundance. Several observational studies on B stars in the LMC (notably the {\it ESO VLT-FLAMES} survey \citealt{evans2005,evans2006}) have shown that at least 60\% of the B stars display the pattern expected from evolutionary models which include the rotational mixing. However, simulations of LMC early B-type stars made by \citet{brott11b} failed to reproduce the remaining 40\%, which are split into two different groups of stars, one containing slowly rotating, nitrogen-enriched objects and another one containing rapidly rotating un-enriched objects. The analysis of \citet{riv12} also indicated a large number of O-type stars in the LMC with large enrichment in nitrogen and with a low rotation rate, supporting the same conclusions as for the B stars in the LMC. These authors stressed, however, that the problem of the group containing the highly-enriched slow-rotating stars was more severe for the O-type star population than for the B-type star one. A possible explanation concerning the presence of the group with high-enriched slow-rotating stars was advanced by \citet{mor06} who have analysed ten slowly rotating galactic early $\beta$-Cephei B-type stars and found that out of four heavily enriched stars, three had a magnetic field.

\begin{figure}[htbp]
\begin{center}
\includegraphics[width=8.cm,bb=25 5 536 406,clip]{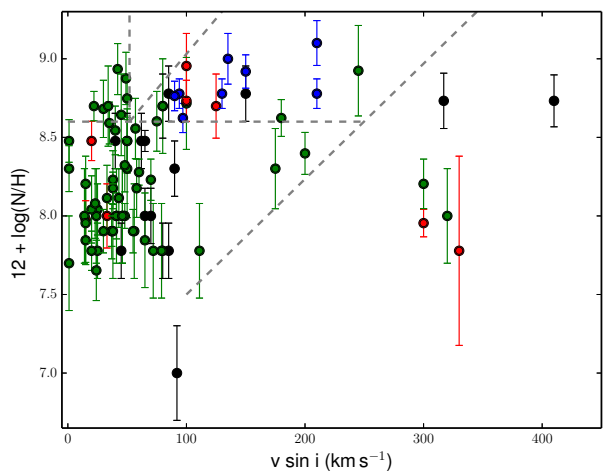}
\caption{Nitrogen surface abundance (in units of 12 + $\log$(N/H)) as a function of projected rotational velocity, the so-called Hunter diagram. O stars in our sample are shown in black, the O-stars of NGC\,2244 \citep{mar12} in red, stars studied by \citet{bouret12} in blue, and the MiMes O-type stars \citep{mar14} in green. Grey lines define the possible locations of the different groups.}\label{fig:Ncontent}
\end{center}
\end{figure}

In the present paper, we determine the N content of nineteen O stars belonging to four Cygnus OB associations. This result is combined and put in perspective with several other investigations of O-type stars in the Galaxy (\citealt{mar12}, \citealt{bouret12}, and \citealt{mar14}). Although this constitutes a set of heterogeneous studies, all the analyses use the same tool to determine the stellar parameters. We show in Fig.~\ref{fig:Ncontent}, the nitrogen content as a function of the projected rotational velocities of 90 O-type stars, the so-called Hunter diagram \citep{hun09}. We stress that we remove from our sample all the binary components as well as a few stars listed in \citet{mar14} that are binaries, e.g. HD\,93250 \citep{san11b} or HD\,193443 (Paper~I). We tentatively define possible groups similar to those introduced by \citet{hun08,hun09}. To do that, the group constituted of stars with intermediate rotation rates (between 50 and 110~\kms) and a high enrichment in nitrogen is assumed to be empty as noted by \citet{brott11b}. In this diagram, there is a clear outlier (HD\,194280) that seems to belong to none of the different groups. We stress though that defining their exact location requires a more sophisticated theoretical work similar to that of \citet{brott11b} for massive stars in the LMC.

\begin{figure}[htbp]
\begin{center}
\includegraphics[width=8.cm,bb=26 9 535 408,clip]{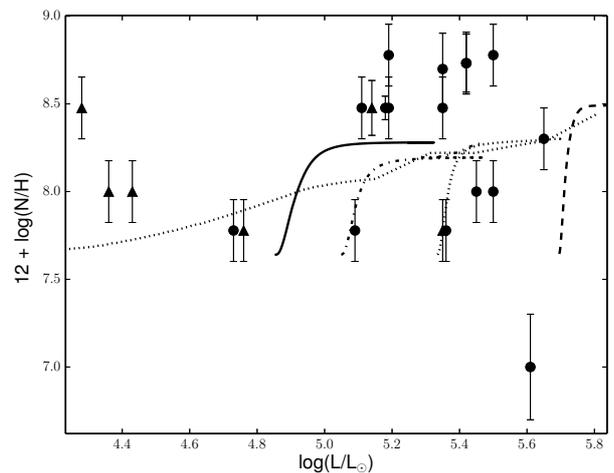}
\caption{Nitrogen surface abundance (in units of 12 + $\log$(N/H)) as a function of stellar luminosity. Dots represent the single O stars in our sample, whilst triangles are the binary components. Evolutionary tracks are (from left to right) computed for stars with 10, 15, 20, and 40~$M_{\odot}$ and initial velocities of about 300~\kms.}\label{N_L_plot}
\end{center}
\end{figure}

In Fig.~\ref{N_L_plot}, we show the N content of our initial targets versus their luminosity. We clearly see a trend of higher N content for more luminous stars. As already stressed by \citet{mar12}, the fast rotators do not appear as outliers. Two clear outliers are shown in this figure: HD\,194280, as we have already mentioned, with a very low nitrogen abundance and, to a lesser extent, the secondary component of HD\,228989. 

\subsection{Stellar masses}

The  differences between the evolutionary masses, i.e. the masses determined from the HR diagram and the evolutionary tracks, and the spectroscopic masses, i.e. those determined from the atmospherical parameters (\teff, \logg, and $L$), are recurrent issues in astrophysics. This so-called "mass discrepancy" refers to a systematic overestimate of the former relative to the latter. \citet{rep04} already noted that for stars with masses smaller than $50~M_{\odot}$ a parallel relation to the 1:1 relation in the $M_{\mathrm{evol}}-M_{\mathrm{spec}}$ diagram could be followed. More recently, \citet{mar12}, from a small sample of O-type stars in NGC~2244 and in Mon~OB2, reported on a clear trend of mass discrepancy for stars with $M< 25~M_{\odot}$, consistent with the results of \citet{rep04}.

In Paper I we also determined the dynamical mass of each component of the binary systems. We present them in Table~\ref{cygnus_param} as a reminder. We can thus compare them to their evolutionary and spectroscopic masses. The evolutionary masses have been estimated for each star (presumably single or individual component) by a bi-interpolation of the evolutionary tracks on \teff\ and on $\log(L/L_{\odot})$, whilst the spectroscopic masses have been computed from $M = g\,R^2/G$, where $G$ is the universal gravitational constant. We stress that these masses were computed from the effective gravities corrected for the effects of the centrifugal forces caused by rotation ($g_{\mathrm{true}}$ in Table~\ref{cygnus_param}) by following the approach of \citet{herrero1992}. We clearly see a real disagreement for the primary star of HD\,228989. First, the spectroscopic mass is smaller than the dynamical mass that is supposed to be the minimum mass. Then, the mass ratio computed from the spectroscopic values indicates that the secondary component is more massive than the primary one which is the opposite of the results provided in Paper I. This assumes that the distance and/or the gravities of both components are poorly estimated. The reason for a poor determination of the \logg\ is that the disentangling process does not perfectly reproduce the wings of the Balmer lines because of the broadness of these lines and because the two individual line profiles are not completely resolved throughout the orbit.  

\begin{figure}[htbp]
\begin{center}
\includegraphics[width=8.cm,bb=24 2 537 407,clip]{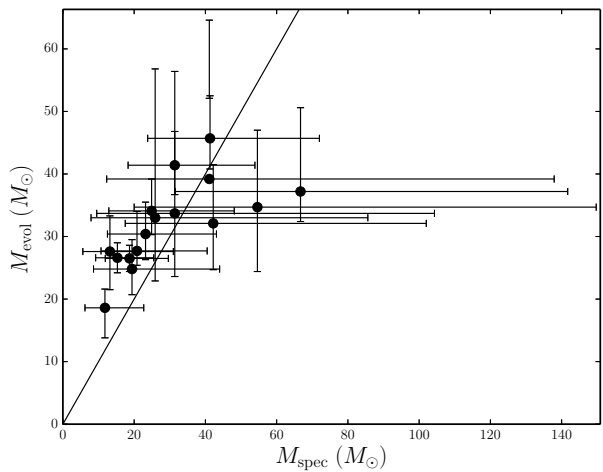}
\caption{Comparison of the evolutionary masses and the spectroscopic masses of the presumably single stars. }\label{fig:mass}
\end{center}
\end{figure}

When we only focus on the single stars, we find that the objects in our sample (Fig.~\ref{fig:mass}) with a mass smaller than $30~M_{\odot}$ present a clear mass discrepancy as already detected for O-type stars belonging to NGC\,2244 and Mon\,OB2 \citep{mar12}. For the stars in our sample with a higher mass, the uncertainties do not allow us to conclude whether or not the mass estimates are significantly different. Two ingredients play an important role in the determination of these uncertainties: the distance and \logg. As we have already mentioned, the distances of the stars are still poorly known. It affects the stellar luminosities and thus the stellar radii since the \teff\ is thought to be well constrained. Concerning \logg, we made several tests to correctly fit the wings of the Balmer lines on {\'e}chelle spectra and also on the Aur{\'e}lie spectra (see Paper I, for a complete list of the data). Our estimation of the stellar gravity thus seems quite robust, therefore minimizing the contribution of the \logg\ to the uncertainty budget of $M_{\mathrm{spec}}$.

\subsection{Mass-loss rates}

We also estimate the mass-loss rates (Table~\ref{cygnus_param}) for most objects in our sample, but the lack of UV spectra does not allow us to determine all the wind parameters of these stars. Often, the terminal velocity, $f$, and the $\beta$ parameters cannot be estimated from only the H$\alpha$ line. Moreover, when no FUSE data were available, we only assumed that the stellar winds are homogeneous ($f = 1.0$).

The mass-loss rates obtained from atmosphere models and included in Table~\ref{cygnus_param} can be compared to the theoretical values quoted by \citet{muijres2012}. These values are given as a function of spectral type, but some of them are missing notably for late main-sequence stars. For most of our objects for which a homogeneous wind was assumed, the agreement between estimated and theoretical values is relatively good. Only few stars (HD\,228841, HD\,193514, HD\,227018, HD\,193117, and HD\,191978) have mass-loss rates which differ by more than about 0.4~dex. Although the quantities $\dot{M}/\sqrt{f}$ remain quite close between the observations and the theory, we can see that the observed mass-loss rates  in general appear slightly larger than the theoretical values. 


\section{Conclusion}
\label{sec:conc}

We analysed by means of the CMFGEN atmosphere code, the fundamental properties of fifteen presumably single stars and of the individual components of four binary systems belonging to different Cygnus OB associations: Cyg\,OB1, Cyg\,OB3, Cyg\,OB8, and Cyg\,OB9. In addition to the optical spectra, we retrieved, for several single stars, the UV spectra (FUSE and/or IUE) to better constrain the wind parameters of these objects. 

The analysis of the individual parameters showed that the luminosities of several objects do not agree with the standard luminosities of stars with similar spectral classification. This is most likely because the distances of these stars are poorly constrained. Therefore, it is more likely that the positions of certain stars should be shifted upwards in the HR diagram. A direct consequence is that the stars in Cyg\,OB8 and Cyg\,OB9 could be younger than  we have determined. However, if the distance used in our analysis is correct, this would mean that Cyg\,OB8 is older than 3~Myrs as quoted by \citet{uyaniker2001}. 

We compared the N content of the stars in our sample to other galactic O-type stars. The resulting Hunter diagram shows the same structure (the same five groups) as for the O and B-type stars in the LMC \citep{hun08,hun09,riv12}. 

By comparing the evolutionary and the spectroscopic masses, we also pointed out a discrepancy between the masses for the stars in our sample. This overestimate of the evolutionary masses relative to the spectroscopic masses is particularly obvious for stars with masses smaller than $30~M_{\odot}$, as  was already noted by \citet{mar12}. Above this value, the masses agree with each other within the uncertainties. Although this sample of stars takes into account objects more evolved than those in the investigation of \citet{mar12}, we obtain similar general conclusions. 


\begin{acknowledgements}
We wish to thank the anonymous referee for his clear report and his constructive remarks that have considerabily improved this paper. This research was supported by the Fonds National de la Recherche Scientifique (F.N.R.S.), the PRODEX XMM/Integral contract (Belspo), and the Communaut\'e fran\c caise de Belgique -- Action de recherche concert\'ee -- A.R.C. -- Acad\'emie Wallonie-Europe. We thank John Hillier for making CMFGEN available and Fabrice Martins for his advice.
\end{acknowledgements}

\bibliography{laurent.bib}

\begin{thebibliography}{58}
\expandafter\ifx\csname natexlab\endcsname\relax\def\natexlab#1{#1}\fi

\bibitem[{{Aerts} {et~al.}(2009){Aerts}, {Puls}, {Godart}, \&
  {Dupret}}]{aerts09}
{Aerts}, C., {Puls}, J., {Godart}, M., \& {Dupret}, M.-A. 2009, \aap, 508, 409

\bibitem[{{Bouret} {et~al.}(2012){Bouret}, {Hillier}, {Lanz}, \&
  {Fullerton}}]{bouret12}
{Bouret}, J.-C., {Hillier}, D.~J., {Lanz}, T., \& {Fullerton}, A.~W. 2012,
  \aap, 544, A67

\bibitem[{{Bouret} {et~al.}(2005){Bouret}, {Lanz}, \& {Hillier}}]{bouret05}
{Bouret}, J.-C., {Lanz}, T., \& {Hillier}, D.~J. 2005, \aap, 438, 301

\bibitem[{{Brott} {et~al.}(2011{\natexlab{a}}){Brott}, {de Mink}, {Cantiello},
  {Langer}, {de Koter}, {Evans}, {Hunter}, {Trundle}, \& {Vink}}]{brott11}
{Brott}, I., {de Mink}, S.~E., {Cantiello}, M., {et~al.} 2011{\natexlab{a}},
  \aap, 530, A115+

\bibitem[{{Brott} {et~al.}(2011{\natexlab{b}}){Brott}, {Evans}, {Hunter}, {de
  Koter}, {Langer}, {Dufton}, {Cantiello}, {Trundle}, {Lennon}, {de Mink},
  {Yoon}, \& {Anders}}]{brott11b}
{Brott}, I., {Evans}, C.~J., {Hunter}, I., {et~al.} 2011{\natexlab{b}}, \aap,
  530, A116+

\bibitem[{{Cardelli} {et~al.}(1989){Cardelli}, {Clayton}, \& {Mathis}}]{car89}
{Cardelli}, J.~A., {Clayton}, G.~C., \& {Mathis}, J.~S. 1989, \apj, 345, 245

\bibitem[{{Evans} {et~al.}(2006){Evans}, {Lennon}, {Smartt}, \&
  {Trundle}}]{evans2006}
{Evans}, C.~J., {Lennon}, D.~J., {Smartt}, S.~J., \& {Trundle}, C. 2006, \aap,
  456, 623

\bibitem[{{Evans} {et~al.}(2005){Evans}, {Smartt}, {Lee}, {Lennon}, {Kaufer},
  {Dufton}, {Trundle}, {Herrero}, {Sim{\'o}n-D{\'{\i}}az}, {de Koter},
  {Hamann}, {Hendry}, {Hunter}, {Irwin}, {Korn}, {Kudritzki}, {Langer},
  {Mokiem}, {Najarro}, {Pauldrach}, {Przybilla}, {Puls}, {Ryans}, {Urbaneja},
  {Venn}, \& {Villamariz}}]{evans2005}
{Evans}, C.~J., {Smartt}, S.~J., {Lee}, J.-K., {et~al.} 2005, \aap, 437, 467

\bibitem[{{Eversberg} {et~al.}(1998){Eversberg}, {Lepine}, \&
  {Moffat}}]{eversberg98}
{Eversberg}, T., {Lepine}, S., \& {Moffat}, A.~F.~J. 1998, \apj, 494, 799

\bibitem[{{Feldmeier} {et~al.}(2003){Feldmeier}, {Oskinova}, \&
  {Hamann}}]{feld03}
{Feldmeier}, A., {Oskinova}, L., \& {Hamann}, W.-R. 2003, \aap, 403, 217

\bibitem[{{Fernie}(1983)}]{fer83}
{Fernie}, J.~D. 1983, \apjs, 52, 7

\bibitem[{{Fraser} {et~al.}(2010){Fraser}, {Dufton}, {Hunter}, \&
  {Ryans}}]{fraser10}
{Fraser}, M., {Dufton}, P.~L., {Hunter}, I., \& {Ryans}, R.~S.~I. 2010, \mnras,
  404, 1306

\bibitem[{{Gonz{\'a}lez} \& {Levato}(2006)}]{gl06}
{Gonz{\'a}lez}, J.~F. \& {Levato}, H. 2006, \aap, 448, 283

\bibitem[{{Grevesse} {et~al.}(2007){Grevesse}, {Asplund}, \& {Sauval}}]{gas07}
{Grevesse}, N., {Asplund}, M., \& {Sauval}, A.~J. 2007, Space Science Reviews,
  130, 105

\bibitem[{{Guarinos}(1991)}]{gua92}
{Guarinos}, J. 1991, PhD thesis, PhD Thesis, Observatoire de Strasbourg,
  France, (1991)

\bibitem[{{Heap} {et~al.}(2006){Heap}, {Lanz}, \& {Hubeny}}]{heap06}
{Heap}, S.~R., {Lanz}, T., \& {Hubeny}, I. 2006, \apj, 638, 409

\bibitem[{{Herrero} {et~al.}(1992){Herrero}, {Kudritzki}, {Vilchez}, {Kunze},
  {Butler}, \& {Haser}}]{herrero1992}
{Herrero}, A., {Kudritzki}, R.~P., {Vilchez}, J.~M., {et~al.} 1992, \aap, 261,
  209

\bibitem[{{Herv{\'e}} {et~al.}(2012){Herv{\'e}}, {Rauw}, {Naz{\'e}}, \&
  {Foster}}]{her12}
{Herv{\'e}}, A., {Rauw}, G., {Naz{\'e}}, Y., \& {Foster}, A. 2012, \apj, 748,
  89

\bibitem[{{Hillier} {et~al.}(2003){Hillier}, {Lanz}, {Heap}, {Hubeny}, {Smith},
  {Evans}, {Lennon}, \& {Bouret}}]{hil03}
{Hillier}, D.~J., {Lanz}, T., {Heap}, S.~R., {et~al.} 2003, \apj, 588, 1039

\bibitem[{{Hillier} \& {Miller}(1998)}]{hm98}
{Hillier}, D.~J. \& {Miller}, D.~L. 1998, \apj, 496, 407

\bibitem[{{Howarth} {et~al.}(1997){Howarth}, {Siebert}, {Hussain}, \&
  {Prinja}}]{how97}
{Howarth}, I.~D., {Siebert}, K.~W., {Hussain}, G.~A.~J., \& {Prinja}, R.~K.
  1997, \mnras, 284, 265

\bibitem[{{Howarth} {et~al.}(2007){Howarth}, {Walborn}, {Lennon}, {Puls},
  {Naz{\'e}}, {Annuk}, {Antokhin}, {Bohlender}, {Bond}, {Donati}, {Georgiev},
  {Gies}, {Harmer}, {Herrero}, {Kolka}, {McDavid}, {Morel}, {Negueruela},
  {Rauw}, \& {Reig}}]{howarth07}
{Howarth}, I.~D., {Walborn}, N.~R., {Lennon}, D.~J., {et~al.} 2007, \mnras,
  381, 433

\bibitem[{{Humphreys}(1978)}]{hum78}
{Humphreys}, R.~M. 1978, \apjs, 38, 309

\bibitem[{{Hunter} {et~al.}(2009){Hunter}, {Brott}, {Langer}, {Lennon},
  {Dufton}, {Howarth}, {Ryans}, {Trundle}, {Evans}, {de Koter}, \&
  {Smartt}}]{hun09}
{Hunter}, I., {Brott}, I., {Langer}, N., {et~al.} 2009, \aap, 496, 841

\bibitem[{{Hunter} {et~al.}(2008){Hunter}, {Lennon}, {Dufton}, {Trundle},
  {Sim{\'o}n-D{\'{\i}}az}, {Smartt}, {Ryans}, \& {Evans}}]{hun08}
{Hunter}, I., {Lennon}, D.~J., {Dufton}, P.~L., {et~al.} 2008, \aap, 479, 541

\bibitem[{{Kharchenko} {et~al.}(2009){Kharchenko}, {Piskunov}, {R{\"o}ser},
  {Schilbach}, {Scholz}, \& {Zinnecker}}]{kha09}
{Kharchenko}, N.~V., {Piskunov}, A.~E., {R{\"o}ser}, S., {et~al.} 2009, \aap,
  504, 681

\bibitem[{{Krelowski} \& {Strobel}(2012)}]{kre12}
{Krelowski}, J. \& {Strobel}, A. 2012, VizieR Online Data Catalog, 1133, 30060

\bibitem[{{Lanz} \& {Hubeny}(2003)}]{lh03}
{Lanz}, T. \& {Hubeny}, I. 2003, \apjs, 146, 417, erratum: 147, 225

\bibitem[{{Mahy} {et~al.}(2013){Mahy}, {Rauw}, {De Becker}, {Eenens}, \&
  {Flores}}]{mah13}
{Mahy}, L., {Rauw}, G., {De Becker}, M., {Eenens}, P., \& {Flores}, C.~A. 2013,
  \aap, 550, A27

\bibitem[{{Ma{\'{\i}}z-Apell{\'a}niz}
  {et~al.}(2004){Ma{\'{\i}}z-Apell{\'a}niz}, {Walborn}, {Galu{\'e}}, \&
  {Wei}}]{mai04}
{Ma{\'{\i}}z-Apell{\'a}niz}, J., {Walborn}, N.~R., {Galu{\'e}}, H.~{\'A}., \&
  {Wei}, L.~H. 2004, \apjs, 151, 103

\bibitem[{{Martins}(2011)}]{mar11}
{Martins}, F. 2011, Bulletin de la Soci{\'e}t{\'e} Royale des Sciences de
  Li{\`e}ge, 80, 29

\bibitem[{{Martins} {et~al.}(2010){Martins}, {Donati}, {Marcolino}, {Bouret},
  {Wade}, {Escolano}, \& {Howarth}}]{martins10}
{Martins}, F., {Donati}, J.-F., {Marcolino}, W.~L.~F., {et~al.} 2010, \mnras,
  407, 1423

\bibitem[{{Martins} {et~al.}(2014){Martins}, {Herv{\'e}}, {Bouret},
  {Marcolino}, {Wade}, {Neiner}, {Alecian}, {Grunhut}, {Petit}, \& {the MiMeS
  collaboration}}]{mar14}
{Martins}, F., {Herv{\'e}}, A., {Bouret}, J.-C., {et~al.} 2014, ArXiv e-prints

\bibitem[{{Martins} {et~al.}(2009){Martins}, {Hillier}, {Bouret}, {Depagne},
  {Foellmi}, {Marchenko}, \& {Moffat}}]{mar09}
{Martins}, F., {Hillier}, D.~J., {Bouret}, J.~C., {et~al.} 2009, \aap, 495, 257

\bibitem[{{Martins} {et~al.}(2012){Martins}, {Mahy}, {Hillier}, \&
  {Rauw}}]{mar12}
{Martins}, F., {Mahy}, L., {Hillier}, D.~J., \& {Rauw}, G. 2012, \aap, 538, A39

\bibitem[{{Martins} \& {Plez}(2006)}]{mar06}
{Martins}, F. \& {Plez}, B. 2006, \aap, 457, 637

\bibitem[{{Martins} {et~al.}(2005{\natexlab{a}}){Martins}, {Schaerer}, \&
  {Hillier}}]{mar05}
{Martins}, F., {Schaerer}, D., \& {Hillier}, D.~J. 2005{\natexlab{a}}, \aap,
  436, 1049

\bibitem[{{Martins} {et~al.}(2005{\natexlab{b}}){Martins}, {Schaerer},
  {Hillier}, {Meynadier}, {Heydari-Malayeri}, \& {Walborn}}]{mar05b}
{Martins}, F., {Schaerer}, D., {Hillier}, D.~J., {et~al.} 2005{\natexlab{b}},
  \aap, 441, 735

\bibitem[{{Meynet} \& {Maeder}(2000)}]{mm00}
{Meynet}, G. \& {Maeder}, A. 2000, \aap, 361, 101

\bibitem[{{Meynet} \& {Maeder}(2003)}]{mm03}
{Meynet}, G. \& {Maeder}, A. 2003, \aap, 404, 975

\bibitem[{{Morel} {et~al.}(2006){Morel}, {Butler}, {Aerts}, {Neiner}, \&
  {Briquet}}]{mor06}
{Morel}, T., {Butler}, K., {Aerts}, C., {Neiner}, C., \& {Briquet}, M. 2006,
  \aap, 457, 651

\bibitem[{{Muijres} {et~al.}(2012){Muijres}, {Vink}, {de Koter}, {M{\"u}ller},
  \& {Langer}}]{muijres2012}
{Muijres}, L.~E., {Vink}, J.~S., {de Koter}, A., {M{\"u}ller}, P.~E., \&
  {Langer}, N. 2012, \aap, 537, A37

\bibitem[{{Naz{\'e}}(2009)}]{naze09}
{Naz{\'e}}, Y. 2009, \aap, 506, 1055

\bibitem[{{Nieva} \& {Przybilla}(2007)}]{nieva07}
{Nieva}, M.~F. \& {Przybilla}, N. 2007, \aap, 467, 295

\bibitem[{{Patriarchi} {et~al.}(2003){Patriarchi}, {Morbidelli}, \&
  {Perinotto}}]{pat03}
{Patriarchi}, P., {Morbidelli}, L., \& {Perinotto}, M. 2003, \aap, 410, 905

\bibitem[{{Patriarchi} {et~al.}(2001){Patriarchi}, {Morbidelli}, {Perinotto},
  \& {Barbaro}}]{pat01}
{Patriarchi}, P., {Morbidelli}, L., {Perinotto}, M., \& {Barbaro}, G. 2001,
  \aap, 372, 644

\bibitem[{{Prinja} {et~al.}(1990){Prinja}, {Barlow}, \& {Howarth}}]{pri90}
{Prinja}, R.~K., {Barlow}, M.~J., \& {Howarth}, I.~D. 1990, \apj, 361, 607

\bibitem[{{Repolust} {et~al.}(2004){Repolust}, {Puls}, \& {Herrero}}]{rep04}
{Repolust}, T., {Puls}, J., \& {Herrero}, A. 2004, \aap, 415, 349

\bibitem[{{Rivero Gonz{\'a}lez} {et~al.}(2012){Rivero Gonz{\'a}lez}, {Puls},
  {Najarro}, \& {Brott}}]{riv12}
{Rivero Gonz{\'a}lez}, J.~G., {Puls}, J., {Najarro}, F., \& {Brott}, I. 2012,
  \aap, 537, A79

\bibitem[{{Ryans} {et~al.}(2002){Ryans}, {Dufton}, {Rolleston}, {Lennon},
  {Keenan}, {Smoker}, \& {Lambert}}]{ryans2002}
{Ryans}, R.~S.~I., {Dufton}, P.~L., {Rolleston}, W.~R.~J., {et~al.} 2002,
  \mnras, 336, 577

\bibitem[{{Sana} {et~al.}(2011){Sana}, {Le Bouquin}, {De Becker}, {Berger}, {de
  Koter}, \& {M{\'e}rand}}]{san11b}
{Sana}, H., {Le Bouquin}, J.-B., {De Becker}, M., {et~al.} 2011, \apjl, 740,
  L43+

\bibitem[{{Sana} {et~al.}(2006){Sana}, {Rauw}, {Naz{\'e}}, {Gosset}, \&
  {Vreux}}]{san06b}
{Sana}, H., {Rauw}, G., {Naz{\'e}}, Y., {Gosset}, E., \& {Vreux}, J.-M. 2006,
  \mnras, 372, 661

\bibitem[{{Sim{\'o}n-D{\'{\i}}az} \& {Herrero}(2007)}]{sim07}
{Sim{\'o}n-D{\'{\i}}az}, S. \& {Herrero}, A. 2007, \aap, 468, 1063

\bibitem[{{Sim{\'o}n-D{\'{\i}}az} {et~al.}(2010){Sim{\'o}n-D{\'{\i}}az},
  {Uytterhoeven}, {Herrero}, {Castro}, \& {Puls}}]{sim10}
{Sim{\'o}n-D{\'{\i}}az}, S., {Uytterhoeven}, K., {Herrero}, A., {Castro}, N.,
  \& {Puls}, J. 2010, Astronomische Nachrichten, 331, 1069

\bibitem[{{Sundqvist} {et~al.}(2011){Sundqvist}, {Puls}, {Feldmeier}, \&
  {Owocki}}]{sun11}
{Sundqvist}, J.~O., {Puls}, J., {Feldmeier}, A., \& {Owocki}, S.~P. 2011, \aap,
  528, A64+

\bibitem[{{Uyan{\i}ker} {et~al.}(2001){Uyan{\i}ker}, {F{\"u}rst}, {Reich},
  {Aschenbach}, \& {Wielebinski}}]{uyaniker2001}
{Uyan{\i}ker}, B., {F{\"u}rst}, E., {Reich}, W., {Aschenbach}, B., \&
  {Wielebinski}, R. 2001, \aap, 371, 675

\bibitem[{{Walborn} \& {Howarth}(2000)}]{wal00}
{Walborn}, N.~R. \& {Howarth}, I.~D. 2000, \pasp, 112, 1446

\bibitem[{{Wesselius} {et~al.}(1982){Wesselius}, {van Duinen}, {de Jonge},
  {Aalders}, {Luinge}, \& {Wildeman}}]{wes82}
{Wesselius}, P.~R., {van Duinen}, R.~J., {de Jonge}, A.~R.~W., {et~al.} 1982,
  \aaps, 49, 427

\end{thebibliography}

\Online
\begin{figure*}[htbp]
\subfigure[HD\,193514]{
\includegraphics[scale=0.400,bb=17 148 569 559,clip]{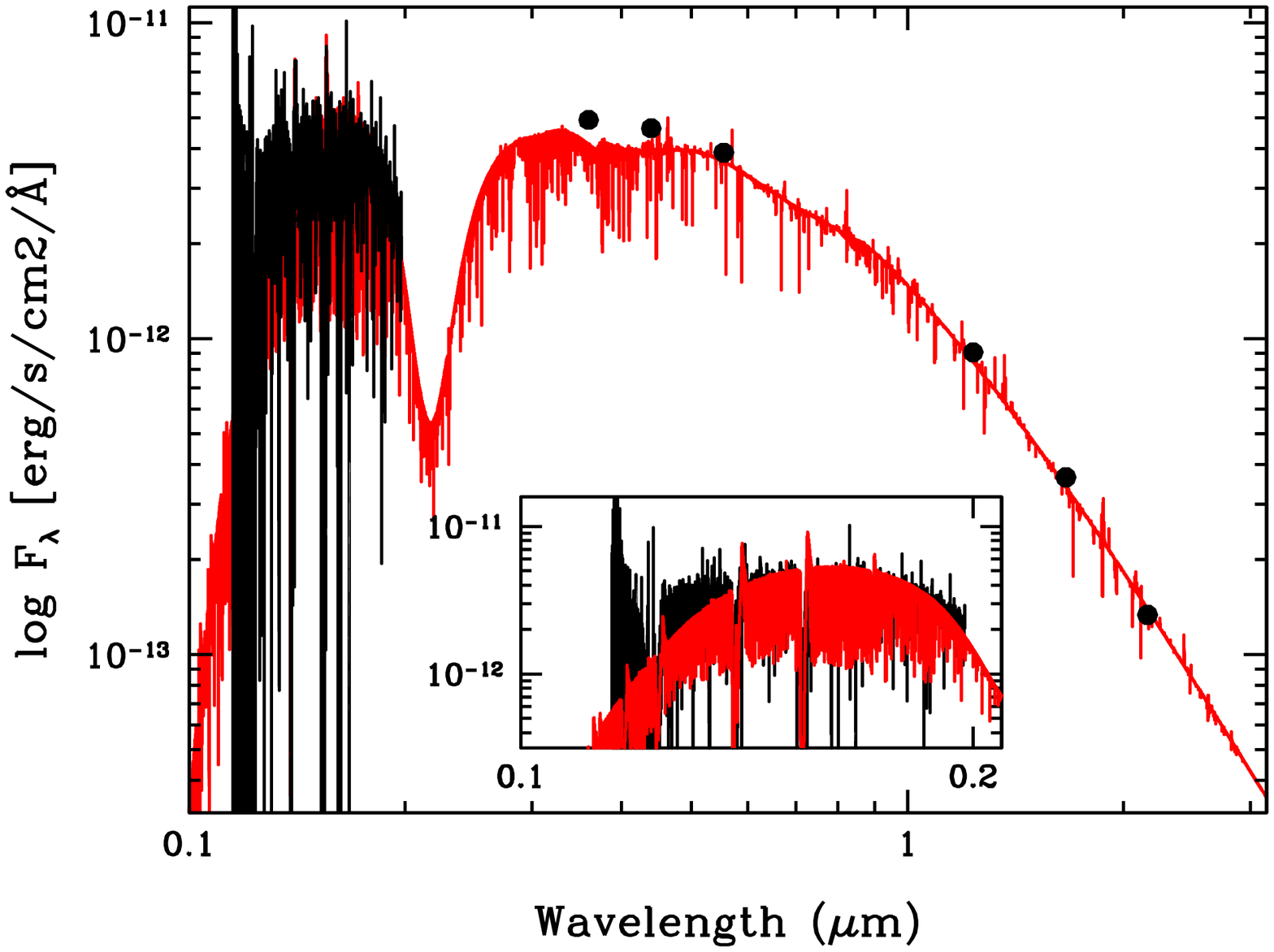}
\label{SED_hd193514}}
\subfigure[HD\,193595]{
\includegraphics[scale=0.400,bb=17 148 569 559,clip]{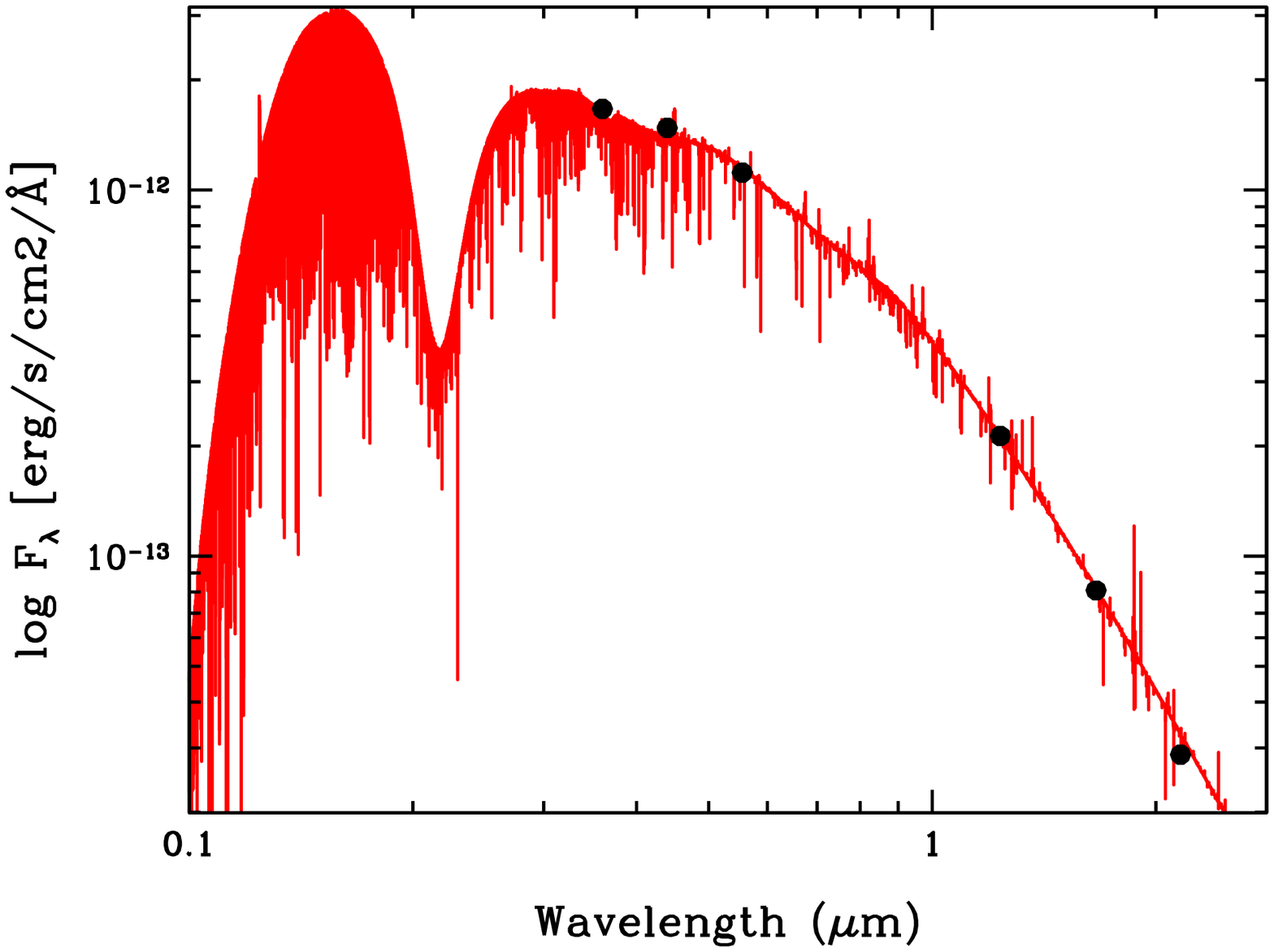}
\label{SED_hd193595}}
\subfigure[HD\,193682]{
\includegraphics[scale=0.400,bb=17 148 569 559,clip]{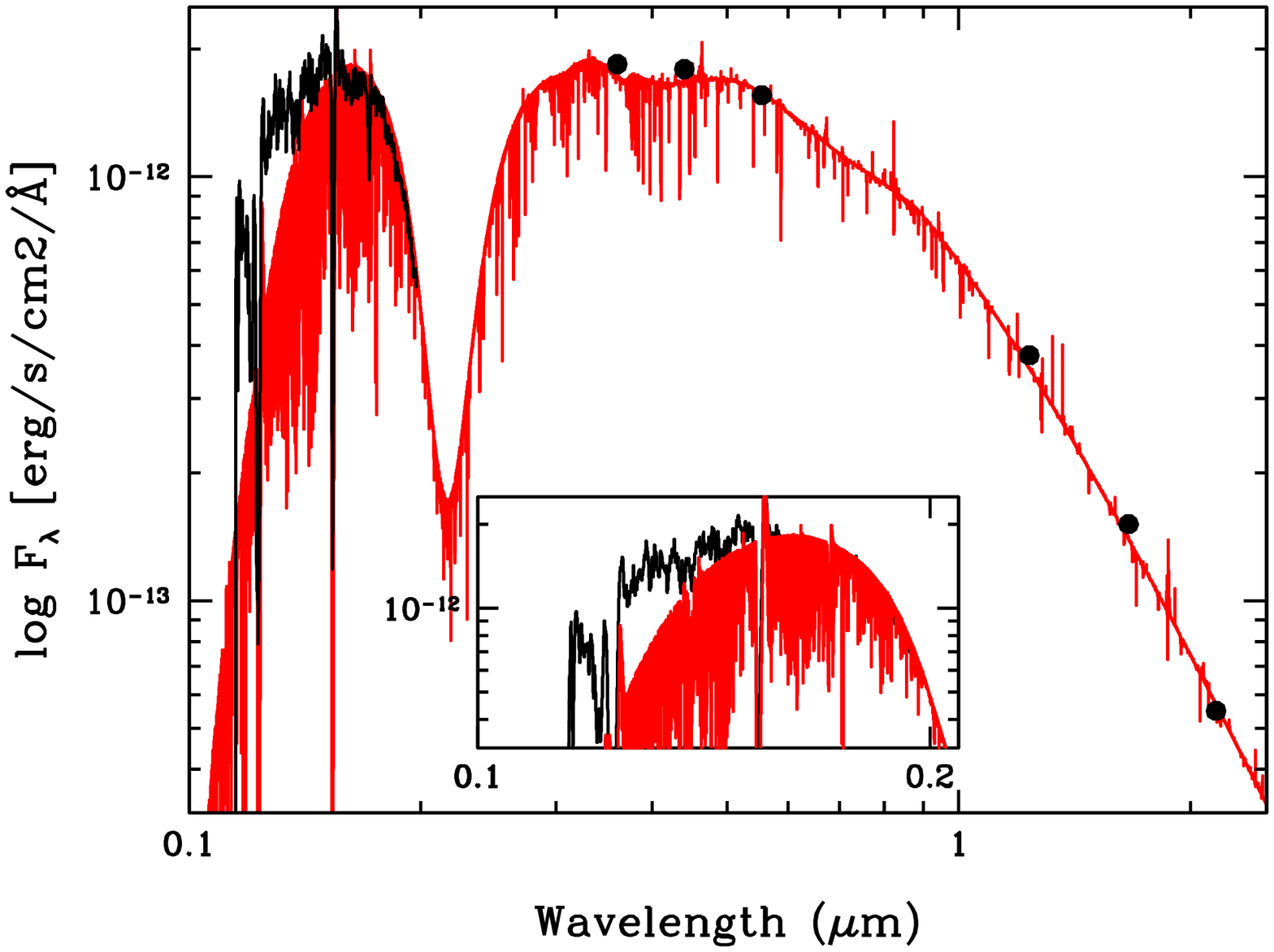}
\label{SED_hd193682}}
\subfigure[HD\,194094]{
\includegraphics[scale=0.400,bb=17 148 569 559,clip]{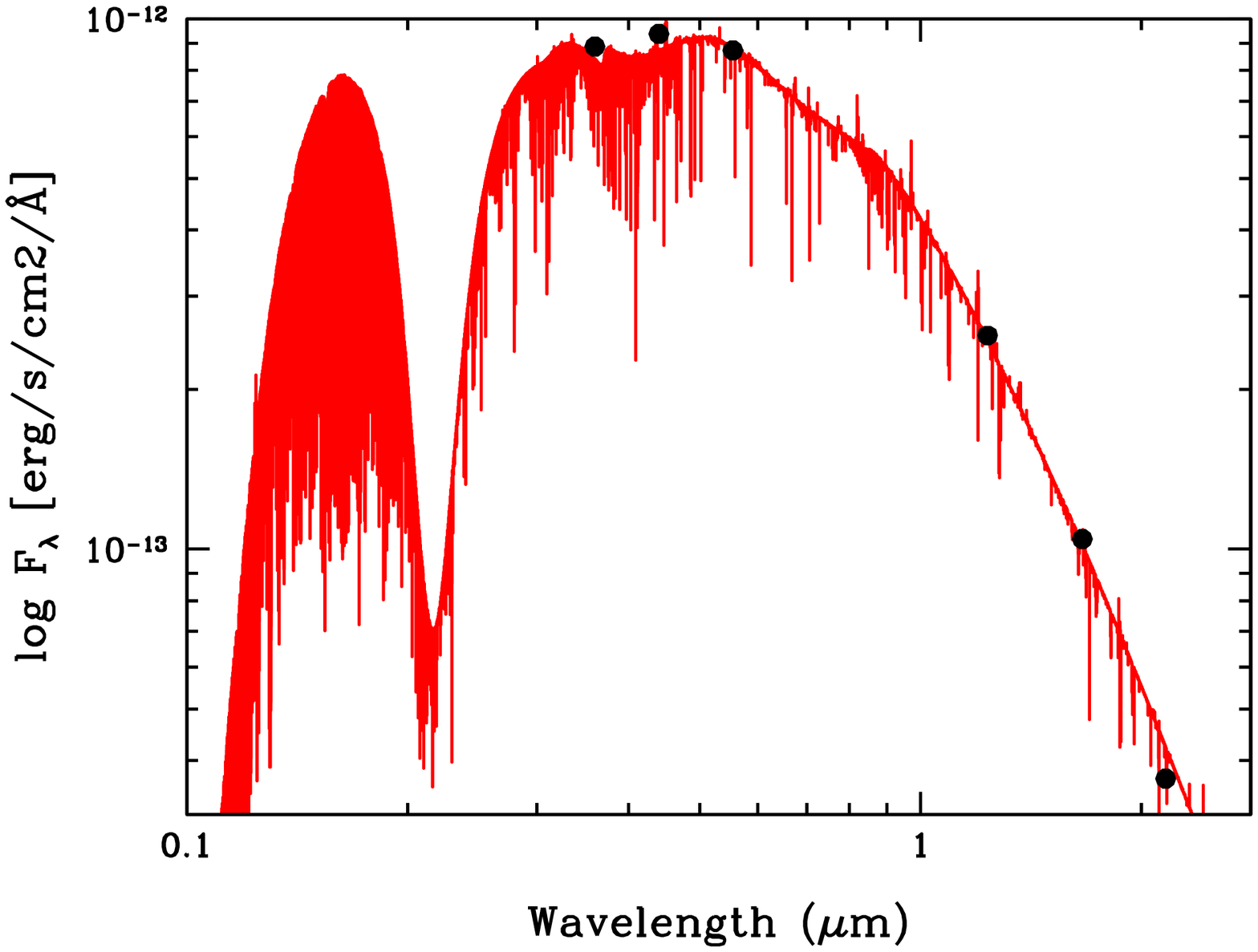}
\label{SED_hd194094}}
\subfigure[HD\,228841]{
\includegraphics[scale=0.400,bb=17 148 569 559,clip]{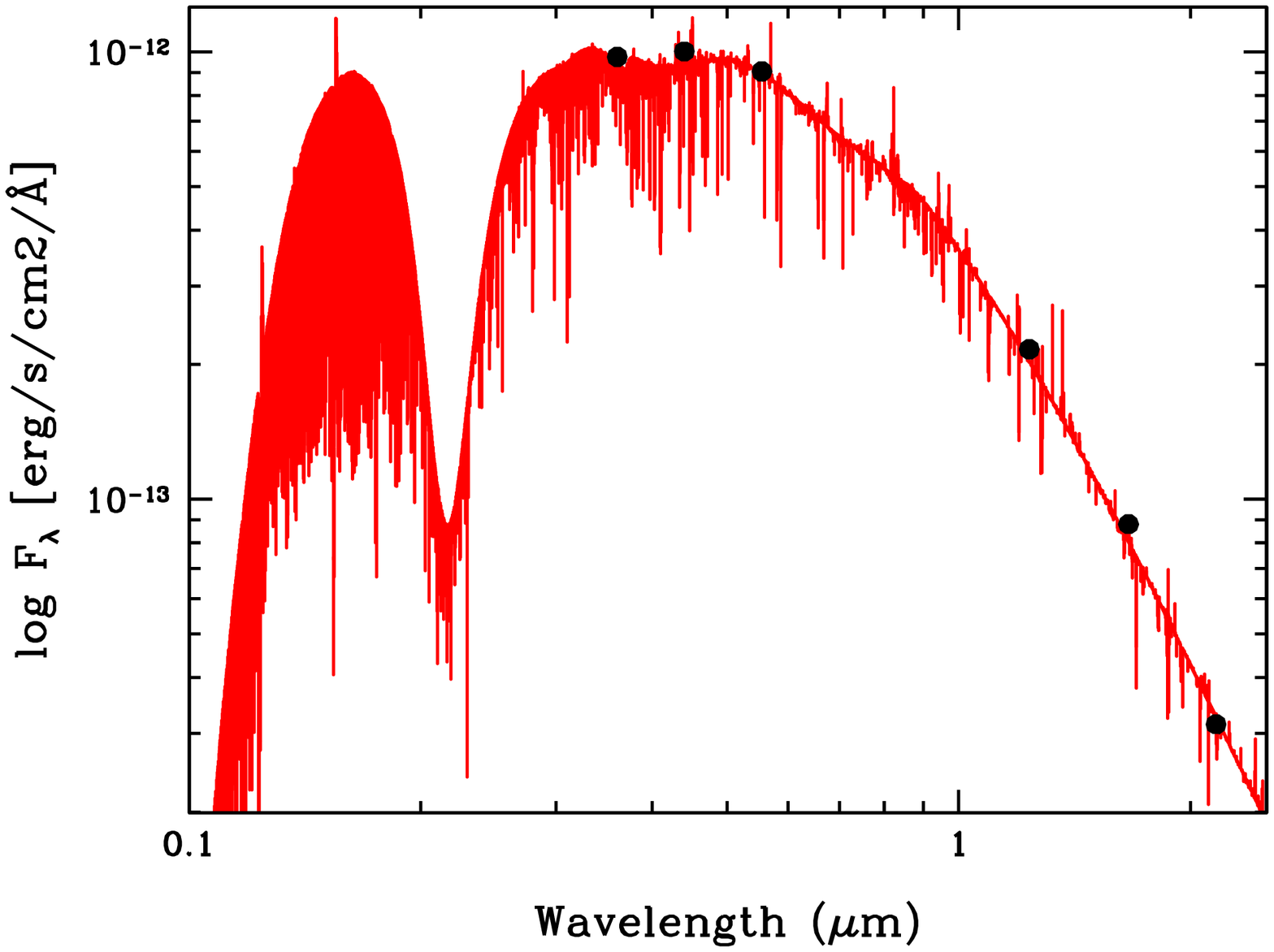}
\label{SED_hd228841}}
\subfigure[HD\,194280]{
\includegraphics[scale=0.400,bb=17 148 569 559,clip]{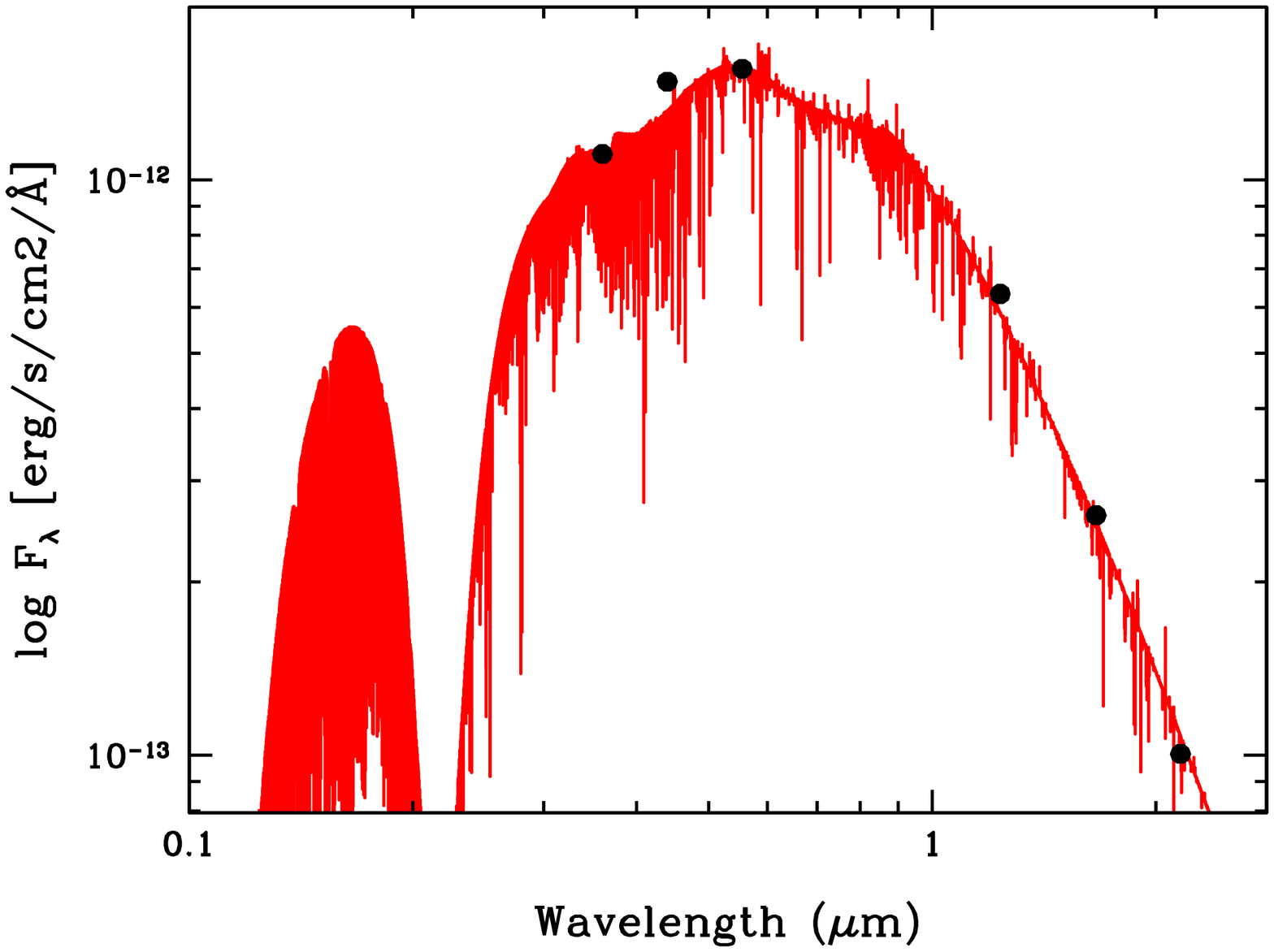}
\label{SED_hd194280}}
\subfigure[HD\,229234]{
\includegraphics[scale=0.400,bb=17 148 569 559,clip]{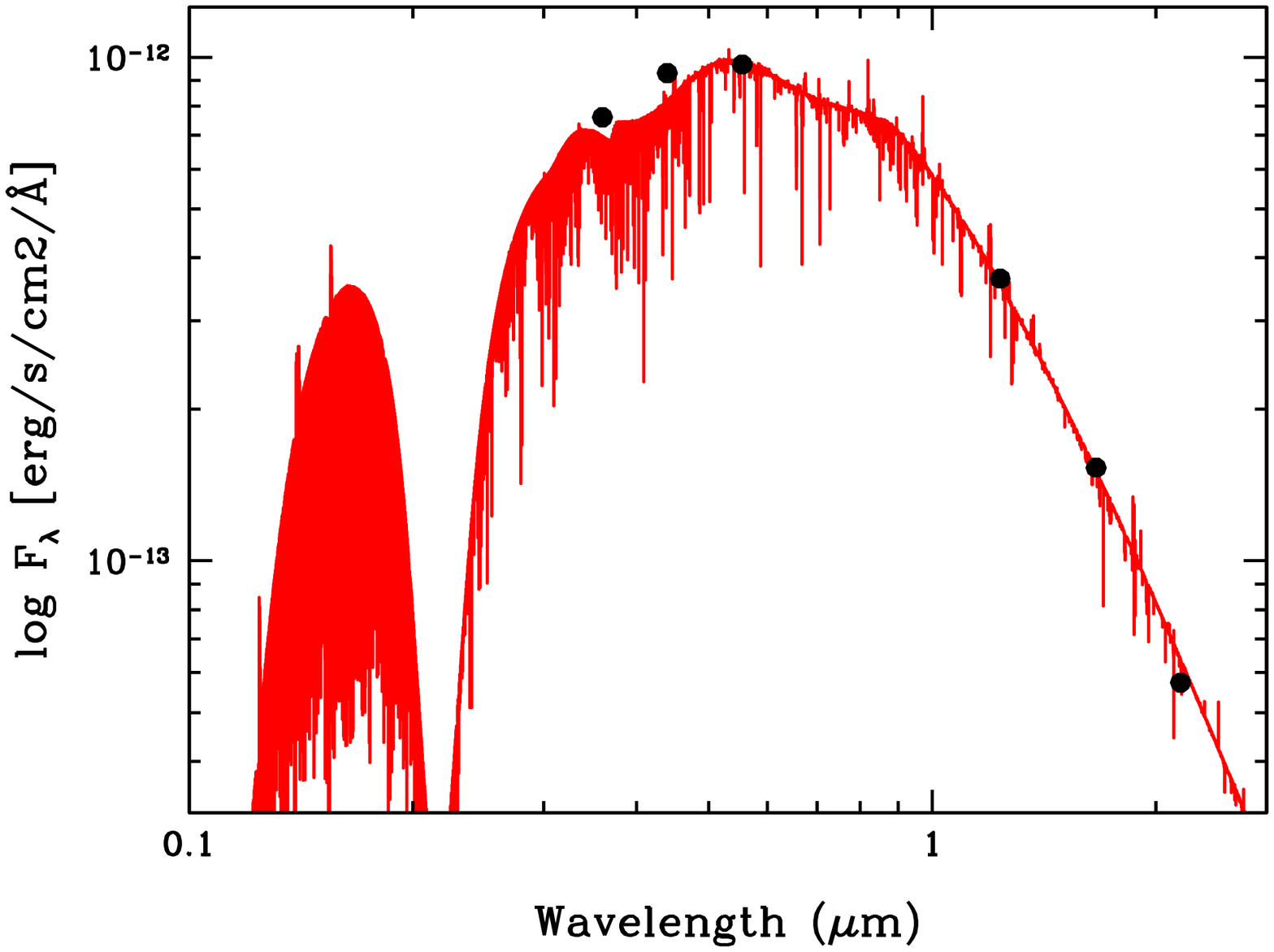}
\label{SED_hd229234}}
\qquad \qquad \qquad \quad \quad
\subfigure[HD\,190864]{
\includegraphics[scale=0.400,bb=17 148 569 559,clip]{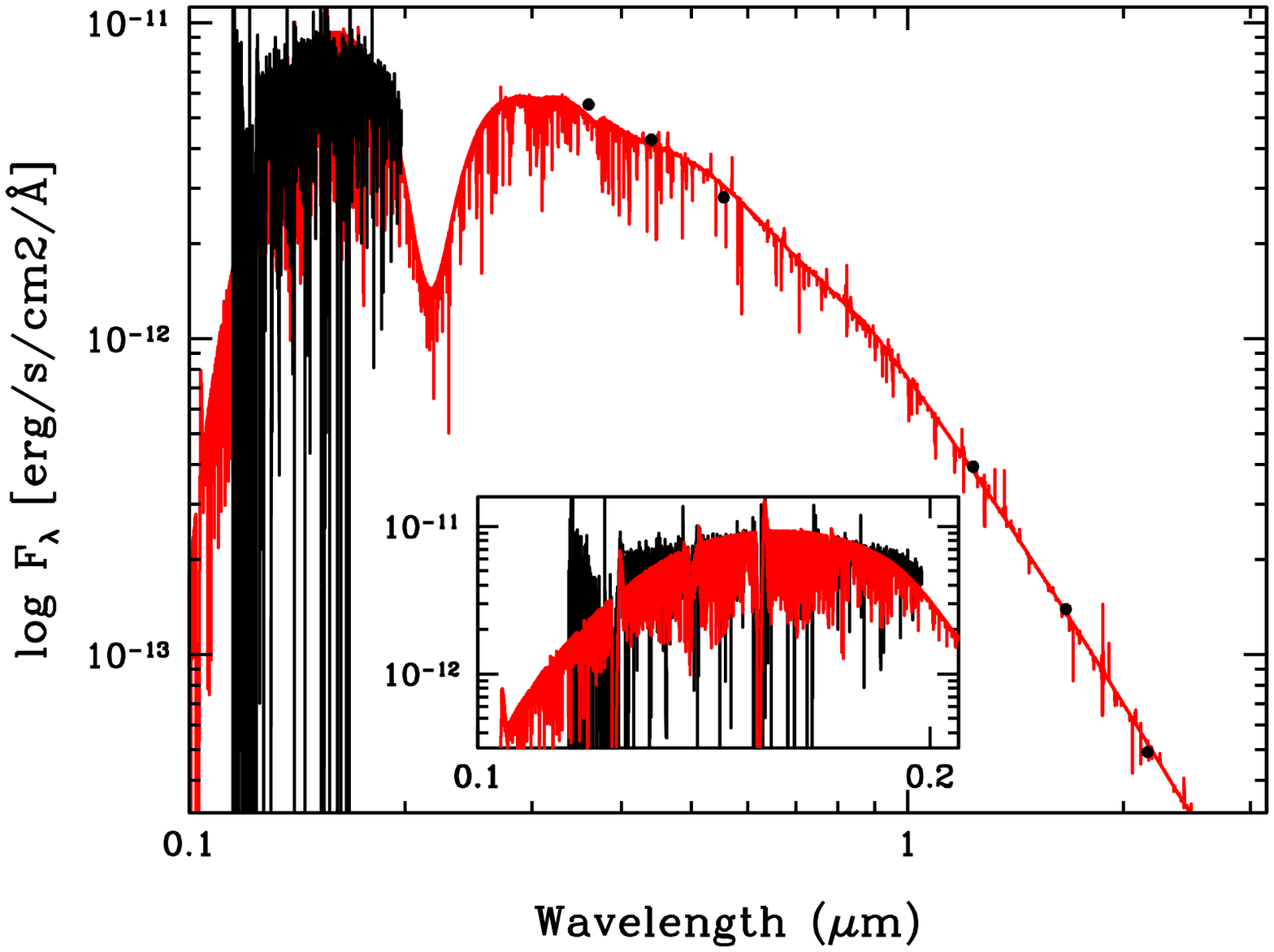}
\label{SED_hd190864}}
\caption{Synthetic SEDs (in red) compared to UBVJHK photometry for the stars in our sample (in black). The distance, E(B-V) and $R_V$ were determined from the best agreement between models and observations.}\label{SED_1}
\end{figure*}
\newpage
\begin{figure*}[htbp]
\subfigure[HD\,227018]{
\includegraphics[scale=0.400,bb=17 148 569 559,clip]{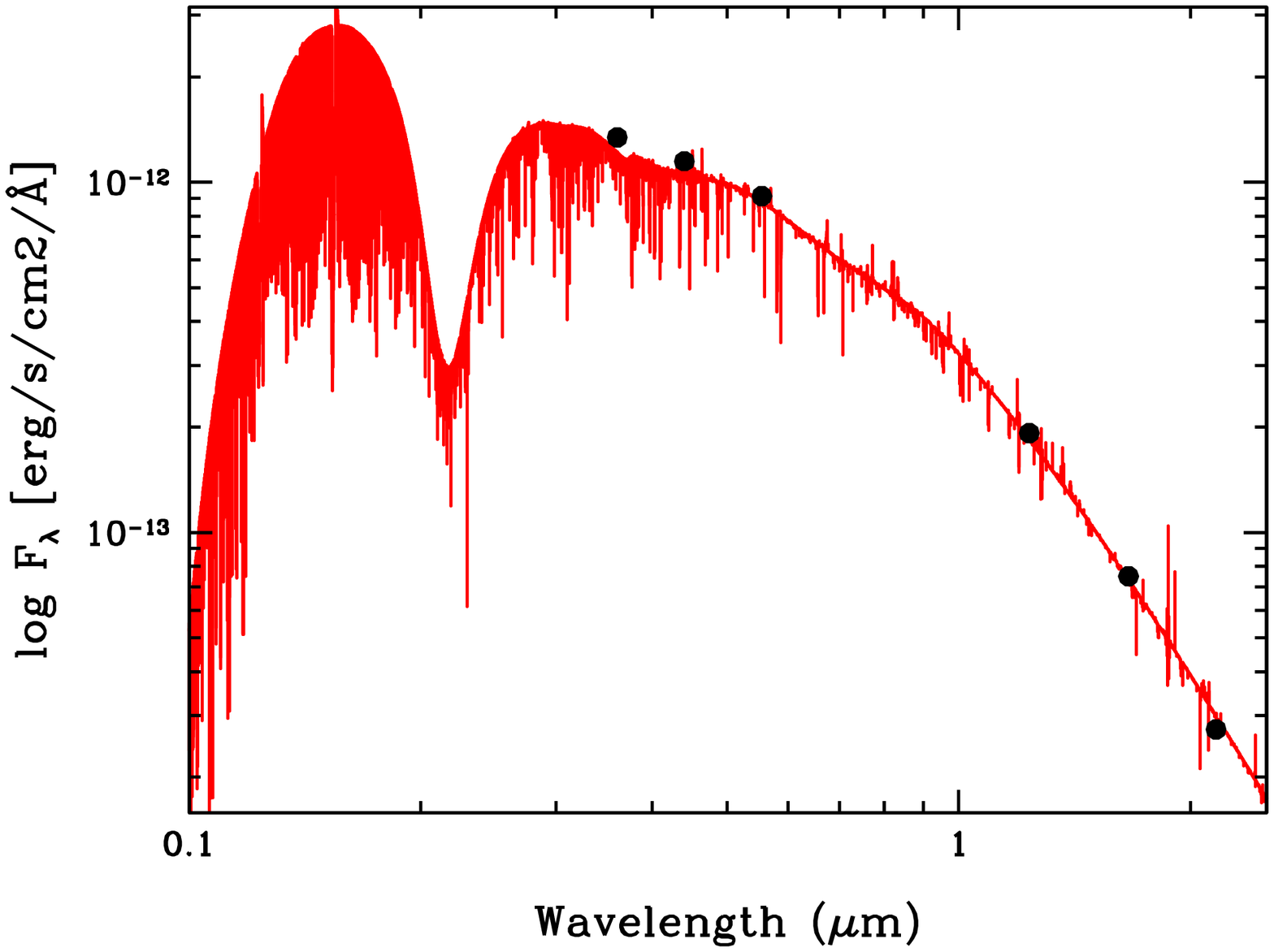}
\label{SED_hd227018}}
\subfigure[HD\,227245]{
\includegraphics[scale=0.400,bb=17 148 569 559,clip]{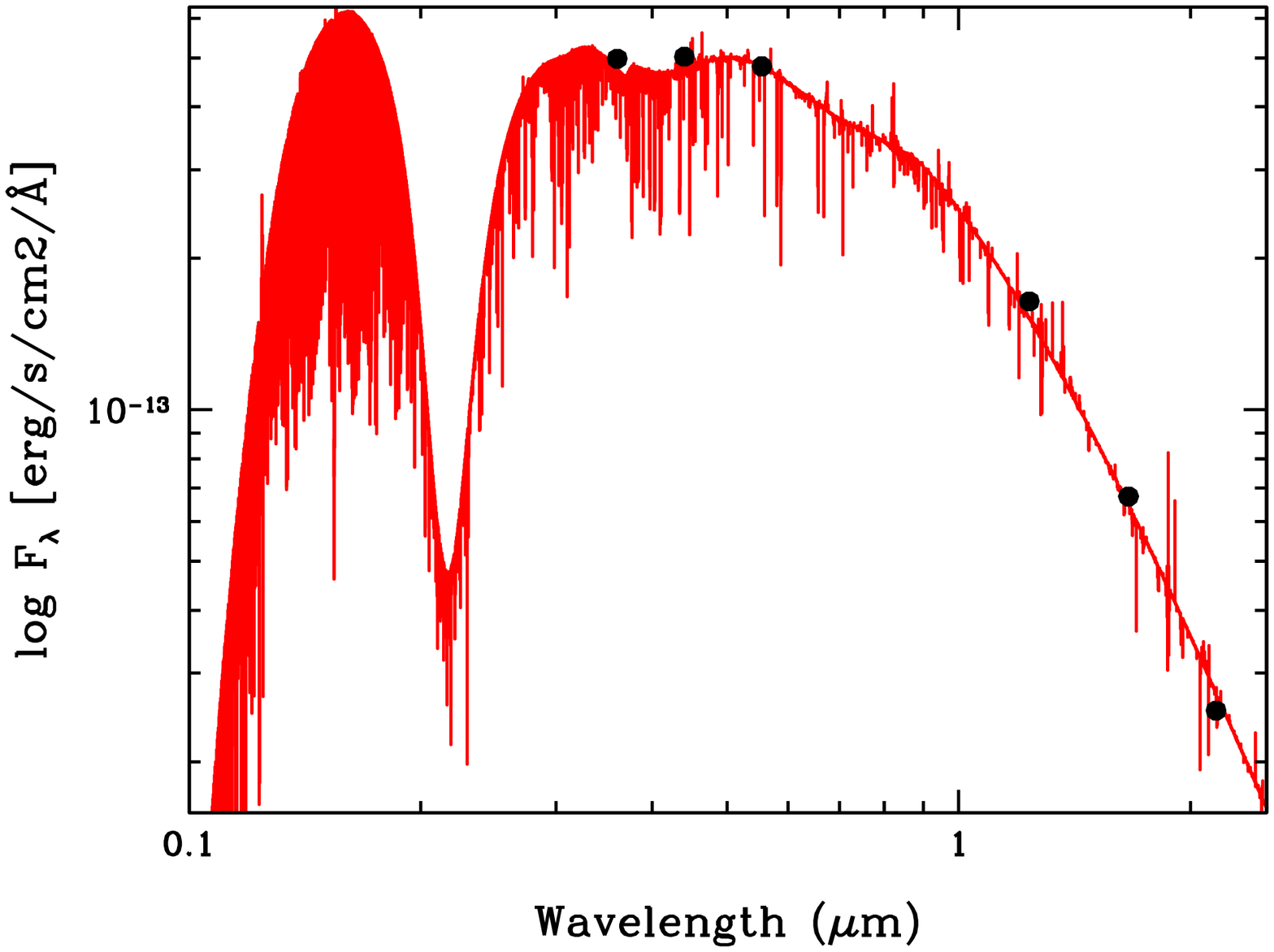}
\label{SED_hd227245}}
\subfigure[HD\,227757]{
\includegraphics[scale=0.400,bb=17 148 569 559,clip]{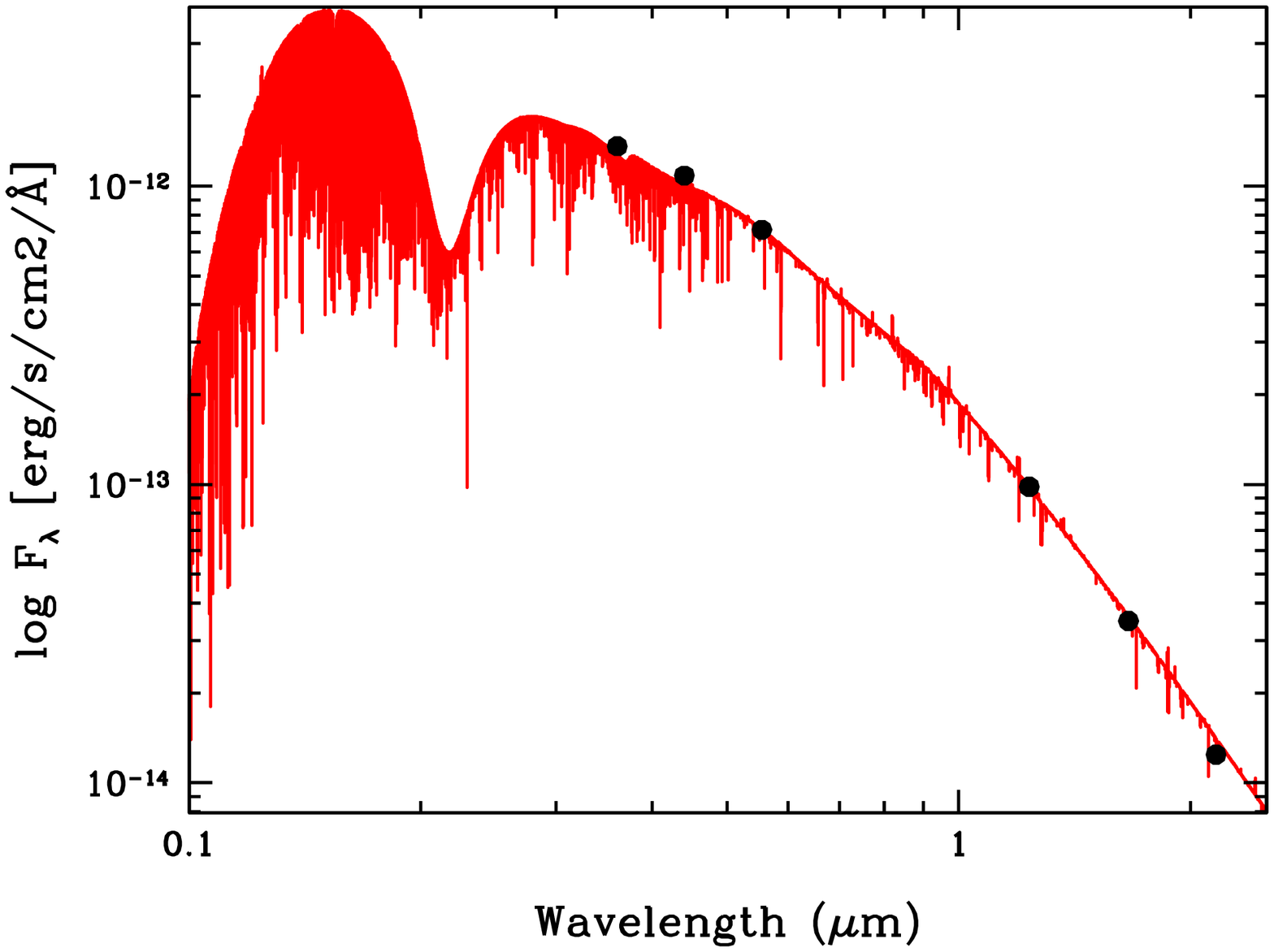}
\label{SED_hd227757}}
\subfigure[HD\,191423]{
\includegraphics[scale=0.400,bb=17 148 569 559,clip]{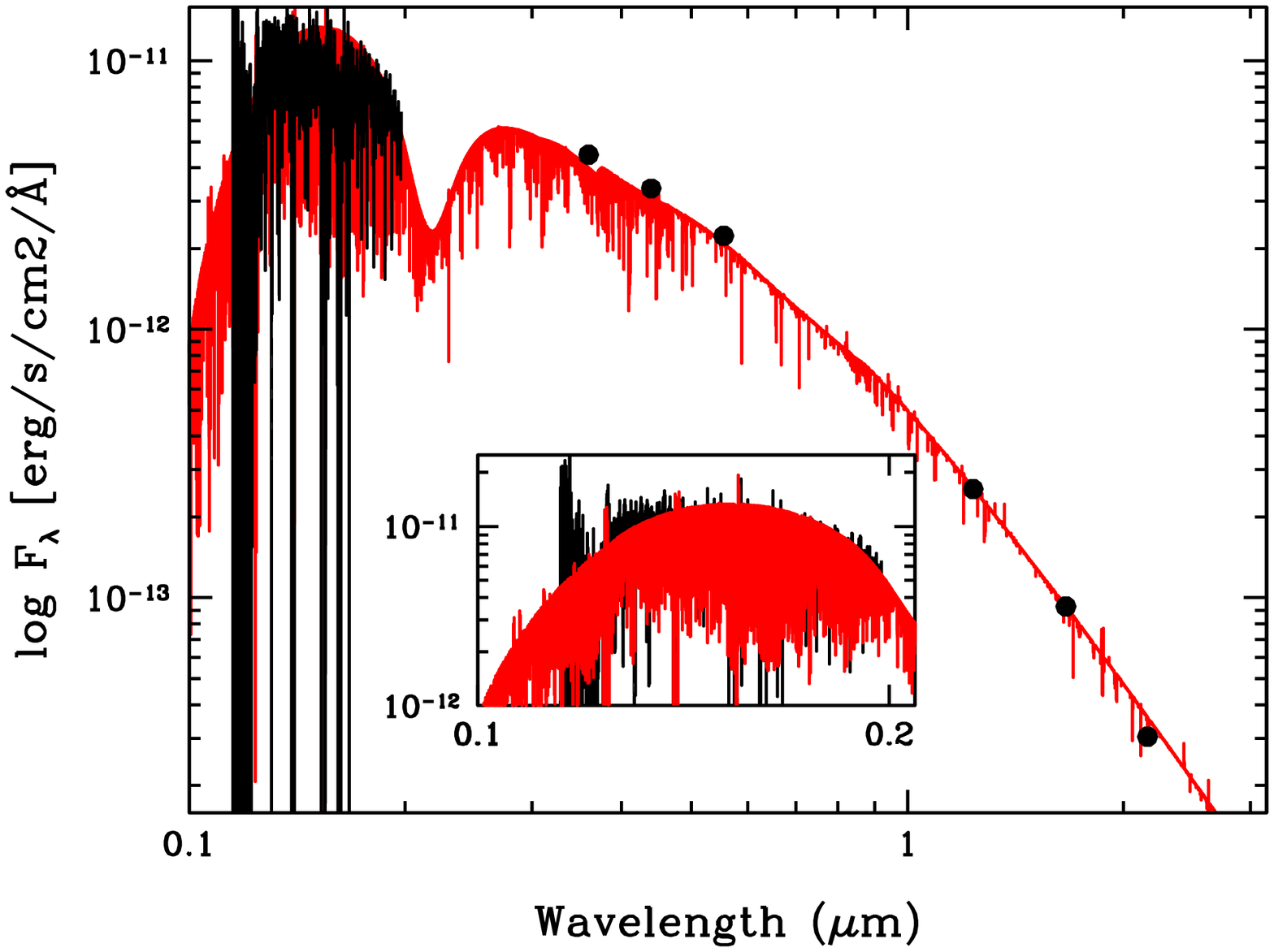}
\label{SED_hd191423}}
\subfigure[HD\,191978]{
\includegraphics[scale=0.400,bb=17 148 569 559,clip]{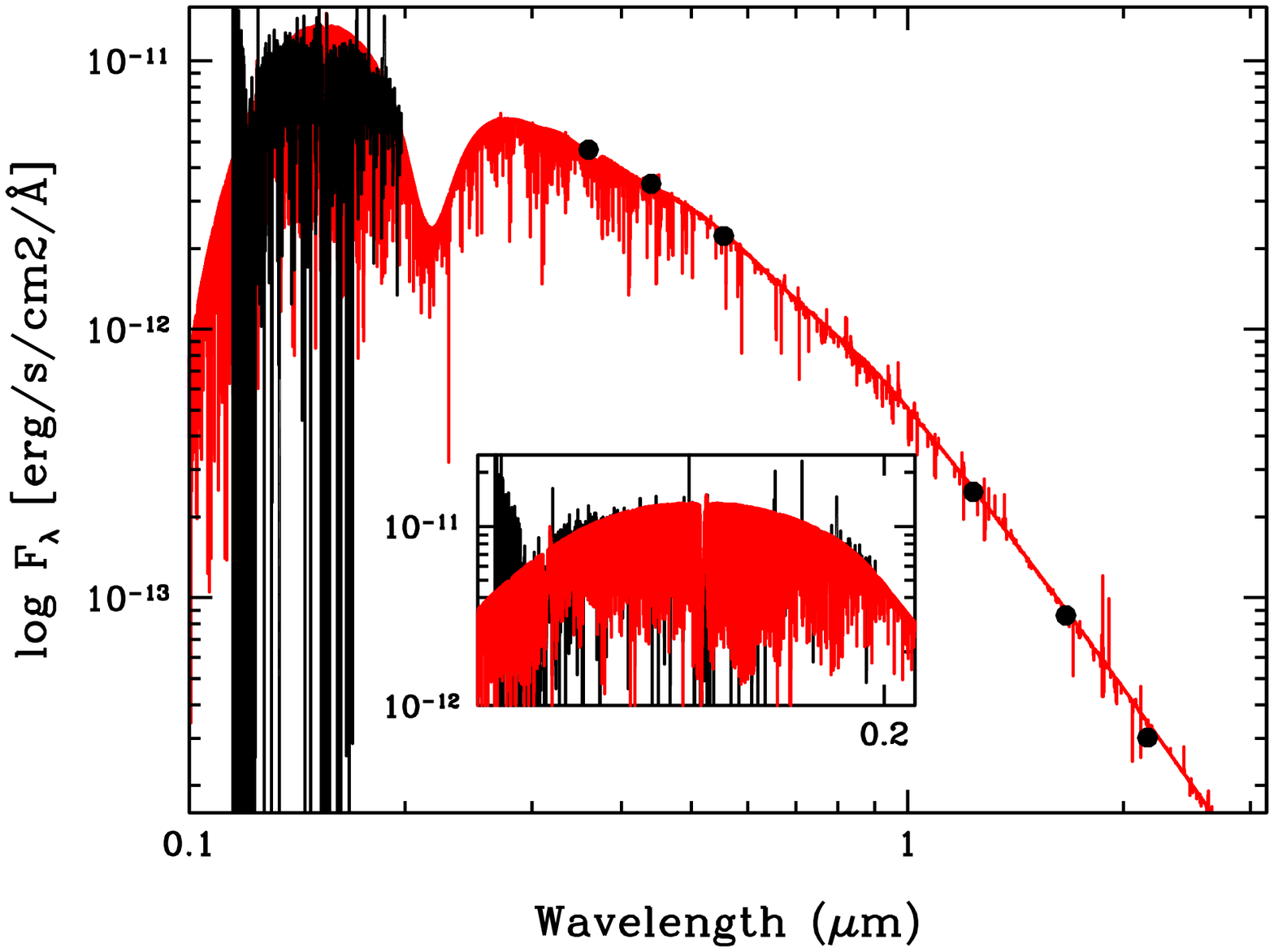}
\label{SED_hd191978}}
\subfigure[HD\,193117]{
\includegraphics[scale=0.400,bb=17 148 569 559,clip]{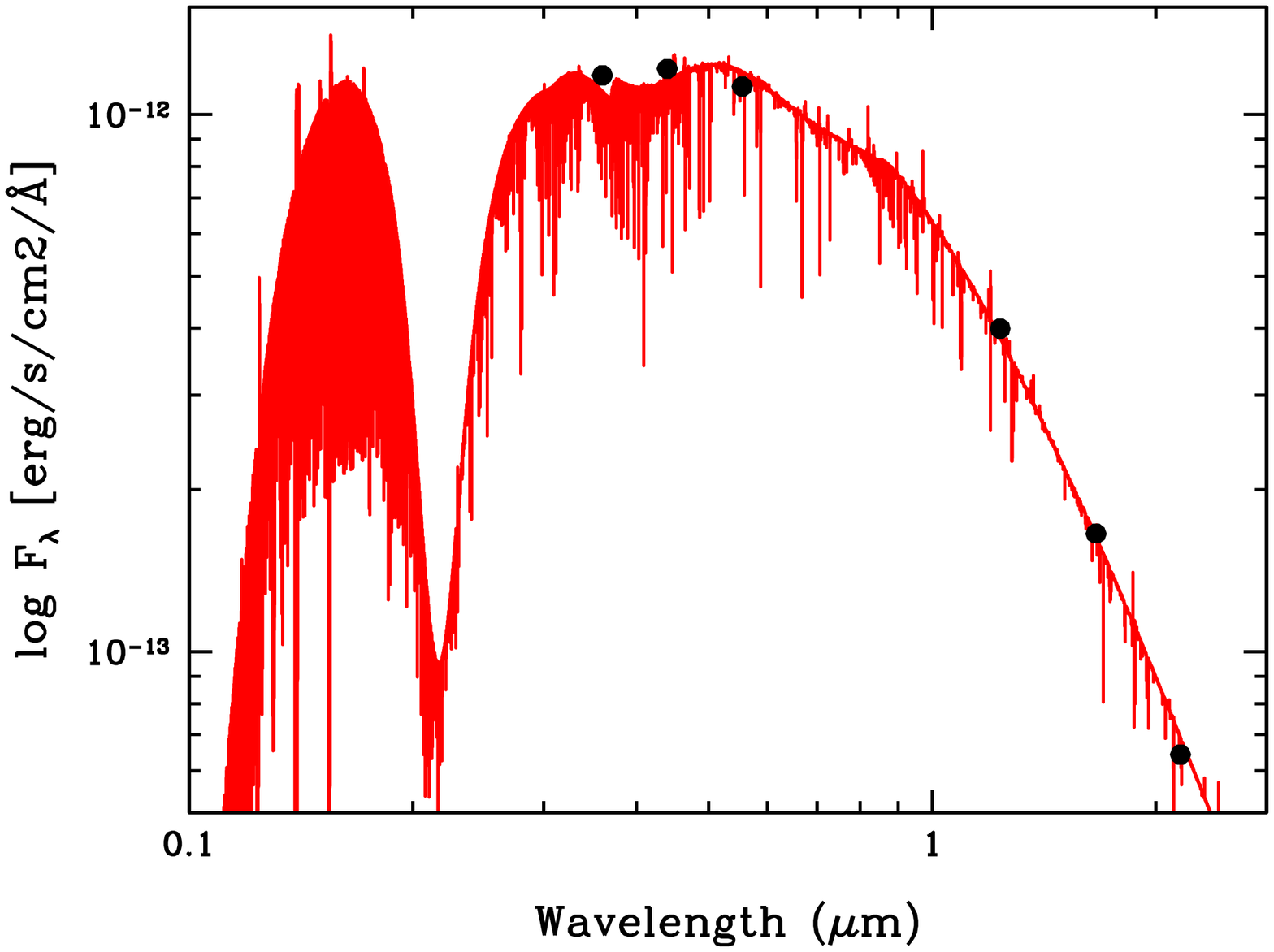}
\label{SED_hd193117}}
\subfigure[HD\,194334]{
\includegraphics[scale=0.400,bb=17 148 569 559,clip]{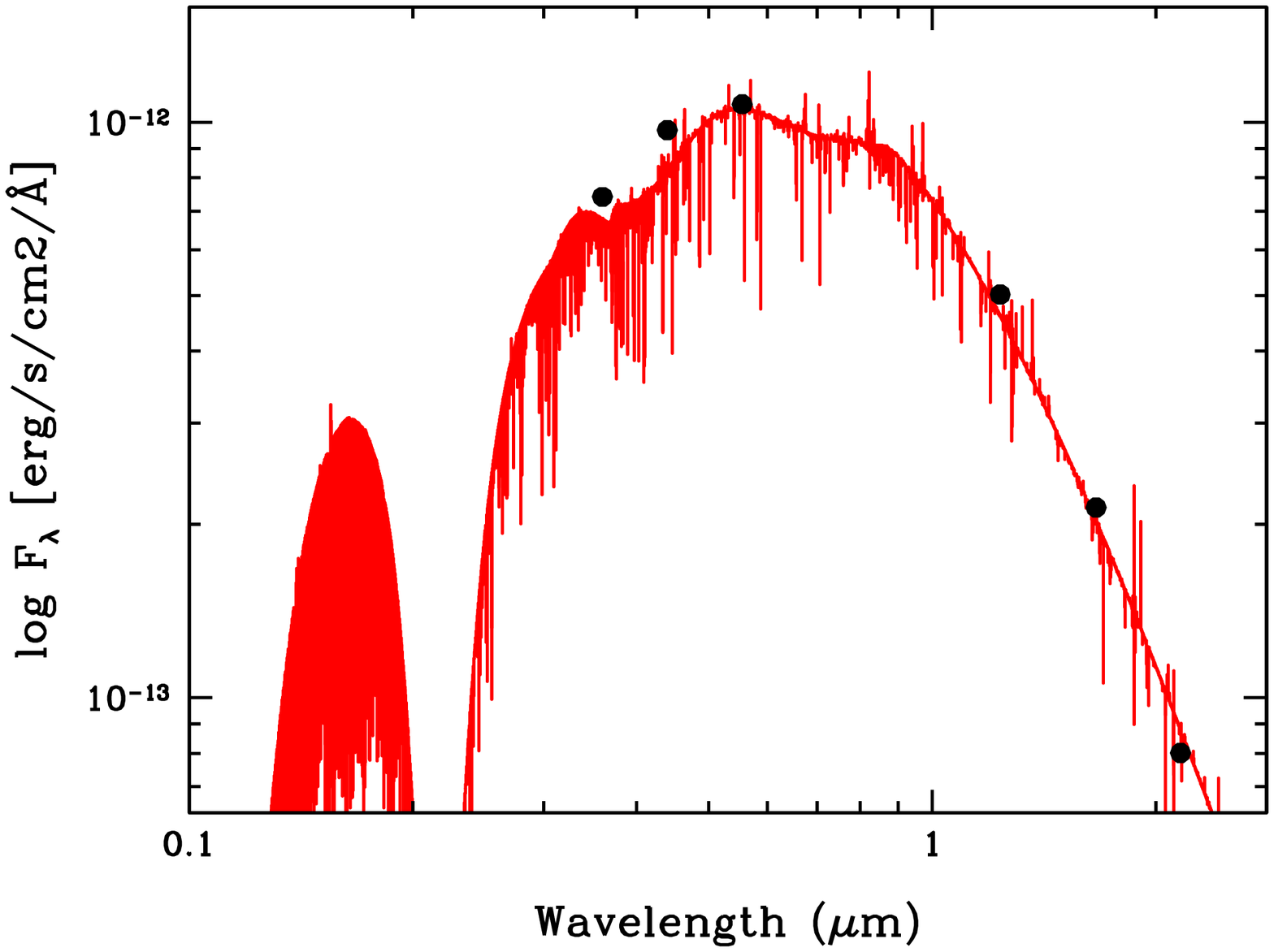}
\label{SED_hd194334}}
\qquad \qquad \qquad \quad \quad
\subfigure[HD\,195213]{
\includegraphics[scale=0.400,bb=17 148 569 559,clip]{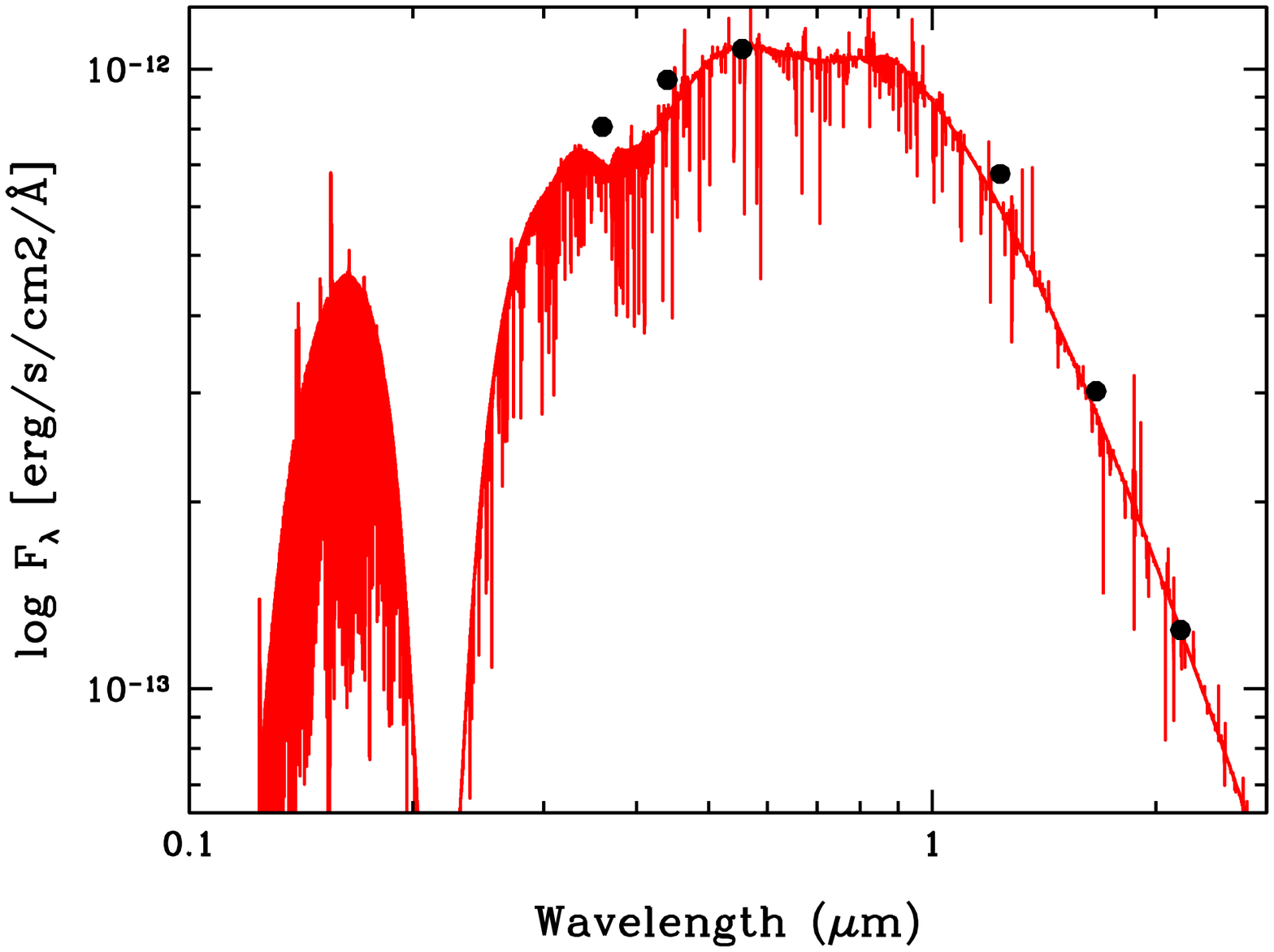}
\label{SED_hd195213}}
\caption{(continued).}\label{SED_2}
\end{figure*}
\newpage
\begin{figure*}[htbp]
\includegraphics[scale=0.9,bb=14 149 572 700,clip]{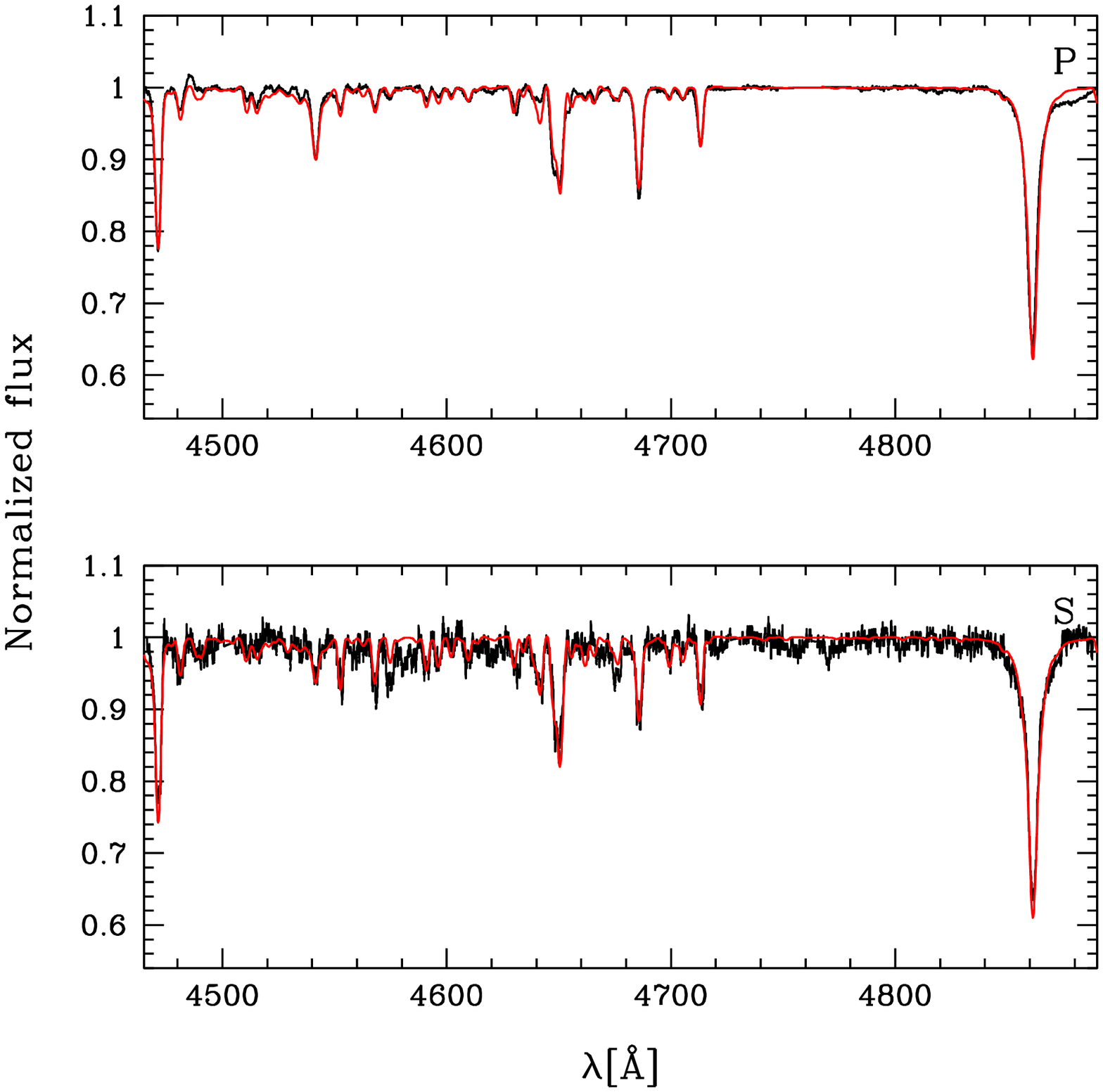}
\label{hd193443_cmfgen}
\caption{Best-fit model for HD\,193443 primary and secondary (red line) compared to disentangled spectra (black line).}
\end{figure*}
\newpage
\begin{figure*}[htbp]
\includegraphics[scale=0.9,bb=23 149 583 700,clip]{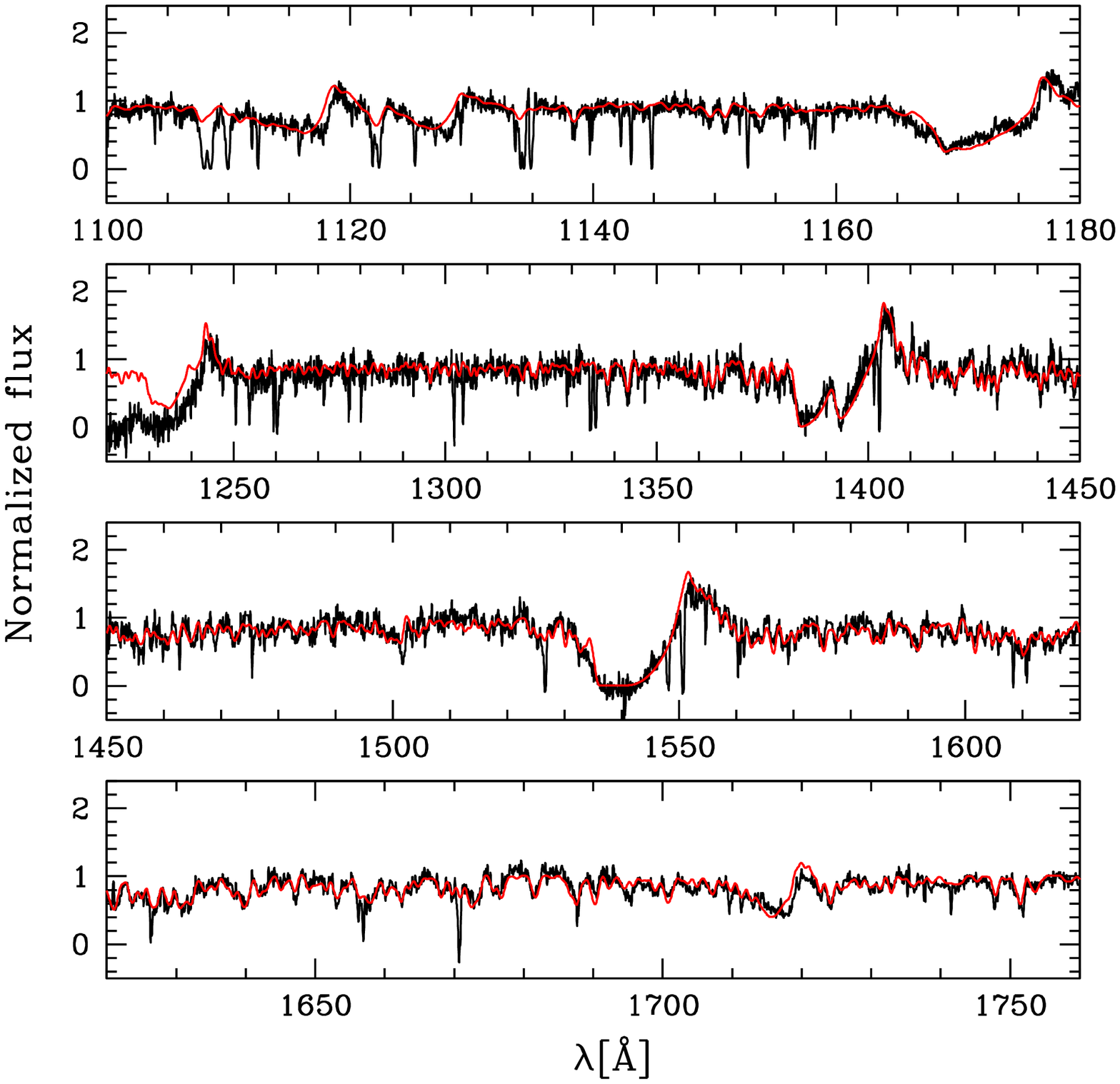}
\label{hd193514_cmfgen}
\caption{Best-fit model for HD\,193514 (red line) compared to FUSE and IUE spectra (black line).}
\end{figure*}
\begin{figure*}[htbp]
\includegraphics[scale=0.9,bb=23 149 583 700,clip]{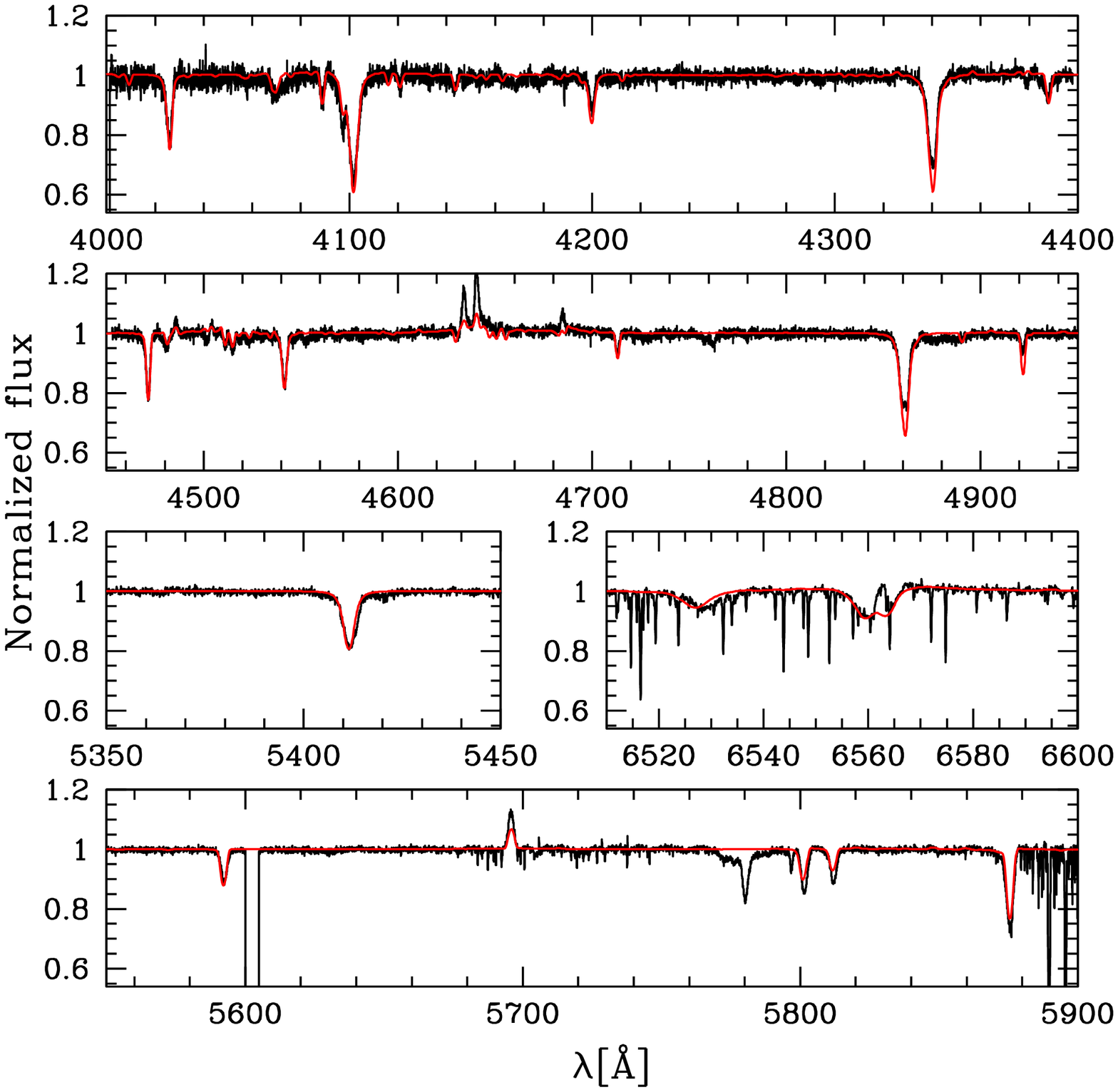}
\label{hd193514_cmfgen}
\caption{Best-fit model for HD\,193514 (red line) compared to Elodie spectrum (black line).}
\end{figure*}
\newpage
\begin{figure*}[htbp]
\includegraphics[scale=0.9,bb=33 158 583 700,clip]{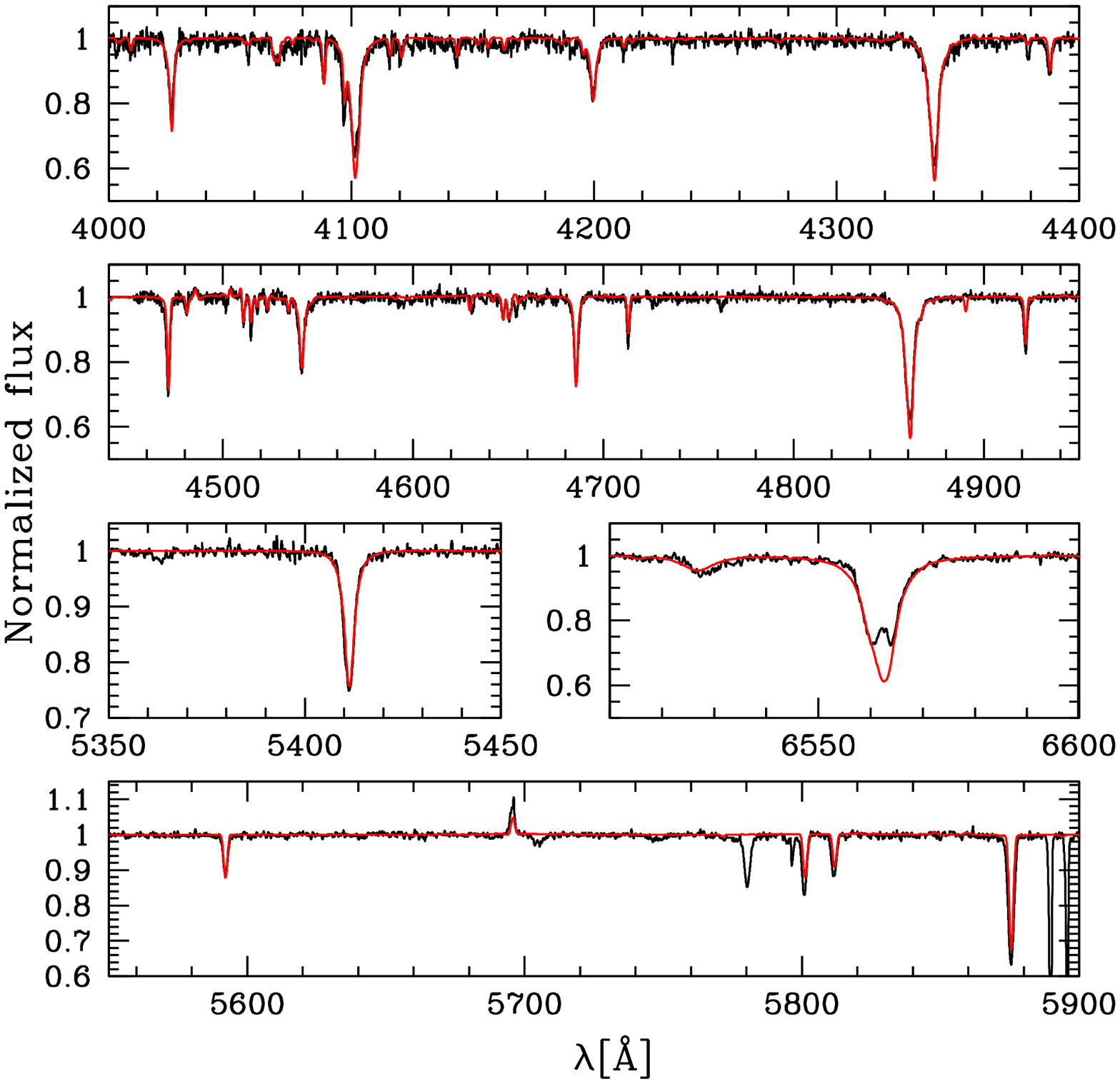}
\label{hd193595_cmfgen}
\caption{Best-fit model for HD\,193595 (red line) compared to  Espresso spectrum (black line).}
\end{figure*}
\newpage
\clearpage
\begin{figure*}[htbp]
\includegraphics[scale=0.9,bb=23 149 583 700,clip]{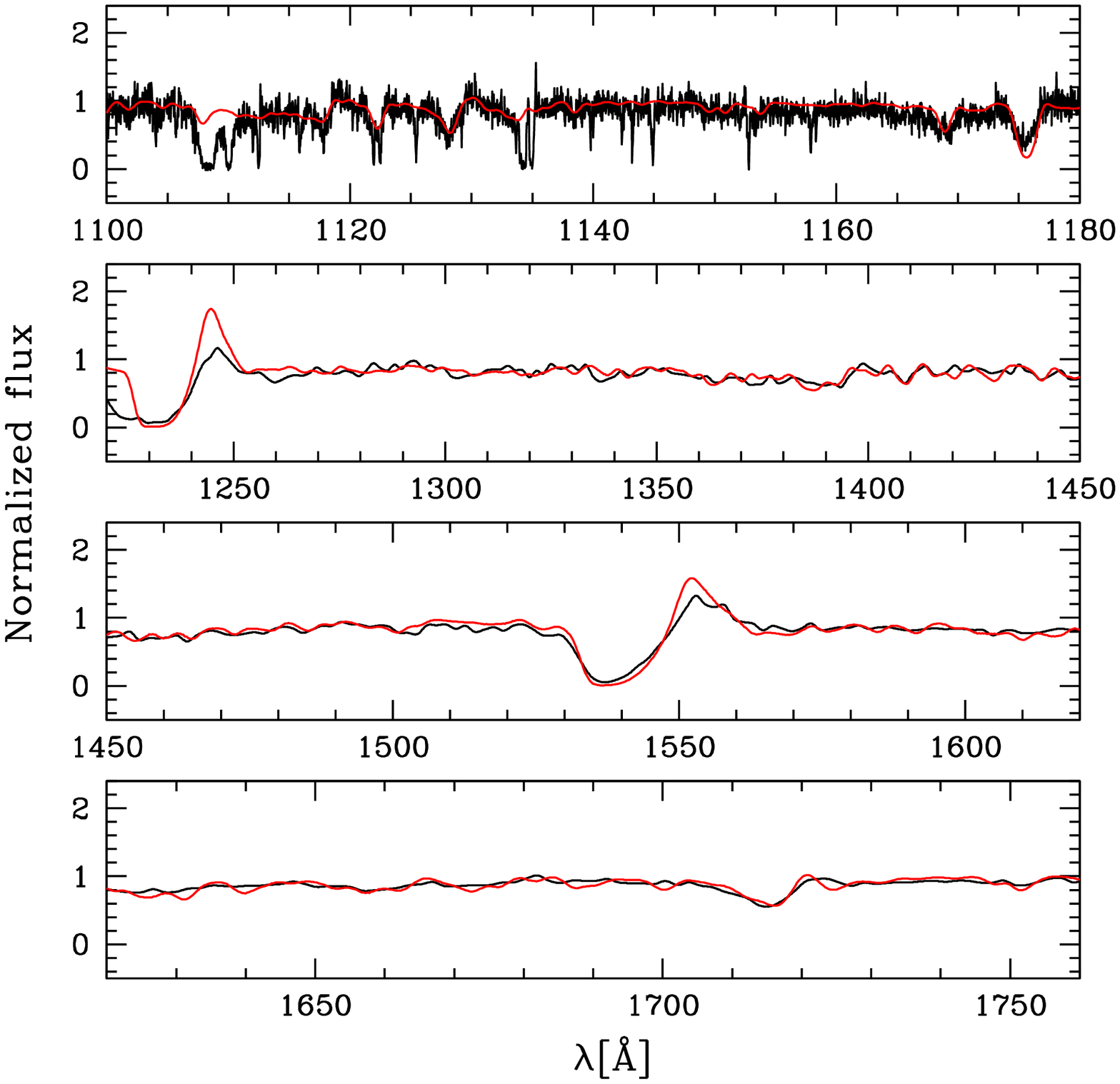}
\label{hd193682_cmfgen}
\caption{Best-fit model for HD\,193682 (red line) compared to FUSE and IUE spectra (black line)}
\end{figure*}
\begin{figure*}[htbp]
\includegraphics[scale=0.9,bb=23 149 583 700,clip]{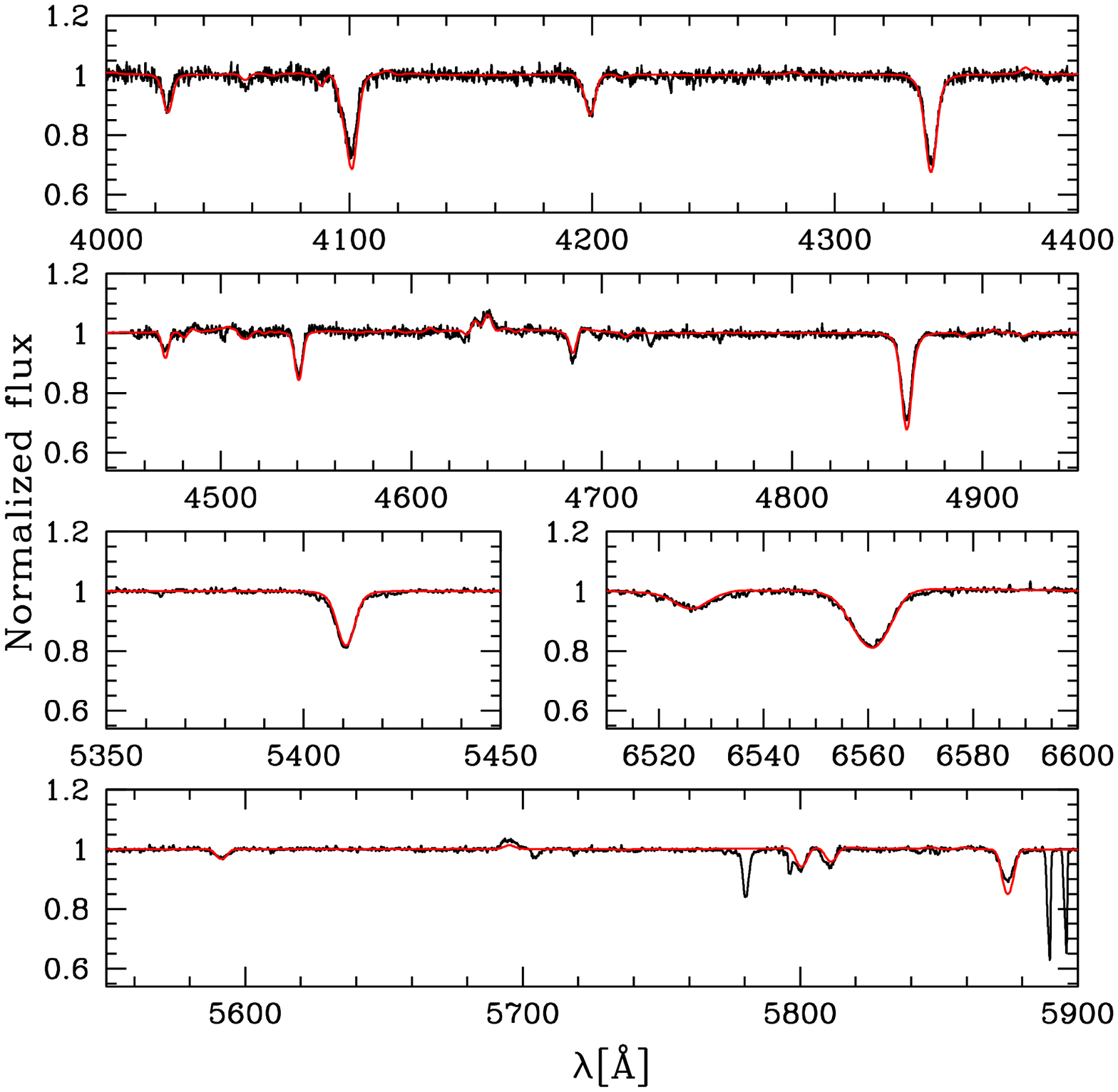}
\label{hd193682_cmfgen}
\caption{Best-fit model for HD\,193682 (red line) compared to Espresso spectrum (black line)}
\end{figure*}
\newpage
\begin{figure*}[htbp]
\includegraphics[scale=0.9,bb=15 158 583 700,clip]{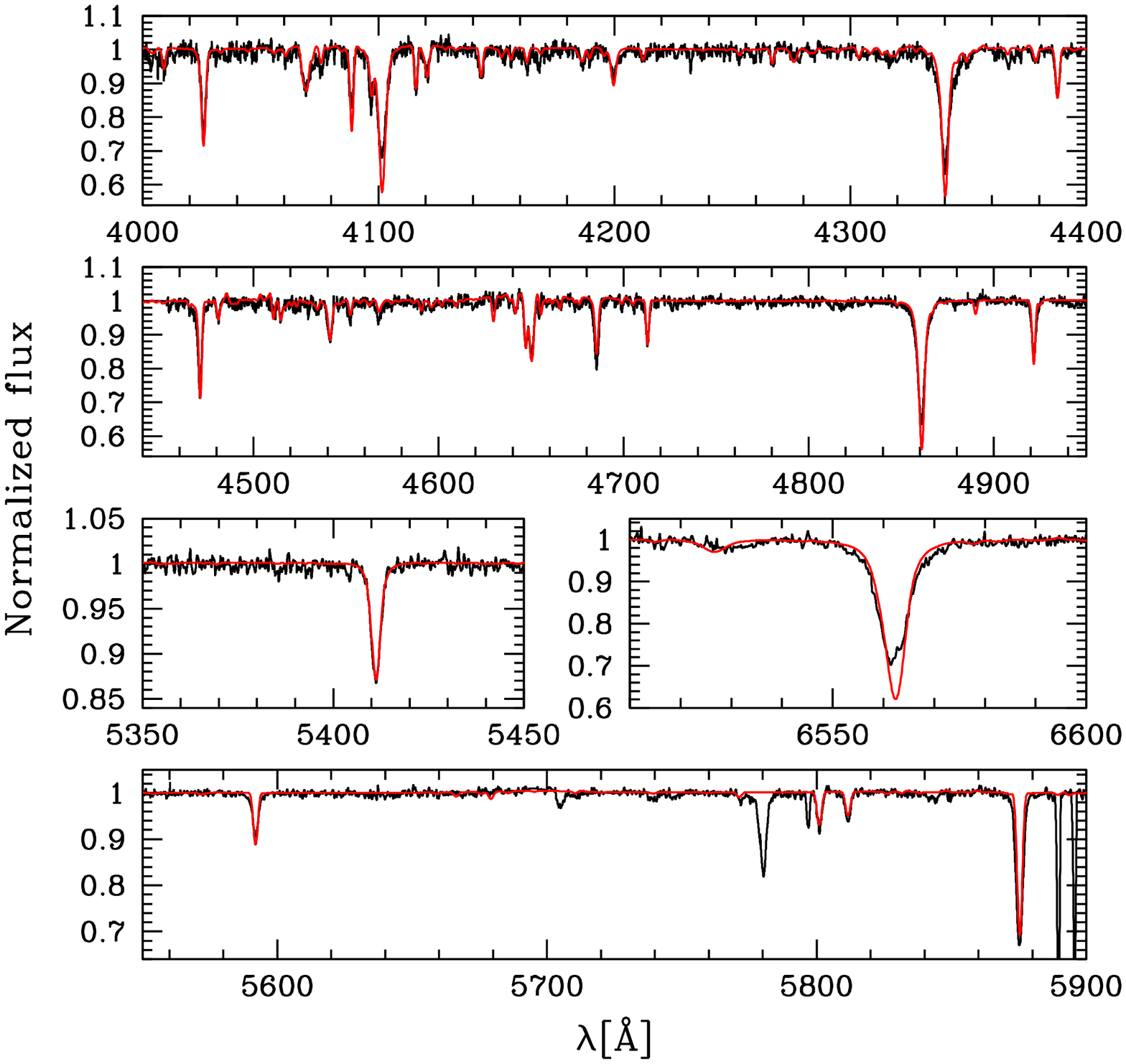}
\label{hd194094_cmfgen}
\caption{Best-fit model for HD\,194094 (red line) compared to Espresso spectrum (black line).}
\end{figure*}
\newpage
\begin{figure*}[htbp]
\includegraphics[scale=0.9,bb=14 158 593 415,clip]{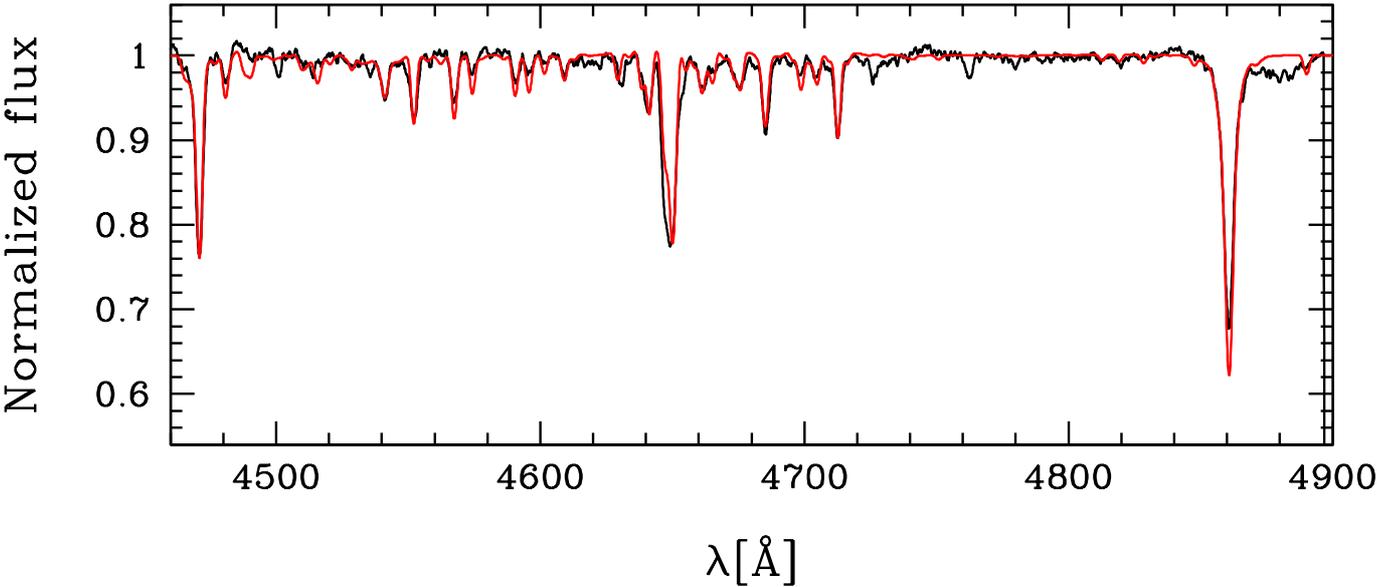}
\label{hd194280_cmfgen}
\caption{Best-fit model for HD\,194280 (red line) compared to Aur{\'e}lie spectrum (black line).}
\end{figure*}
\newpage
\begin{figure*}[htbp]
\includegraphics[scale=0.9,bb=15 158 583 700,clip]{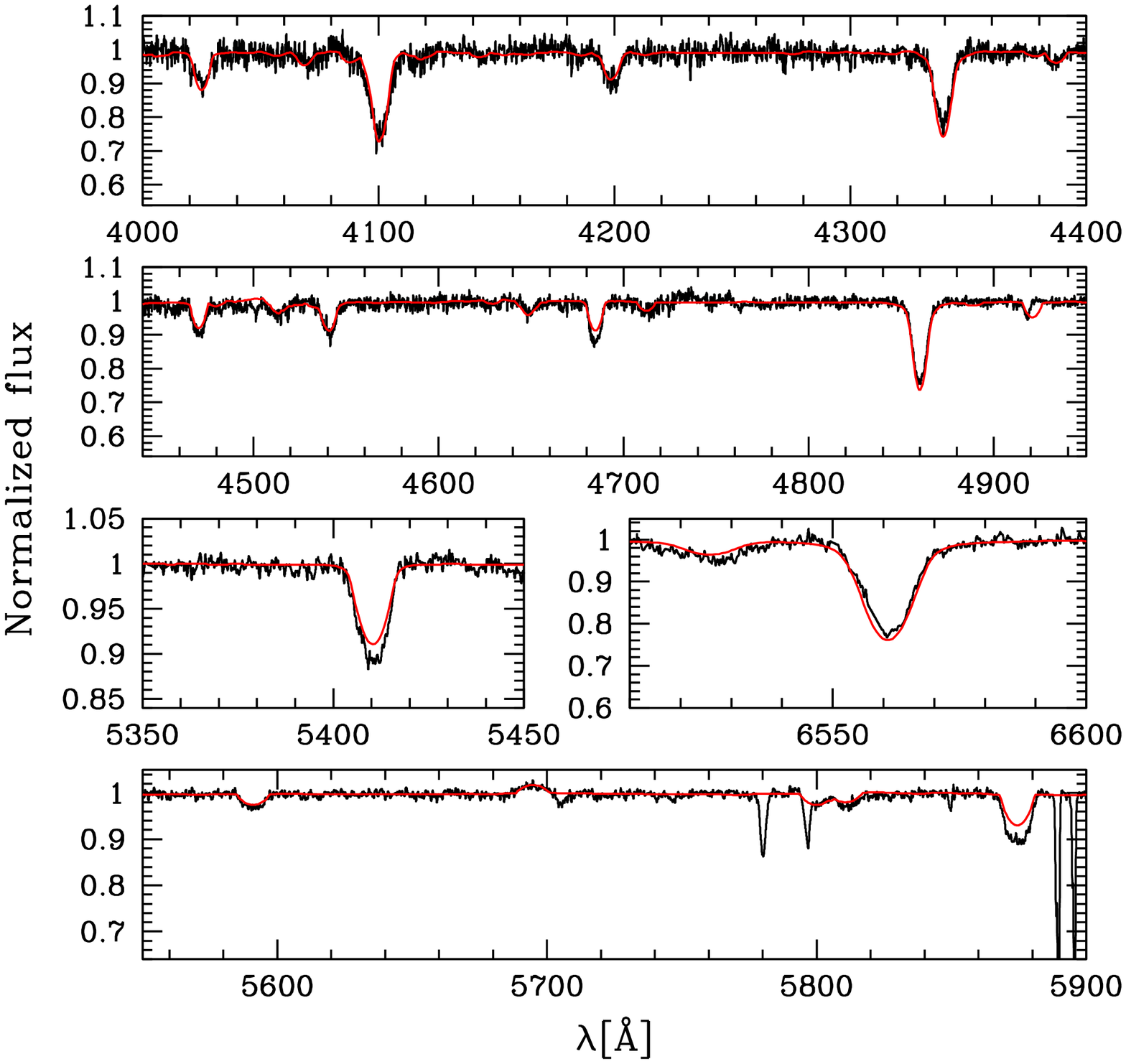}
\label{hd228841_cmfgen}
\caption{Best-fit model for HD\,228841 (red line) compared to Espresso spectrum (black line).}
\end{figure*}
\newpage
\begin{figure*}[htbp]
\includegraphics[scale=0.9,bb=14 149 572 700,clip]{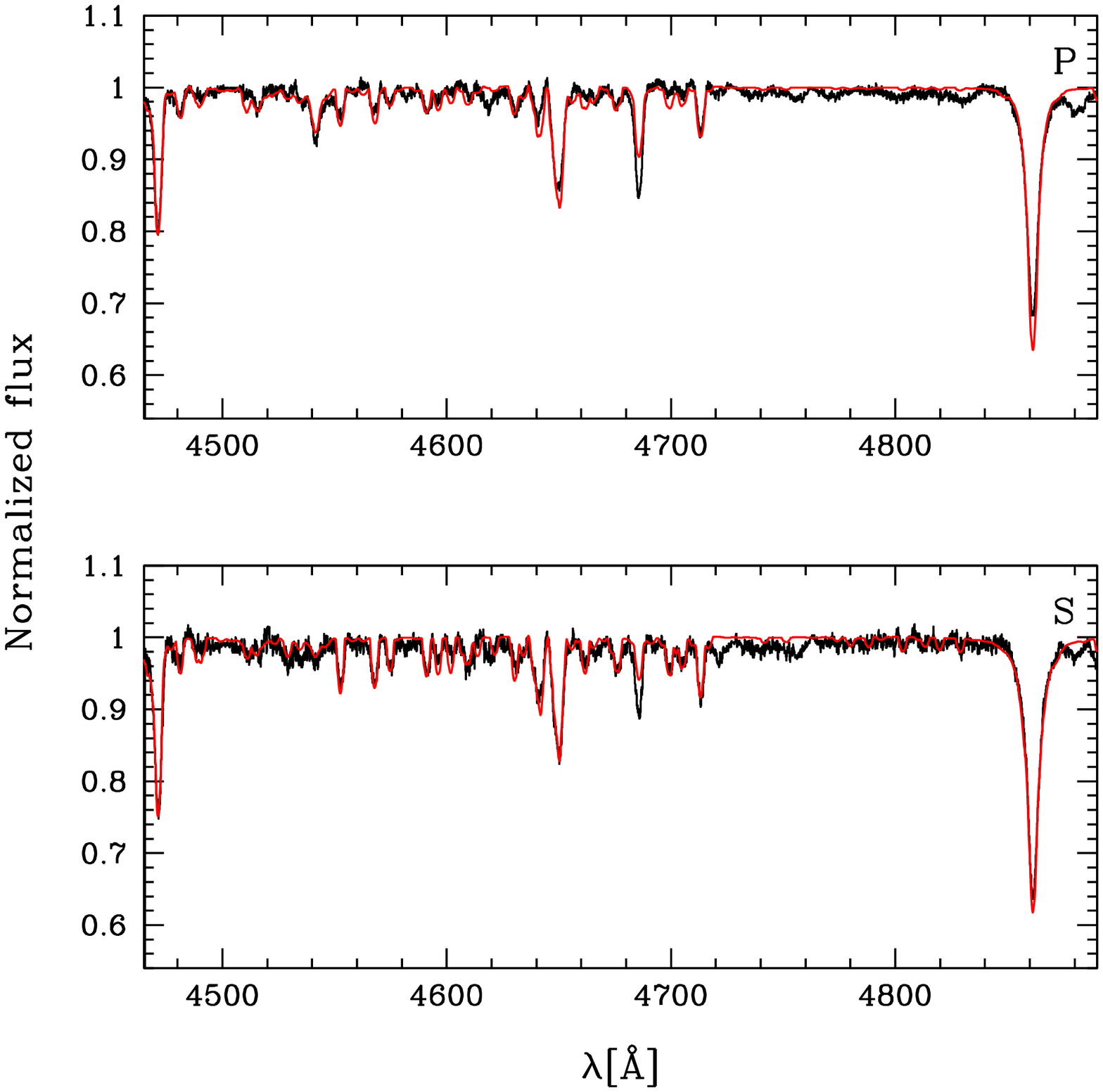}
\label{hd228989_cmfgen}
\caption{Best-fit model for HD\,228989 primary and secondary (red line) compared to disentangled spectra (black line).}
\end{figure*}
\newpage
\begin{figure*}[htbp]
\includegraphics[scale=0.9,bb=15 149 583 700,clip]{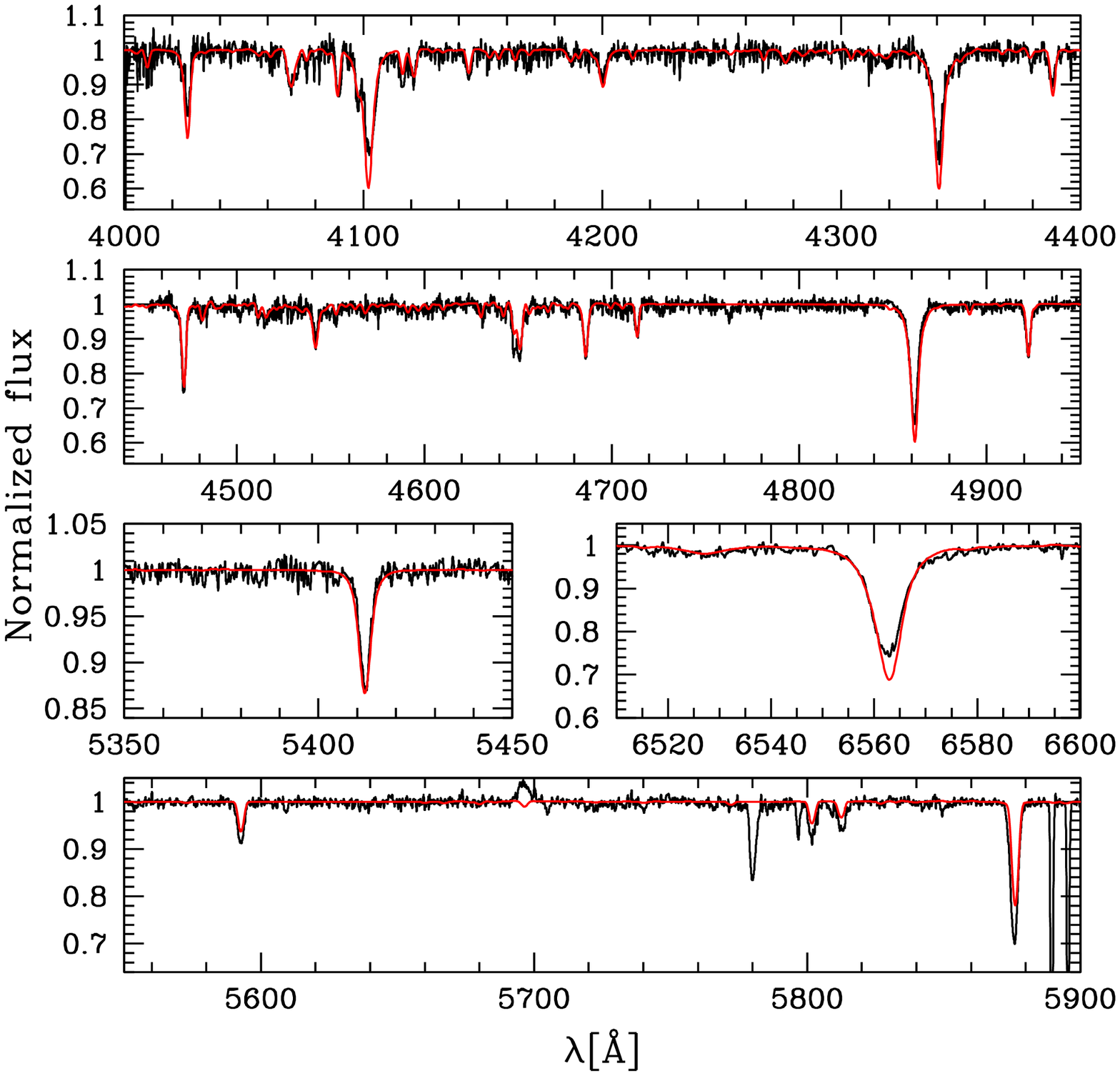}
\label{hd229234_cmfgen}
\caption{Best-fit model for HD\,229234 (red line) compared to Espresso spectrum (black line).}
\end{figure*}
\newpage
\begin{figure*}[htbp]
\includegraphics[scale=0.9,bb=23 149 583 700,clip]{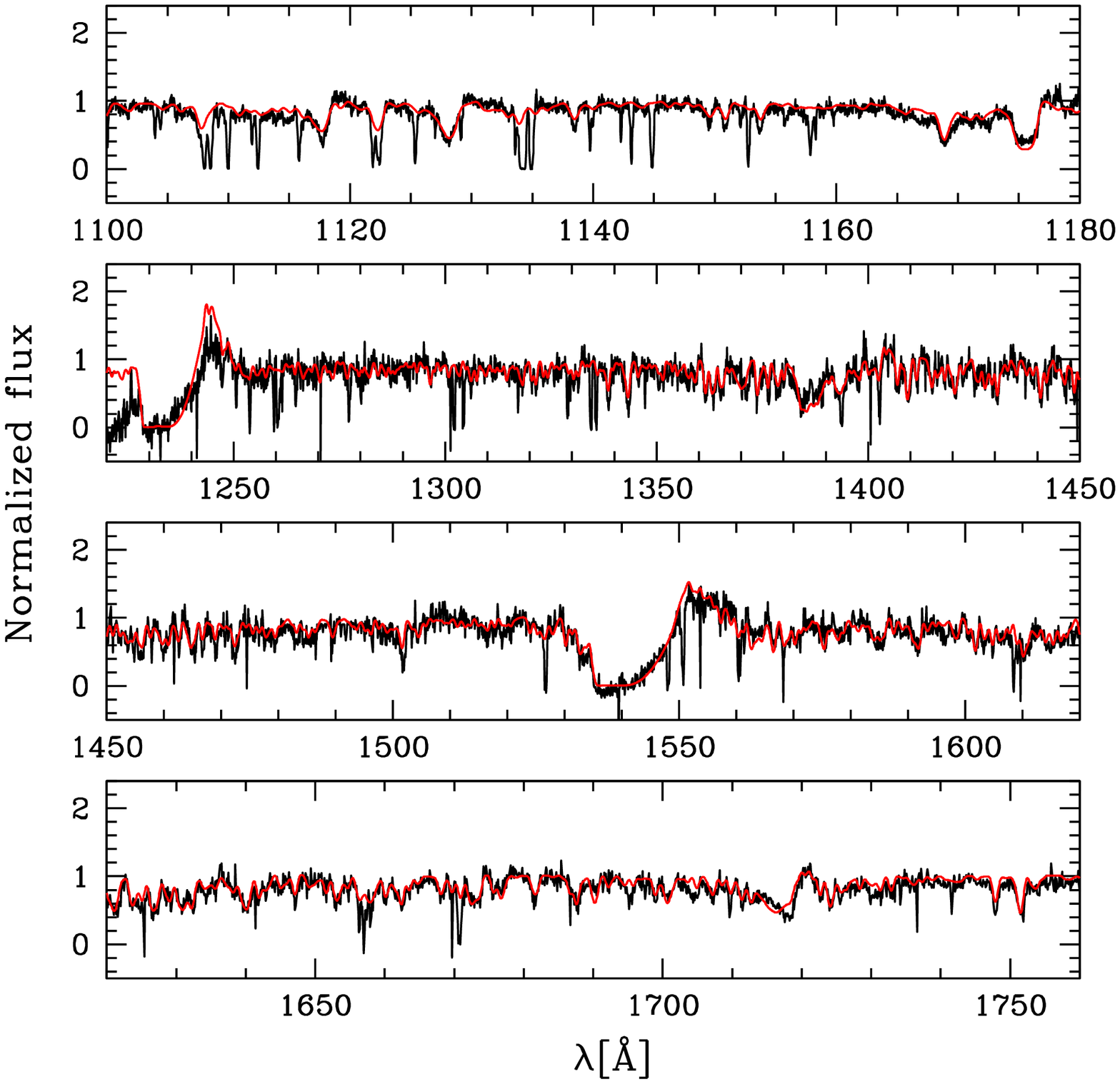}
\label{hd190864_cmfgen}
\caption{Best-fit model for HD\,190864 (red line) compared to FUSE and IUE spectra (black line).}
\end{figure*}
\begin{figure*}[htbp]
\includegraphics[scale=0.9,bb=23 149 583 700,clip]{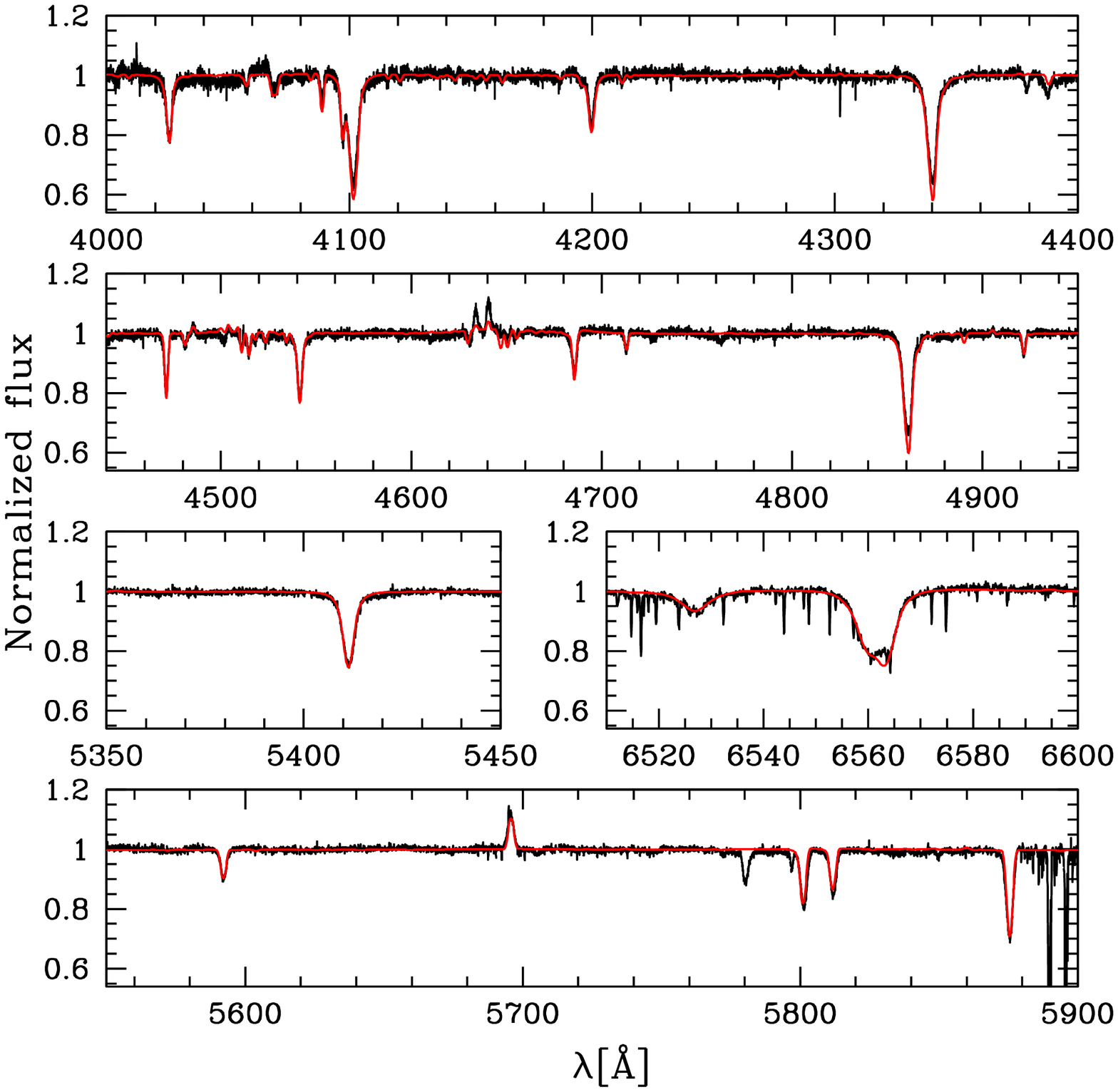}
\label{hd190864_cmfgen}
\caption{Best-fit model for HD\,190864 (red line) compared to Elodie spectrum (black line).}
\end{figure*}
\newpage
\begin{figure*}[htbp]
\subfigure[HD\,227018]{
\includegraphics[scale=0.9,bb=14 158 593 415,clip]{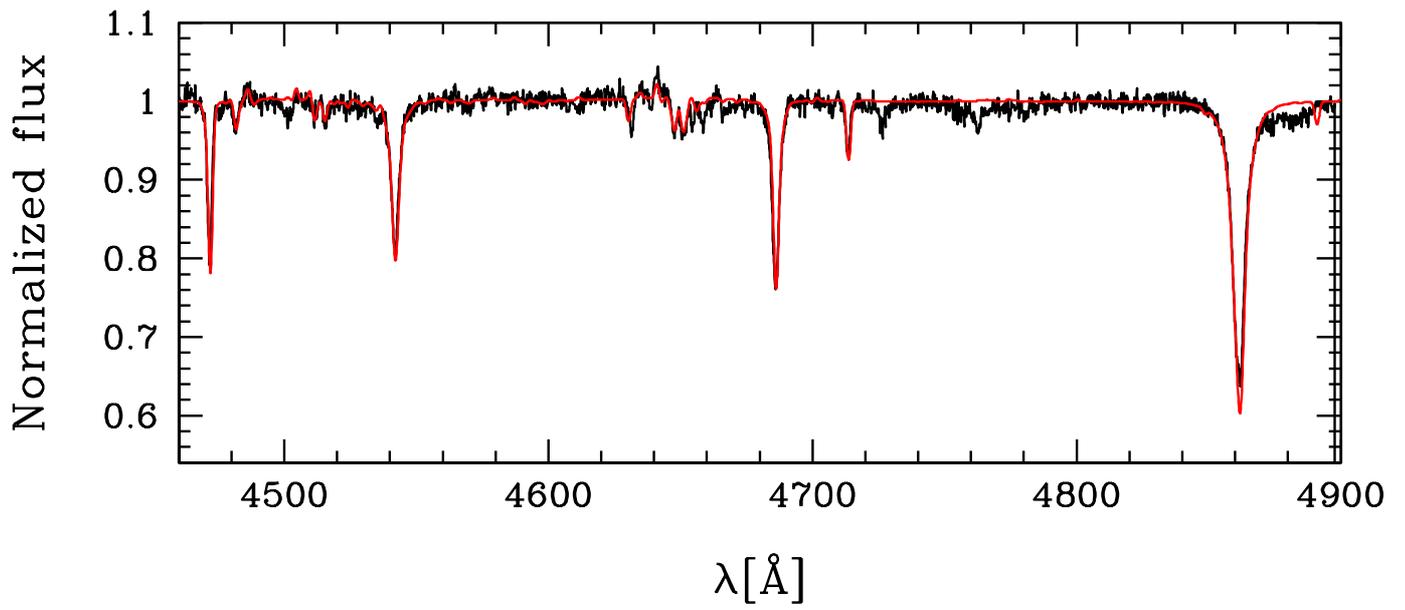}
\label{hd227018_cmfgen}}
\caption{Best-fit model for HD\,227018 (red line) compared to Aur{\'e}lie spectrum (black line).}
\end{figure*}
\newpage
\begin{figure*}[htbp]
\subfigure[HD\,227245]{
\includegraphics[scale=0.9,bb=33 158 583 700,clip]{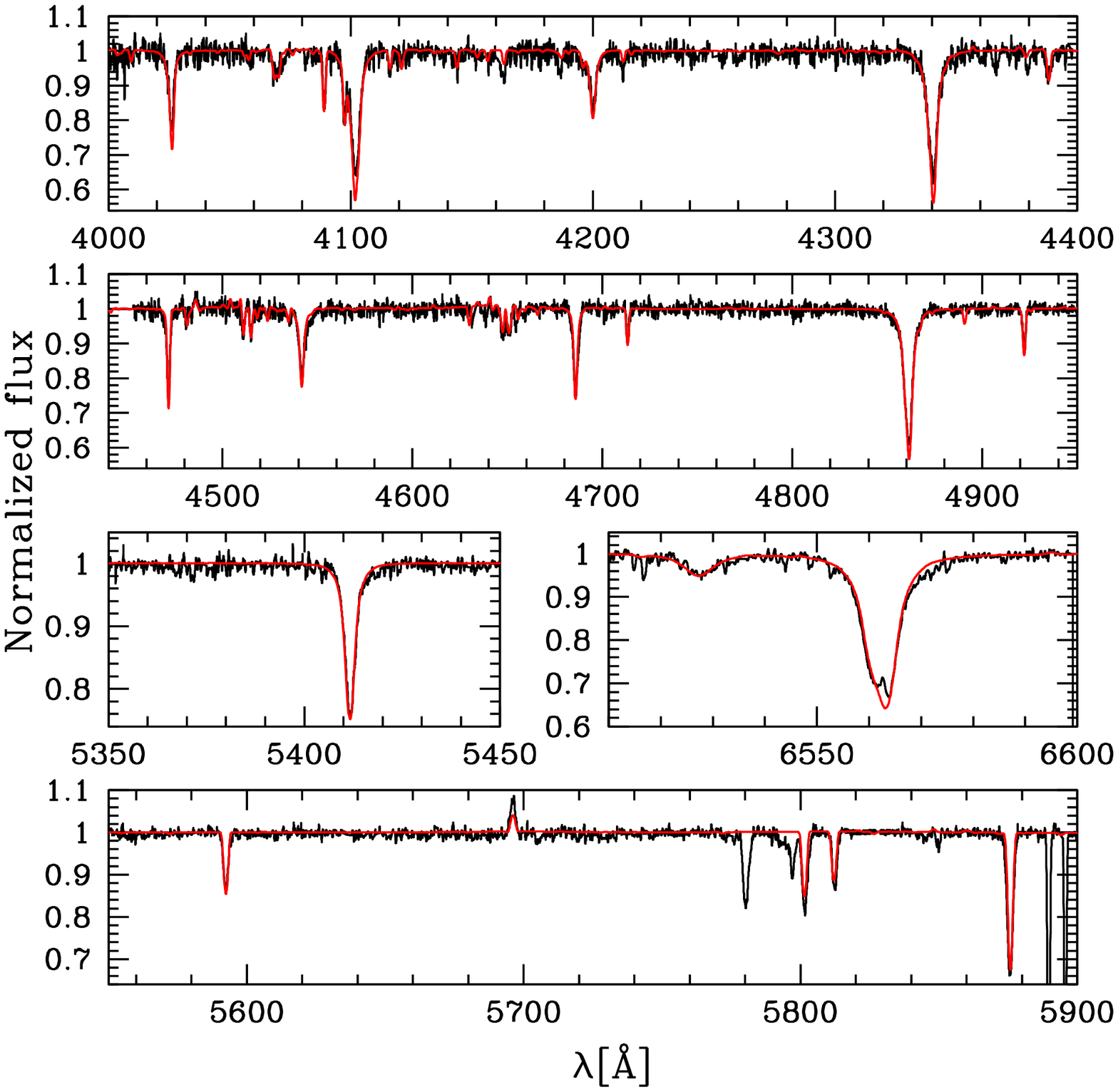}
\label{hd227245_cmfgen}}
\caption{Best-fit model for HD\,227245 (red line) compared to Espresso spectrum (black line).}
\end{figure*}
\newpage
\begin{figure*}[htbp]
\subfigure[HD\,227757]{
\includegraphics[scale=0.9,bb=14 158 583 700,clip]{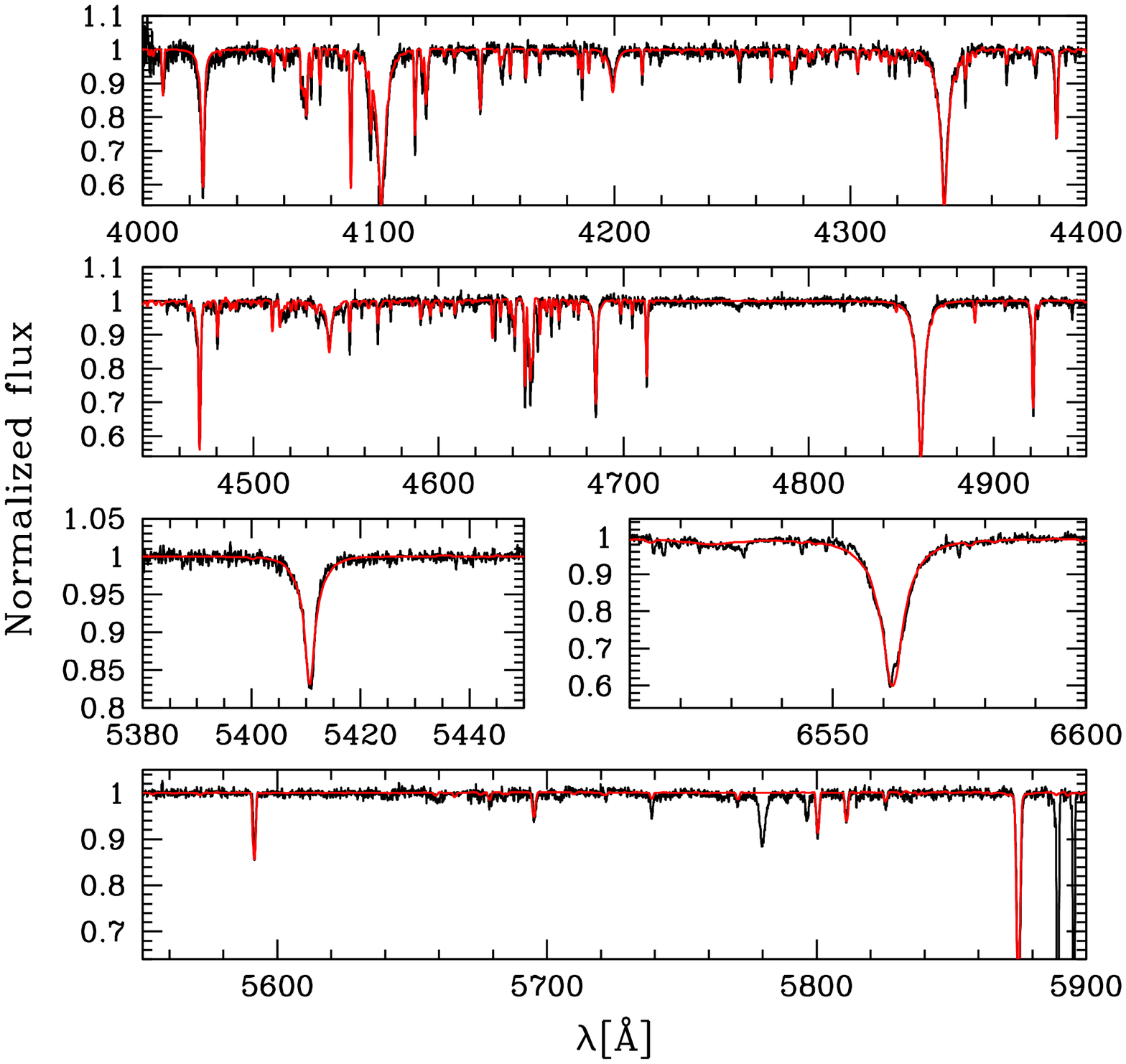}
\label{hd227757_cmfgen}}
\caption{Best-fit model for HD\,227757 (red line) compared to Espresso spectrum (black line).}
\end{figure*}
\newpage
\begin{figure*}[htbp]
\includegraphics[scale=0.9,bb=23 149 583 700,clip]{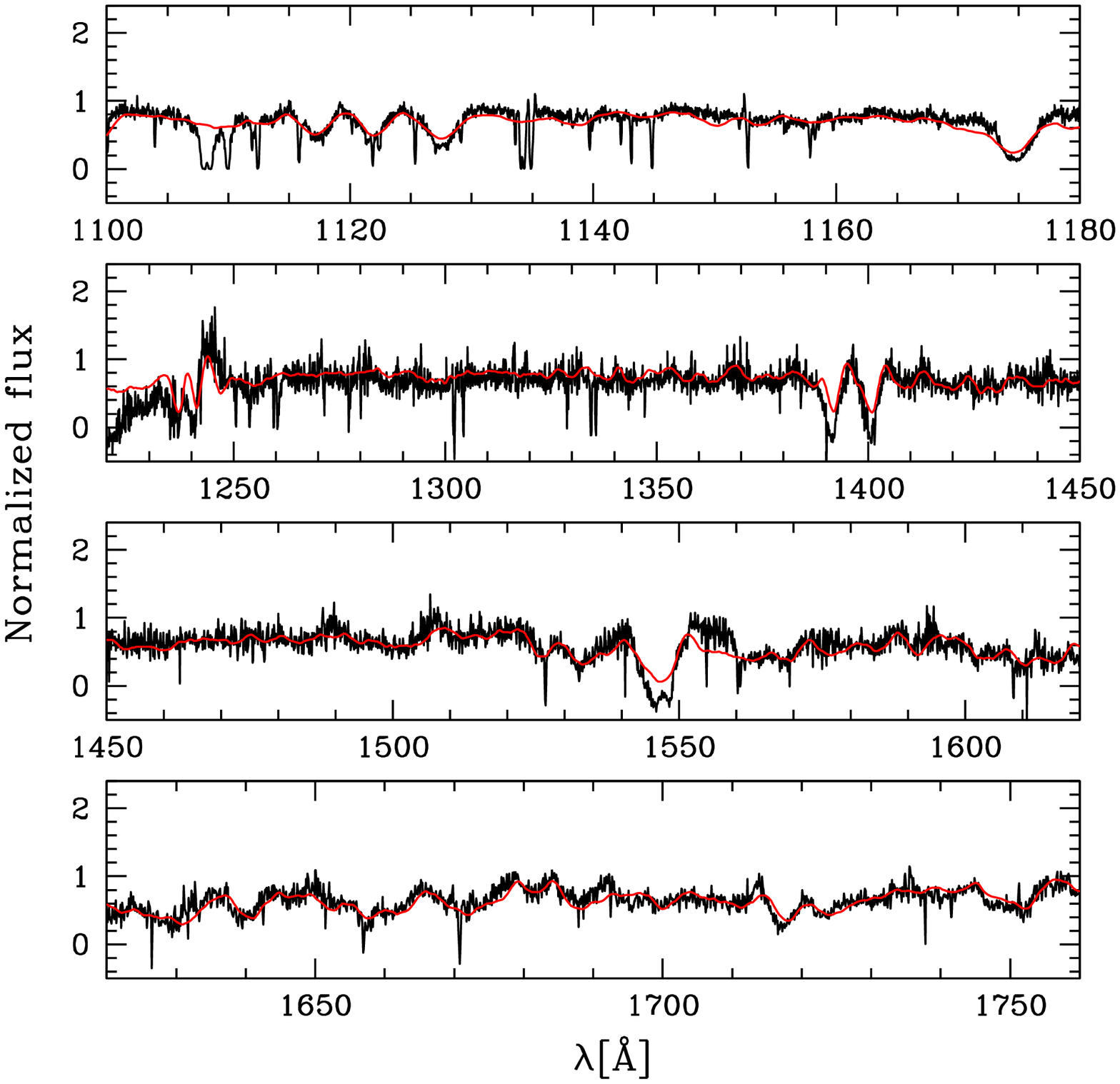}
\label{hd191423_cmfgen}
\caption{Best-fit model for HD\,191423 (red line) compared to FUSE and IUE spectra (black line).}
\end{figure*}
\begin{figure*}[htbp]
\includegraphics[scale=0.9,bb=23 149 583 700,clip]{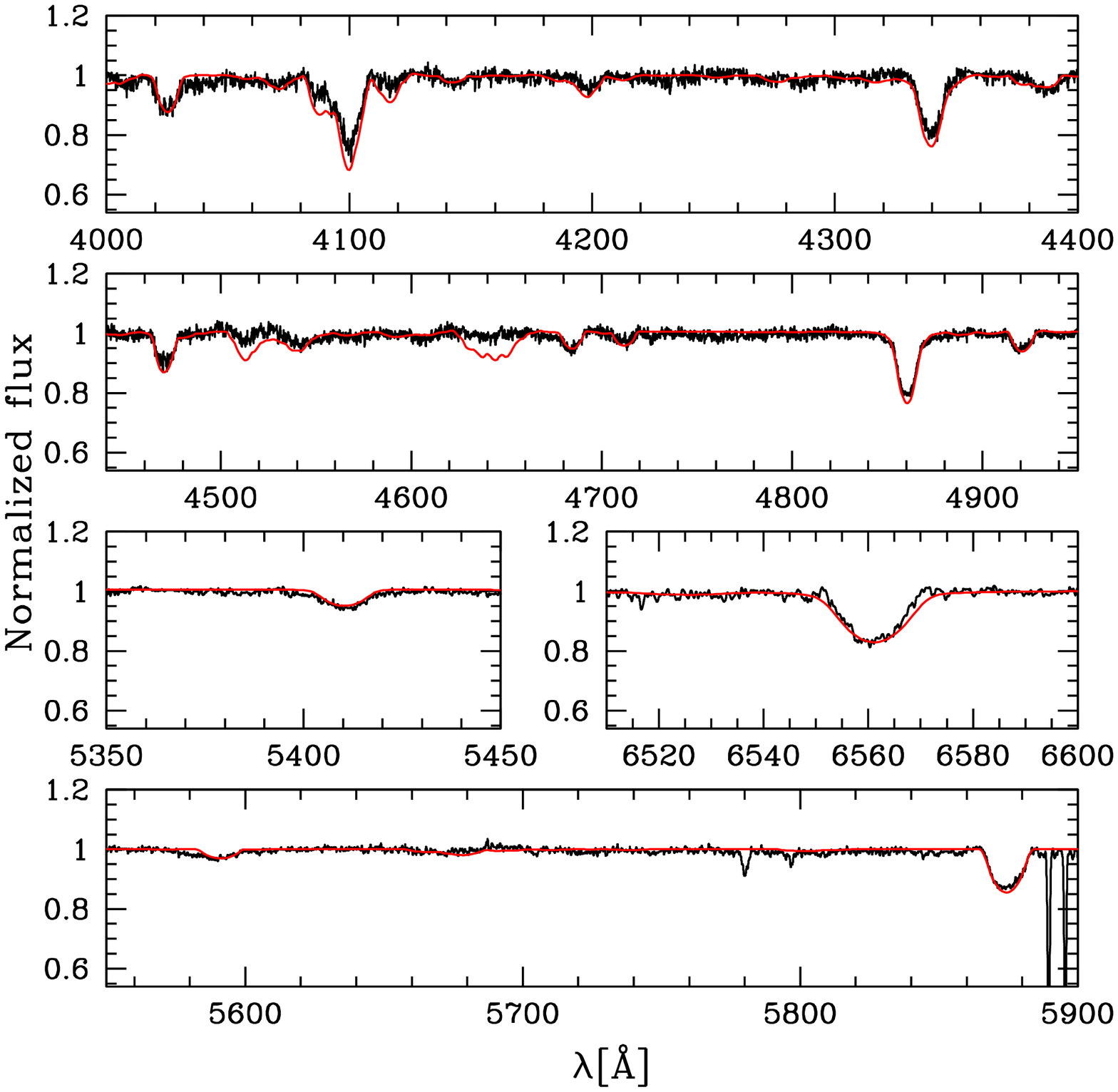}
\label{hd191423_cmfgen}
\caption{Best-fit model for HD\,191423 (red line) compared to Espresso spectrum (black line).}
\end{figure*}
\newpage
\begin{figure*}[htbp]
\includegraphics[scale=0.9,bb=23 149 583 700,clip]{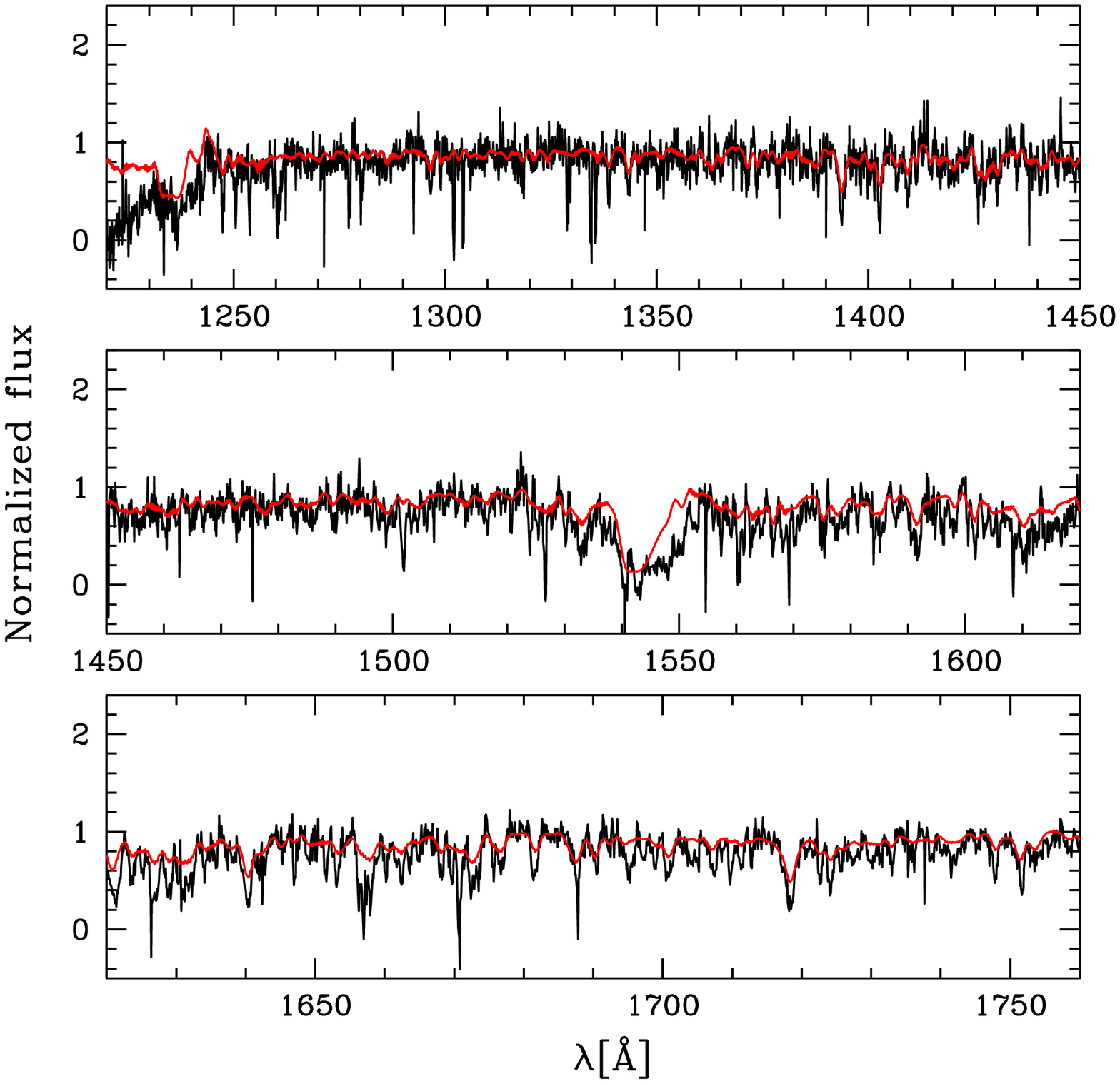}
\label{hd191978_cmfgen}
\caption{Best-fit model for HD\,191978 (red line) compared to IUE spectrum (black line).}
\end{figure*}
\begin{figure*}[htbp]
\includegraphics[scale=0.9,bb=23 149 583 700,clip]{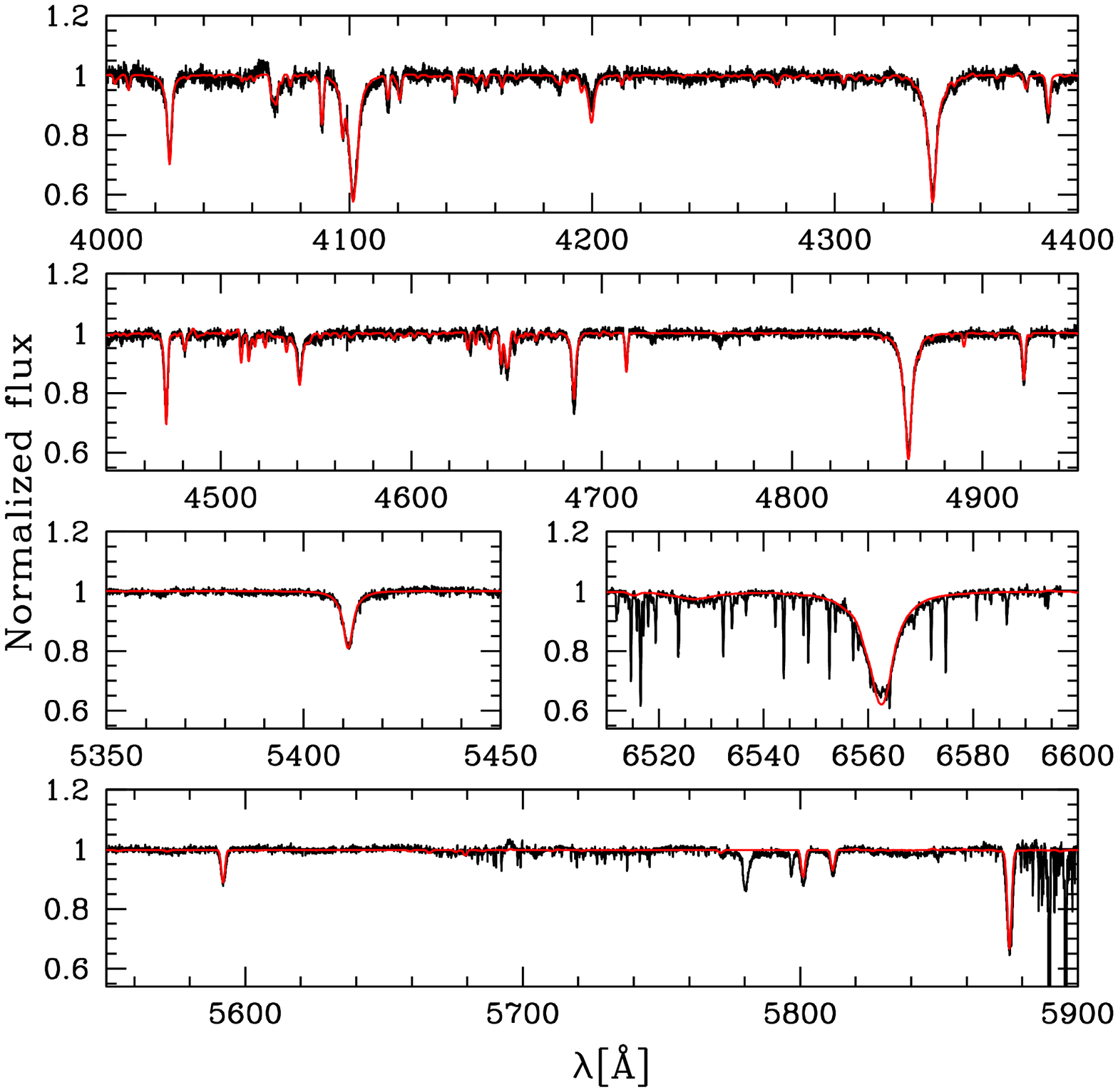}
\label{hd191978_cmfgen}
\caption{Best-fit model for HD\,191978 (red line) compared to Elodie spectrum (black line).}
\end{figure*}
\newpage
\begin{figure*}[htbp]
\includegraphics[scale=0.9,bb=15 158 583 700,clip]{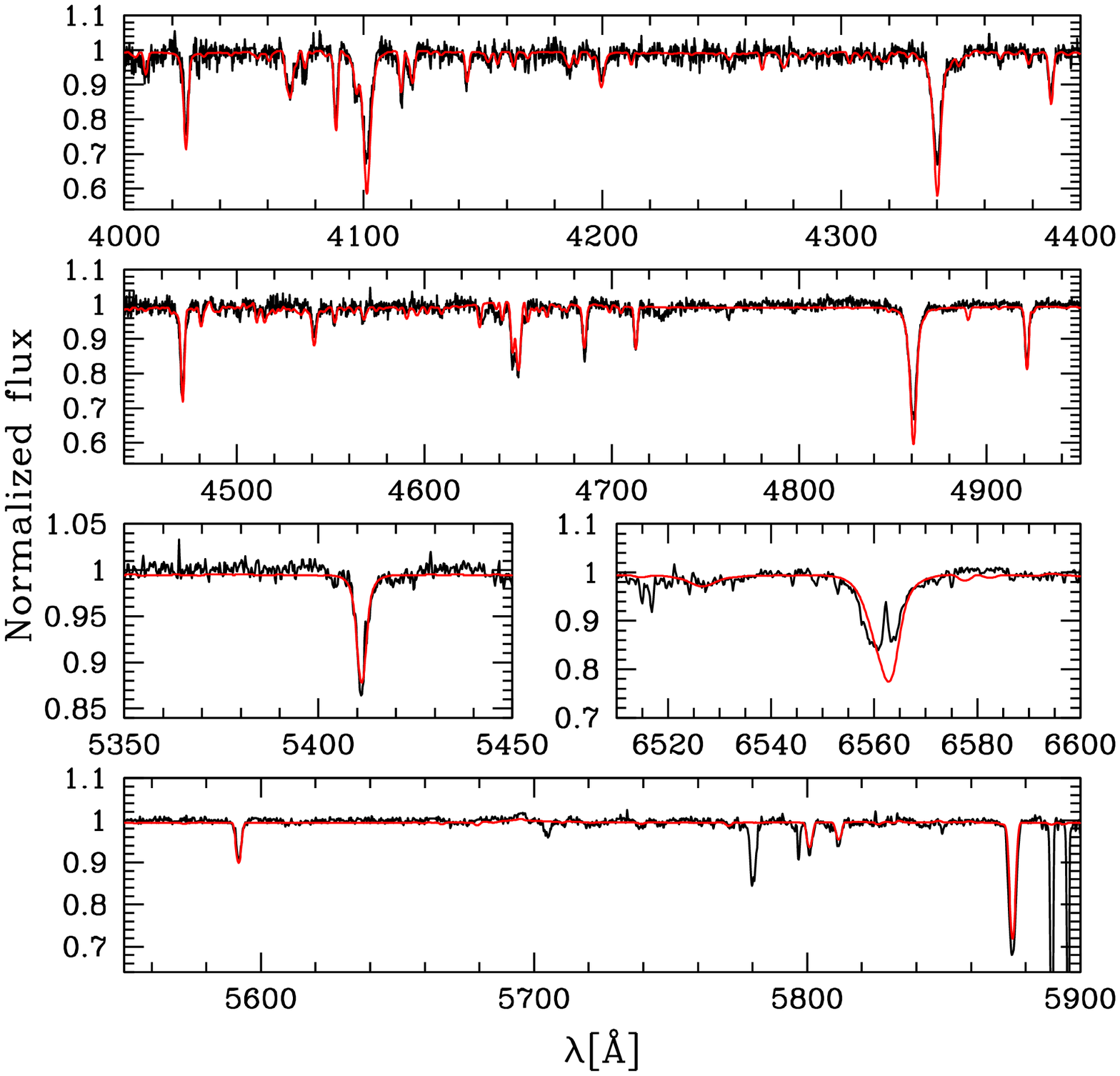}
\label{hd193117_cmfgen}
\caption{Best-fit model for HD\,193117 (red line) compared to Espresso spectrum (black line).}
\end{figure*}
\clearpage
\newpage
\begin{figure*}[htbp]
\includegraphics[scale=0.9,bb=33 158 583 700,clip]{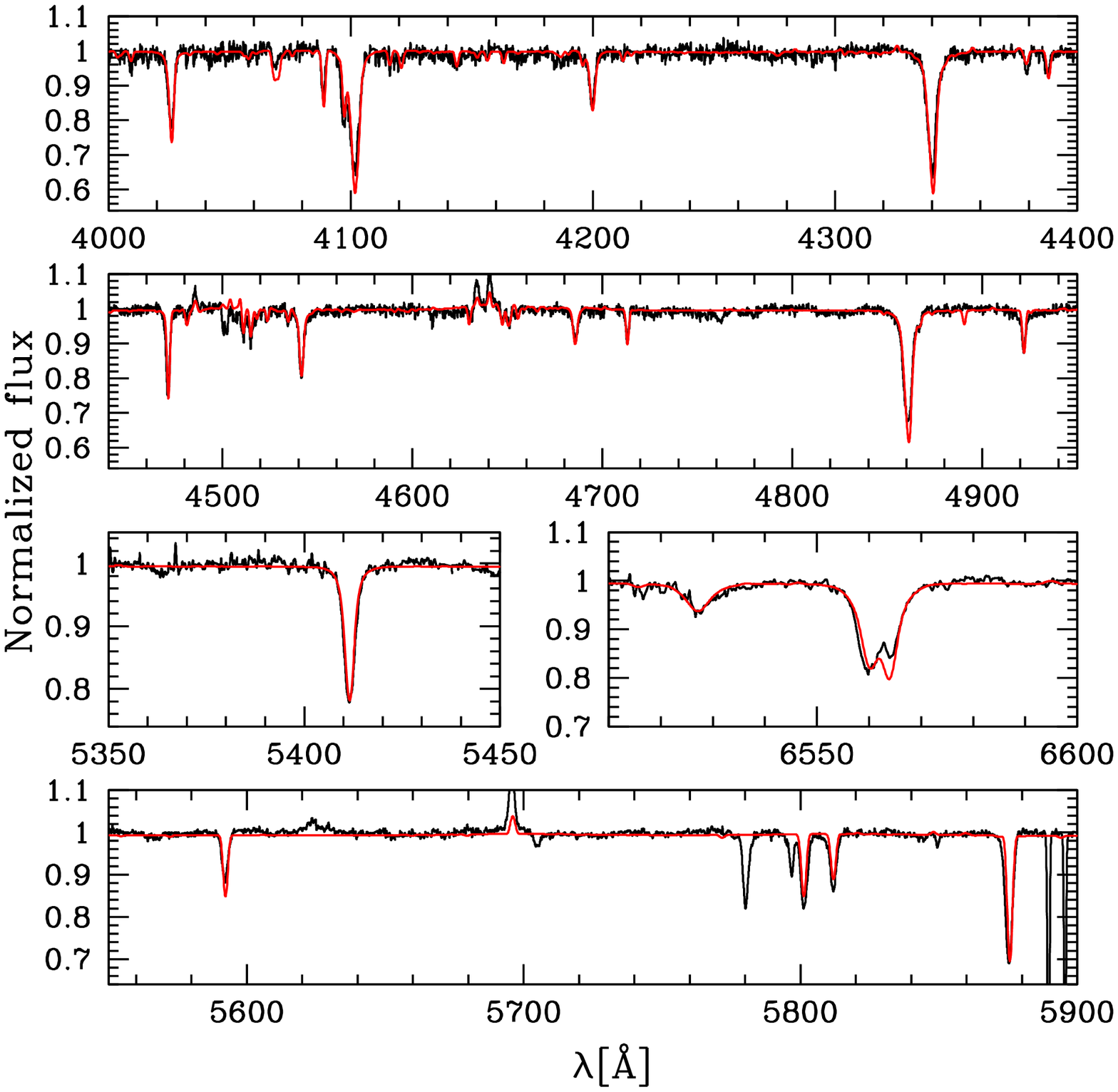}
\label{hd194334_cmfgen}
\caption{Best-fit model for HD\,194334 (red line) compared to Espresso spectrum (black line).}
\end{figure*}
\newpage
\begin{figure*}[htbp]
\includegraphics[scale=0.9,bb=14 149 572 700,clip]{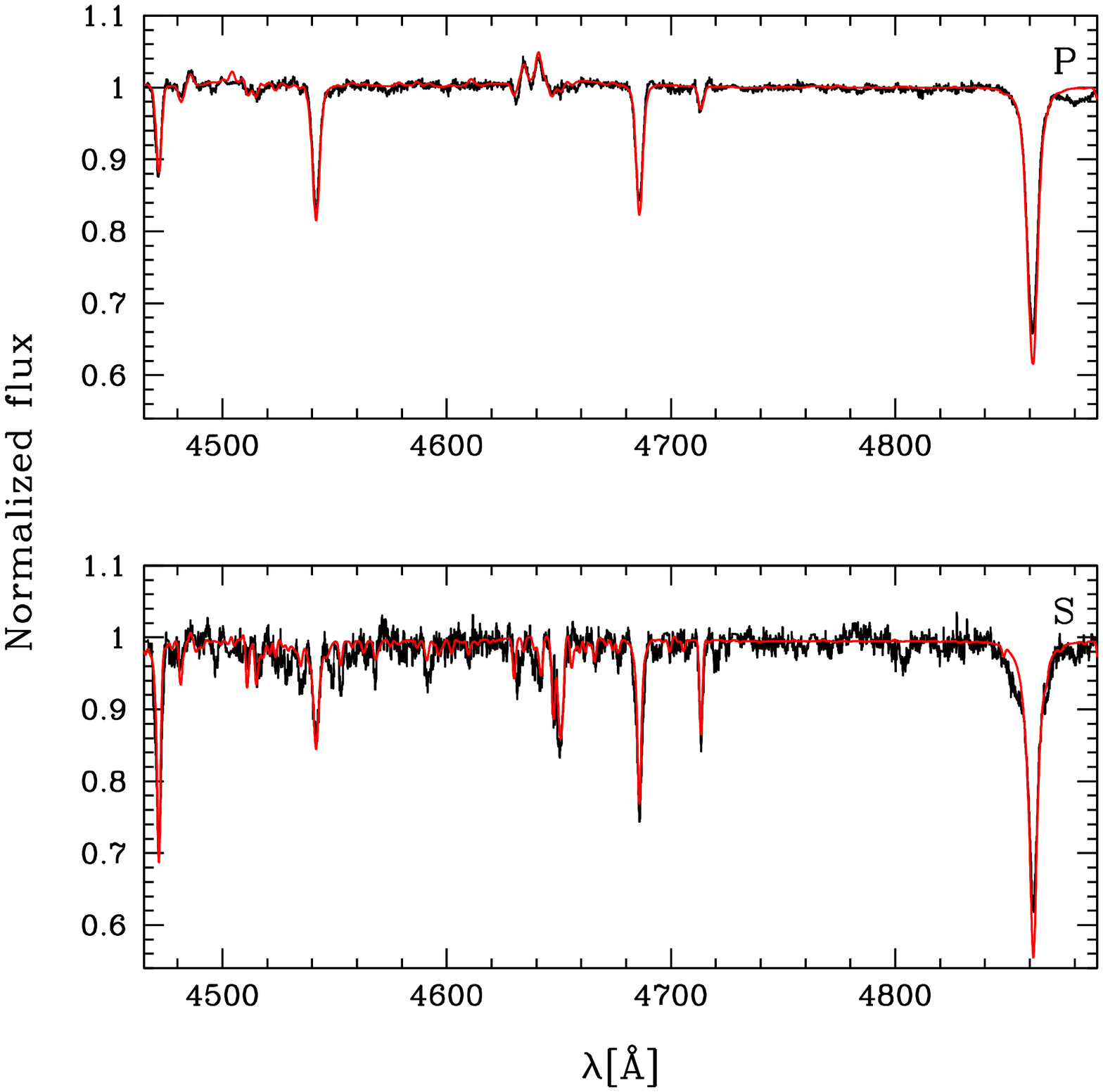}
\label{hd194649_cmfgen}
\caption{Best-fit model for HD\,194649 primary and secondary (red line) compared to disentangled spectra (black line).}
\end{figure*}
\newpage
\begin{figure*}[htbp]
\centering
\includegraphics[scale=0.9,bb=15 158 583 700,clip]{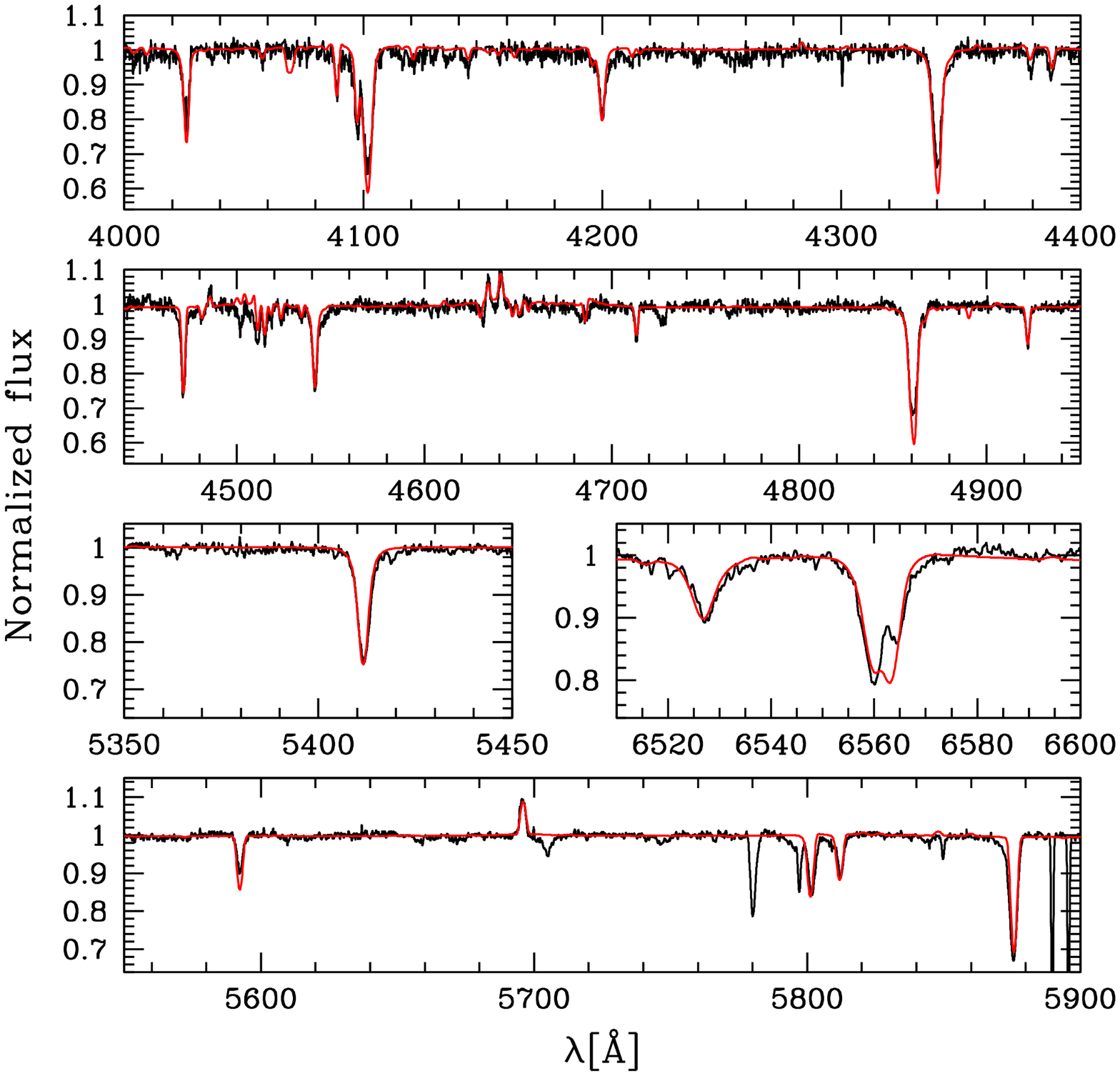}
\label{hd195213_cmfgen}
\caption{Best-fit model for HD\,195213 (red line) compared to Espresso spectrum (black line).}
\end{figure*}
\end{document}